%% file: note2737.tex
\RequirePackage[switch]{lineno_babar_revtex}
\documentclass[twocolumn,showpacs,aps,floatfix,prd,nofootinbib,letterpaper]{revtex4}
\catcode`\@=11\relax
\def\@makecol{%
 \setbox\@outputbox\vbox{%
  \boxmaxdepth\@maxdepth
 \protected@write\@auxout{}{%
 \string\@LN@col{\@ifnum{\pagegrid@cur=\@ne}{1}{2}}
      }%
  \@tempdima\dp\@cclv
  \unvbox\@cclv
  \vskip-\@tempdima
 }%
 \xdef\@freelist{\@freelist\@midlist}\global\let\@midlist\@empty
 \@combinefloats
 \@combineinserts\@outputbox\footins
  \set@adj@colht\dimen@
  \count@\vbadness
  \vbadness\@M
  \setbox\@outputbox\vbox to\dimen@{%
   \@texttop
   \dimen@\dp\@outputbox
   \unvbox\@outputbox
   \vskip-\dimen@
   \@textbottom
  }%
  \vbadness\count@
 \global\maxdepth\@maxdepth
}%
\def\balance@two#1#2{%
\outputdebug@sw{{\tracingall\scrollmode\showbox#1\showbox#2}}{}%
 \setbox\@ne\vbox{%
  \@ifvoid#1{}{%
   \unvcopy#1\recover@footins
   \@ifvoid#2{}{\marry@baselines}%
  }%
  \@ifvoid#2{}{%
   \unvcopy#2\recover@footins
  }%
 }%
 \dimen@\ht\@ne\divide\dimen@\tw@
 \dimen@i\dimen@
 \vbadness\@M
 \vfuzz\maxdimen
 \loopwhile{%
  \dimen@i=.5\dimen@i
  \outputdebug@sw{\saythe\dimen@\saythe\dimen@i\saythe\dimen@ii}{}%
  \setbox\z@\copy\@ne\setbox\tw@\vsplit\z@ to\dimen@
  \setbox\z@ \vbox{%
 \protected@write\@auxout{}{%
 \string\@LN@col{\@ifnum{\pagegrid@cur=\@ne}{1}{2}}
      }%
   \unvcopy\z@
   \setbox\z@\vbox{\unvbox\z@ \setbox\z@\lastbox\aftergroup\vskip\aftergroup-\expandafter}\the\dp\z@\relax
  }%
  \setbox\tw@\vbox{%
   \unvcopy\tw@
   \setbox\z@\vbox{\unvbox\tw@\setbox\z@\lastbox\aftergroup\vskip\aftergroup-\expandafter}\the\dp\z@\relax
  }%
  \dimen@ii\ht\tw@\advance\dimen@ii-\ht\z@
  \@ifdim{\dimen@i>.5\p@}{%
   \advance\dimen@\@ifdim{\dimen@ii<\z@}{}{-}\dimen@i
   \true@sw
  }{%
   \@ifdim{\dimen@ii<\z@}{%
    \advance\dimen@\tw@\dimen@i
    \true@sw
   }{%
    \false@sw
   }%
  }%
 }%
 \outputdebug@sw{\saythe\dimen@\saythe\dimen@i\saythe\dimen@ii}{}%
\@ifdim{\ht\z@=\z@}{%
\@ifdim{\ht\tw@=\z@}{%
\true@sw
}{%
\false@sw
}%
}{%
\true@sw
}%
{%
}{%
\ltxgrid@info{Unsatifactorily balanced columns: giving up}%
\setbox\tw@\box#1%
\setbox\z@ \box#2%
}%
 \setbox\tw@\vbox{\unvbox\tw@\vskip\z@skip}%
 \setbox\z@ \vbox{\unvbox\z@ \vskip\z@skip}%
 \set@colroom
\dimen@\ht\z@\@ifdim{\dimen@<\ht\tw@}{\dimen@\ht\tw@}{}%
\@ifdim{\dimen@>\@colroom}{\dimen@\@colroom}{}%
 \outputdebug@sw{\saythe{\ht\z@}\saythe{\ht\tw@}\saythe\@colroom\saythe\dimen@}{}%
\setbox#1\vbox to\dimen@{\unvbox\tw@\unskip\raggedcolumn@skip}%
\setbox#2\vbox to\dimen@{\unvbox\z@ \unskip\raggedcolumn@skip}%
\outputdebug@sw{{\tracingall\scrollmode\showbox#1\showbox#2}}{}%
}%
\catcode`\@=12\relax

\usepackage{graphicx}
\usepackage{dcolumn}
\usepackage{array, longtable}
\usepackage{epsfig}
\usepackage{amsmath}
\usepackage{colordvi}
\usepackage{hhline}
\usepackage{color}

\newcommand{\BaBarYear}{20}
\newcommand{\BaBarNumber}{004}
\newcommand{\SLACPubNumber}{17587}

 \newcommand{\BaBarType}      {PUB}  

\input babarsym

\def\Ecm          {\ensuremath {E_{\rm c.m.}}\xspace}

\def\mgg  {\ensuremath {m(\gamma\gamma)}\xspace}

\long\def\inst#1{\par\nobreak\kern 4pt\nobreak
    {\it #1}\par\vskip 10pt plus 3pt minus 3pt}

\begin{document}

\begin{flushleft}
\babar-\BaBarType-\BaBarYear/\BaBarNumber \\
SLAC-PUB-\SLACPubNumber \\
\end{flushleft}


\title{\large \bf
\boldmath
Study of the reactions  $\epem\to2(\pipi)\pi^0\pi^0\pi^0$ and
$2(\pipi)\pi^0\pi^0\eta$  at
 center-of-mass energies from threshold to 4.5 GeV using initial-state
radiation
} 

\input 
authors_jan2021_frozen


\begin{abstract}
We study the processes $\epem\to
2(\pipi)\ppz\piz\gamma$ and $2(\pipi)\ppz\eta\gamma$ in which an energetic
photon is radiated from the initial state. 
The data were collected with the \babar~ detector at SLAC.
 About 14\,000 and 4700 events, respectively, are
selected from a data sample corresponding to an integrated
luminosity of 469~\invfb.
The invariant mass of the hadronic final state defines the effective \epem
center-of-mass energy.
The center-of-mass energies range from threshold to 4.5~\gev.
From the mass spectra, 
the first ever measurements of
the $\epem\to 2(\pipi)\ppz\piz$ and  the $\epem\to2(\pipi)\ppz\eta$  cross sections are  performed.
The contributions from  $\omega\pipi\ppz$, $\eta2(\pipi)$, and other
intermediate  states are presented. 
We observe the $J/\psi$ and $\psi(2S)$ in most of these final states and
measure the
corresponding branching fractions, many of them for the first time.
\end{abstract}


\vfill
\newpage
\maketitle

\setcounter{footnote}{0}


\section{Introduction}
\label{sec:Introduction}

 The Standard Model (SM) calculation of the muon anomalous
magnetic moment ($g_\mu-2$) requires input
from experimental \epem hadronic cross section
data in order to account for hadronic vacuum polarization (HVP) terms.
In particular, the calculation is most sensitive to the low-energy
region, from the hadronic threshold to about 2 GeV,
where the inclusive hadronic cross section cannot be measured
reliably and a sum of exclusive states must be used.
Despite the large data set accumulated in the past years and the
analysis studies performed, there is still a $\sim$3.5 sigma discrepancy
between the SM calculation  and the experimental
value ~\cite{dehz}. 
Not all exclusive states have yet been measured, 
and new measurements will improve the reliability of the calculation.
Finally, these studies provide information on the resonant spectroscopy.

Electron-positron annihilation events with initial-state radiation
(ISR) are useful to study processes over a wide range of energies
below the nominal \epem center-of-mass (c.m.) energy (\Ecm),
as proposed in Ref.~\cite{baier}.
Studies of the ISR processes $e^+e^-\to\mumu\gamma$~\cite{Druzhinin1,isr2pi}
and $\epem \to X_h\gamma$,  
using data from the \babar\ experiment at SLAC,
have been previously reported.
Here $X_h$ represents any of several exclusive hadronic final states.
The $X_h$ studied to date include:
charged hadron pairs $\pip\pim$~\cite{isr2pi}, $\Kp\Km$~\cite{isr2k}, and
$p\overline{p}$~\cite{isr2p};
four or six charged mesons~\cite{isr4pi,isr2k2pi,isr6pi};
charged mesons plus one or two or three \piz
mesons~\cite{isr2k2pi,isr6pi,isr3pi,isr5pi,isr2pi2pi0,isr2pi3pi0}; 
a \KS meson plus charged and neutral mesons~\cite{isrkkpi}; 
and channels with  \KL mesons~\cite{isrkskl}. 

In this paper, we report the first measurements of the
$2(\pipi)3\piz$ and $2(\pipi)2\piz\eta$ channels.
The final states are produced in conjunction with
a hard photon, assumed  to result from ISR.
To reduce background from $\Upsilon(4S)$ decays, the analysis is restricted to
the c.m.\@ energy below 4.5~\gev. 
As part of the analysis, we search for and observe
intermediate states, including the $\eta$,
$\omega$, and $\rho$  resonances. In the charmonium region, we observe
$J/\psi$  and $\psi(2S)$ signals in the studied final states and the
corresponding  branching fractions are measured.


%

\section{\boldmath The \babar\ detector and dataset}
\label{sec:babar}

The data used in this analysis were collected with the \babar\ detector at
the \pep2\ asymmetric-energy \epem\ storage ring. 
The total integrated luminosity used is 468.6~\invfb~\cite{lumi}, 
which includes data collected at the $\Upsilon(4S)$
resonance (424.7~\invfb) and at a c.m.\ energy 40~\mev below this
resonance (43.9~\invfb).

The \babar\ detector is described in detail elsewhere~\cite{babar}. 
Charged particles are reconstructed using a \babar\ tracking system,
which is comprised of a silicon vertex tracker (SVT) and a drift
chamber (DCH), both located 
inside a 1.5 T solenoid.
Separation of pions and kaons is accomplished by means of a detector of
internally reflected Cherenkov light (DIRC) and energy-loss measurements in
the SVT and DCH. 
Photons  are detected in an electromagnetic calorimeter (EMC).  
Muon identification is provided by an instrumented flux return.

To evaluate the detector acceptance and efficiency, 
we have developed a special package of Monte Carlo (MC) simulation programs for
radiative processes based on 
the approach of K\"uhn  and Czy\.z~\cite{kuehn2}.  
Multiple collinear soft-photon emission from the initial \epem state 
is implemented with a structure function technique~\cite{kuraev,strfun}, 
while additional photon radiation from final-state particles is
simulated using the PHOTOS package~\cite{PHOTOS}.  
The precision of the radiative simulation is such that it contributes less than 1\% to
the uncertainty in the measured hadronic cross sections.

To evaluate the detection efficiency we simulate $\epem\to
2(\pipi)\ppz\piz\gamma$ events assuming production 
through the $\omega(782)\piz\eta$ and $\pipi\ppz\eta$ intermediate channels,
with decay of  the $\omega$ to three pions and
decay of the $\eta$ to all  its measured decay modes~\cite{PDG}, from
which decays to three pions are used in present analysis.

A sample of 100-200k simulated events is generated  for each
signal reaction  and
processed through the detector response simulation, based on the GEANT4 package~\cite{GEANT4}. These
events are reconstructed using
the same software chain as the data. Variations in the detector
conditions are taken into account.
The simulation includes random trigger events to account for
the observed distributions of
the background tracks and photons.
Most of the experimental events contain additional soft photons due to machine background
or interactions in the detector material, which are properly modeled
in the simulation.

For the purpose of background estimation,  large samples of events from the
main relevant ISR processes ($4\pi\gamma$, $5\pi\gamma$, $\omega\eta\gamma$,
and  $2(\pipi)\ppz\gamma$) are simulated.  
The background from the relevant
 non-ISR processes, namely $\epem\to\qqbar$ $(q=u, d, s)$ and
 $\epem\to\tau^+\tau^-$,  are generated  using the \textsc{jetset}~\cite{jetset}
and \textsc{koralb}~\cite{koralb} programs,
respectively.
The cross sections for the above processes
are  known with an accuracy about or better than
10\%, which is sufficient for the present purpose.

\begin{figure}[b]
\begin{center}
\includegraphics[width=0.95\linewidth]{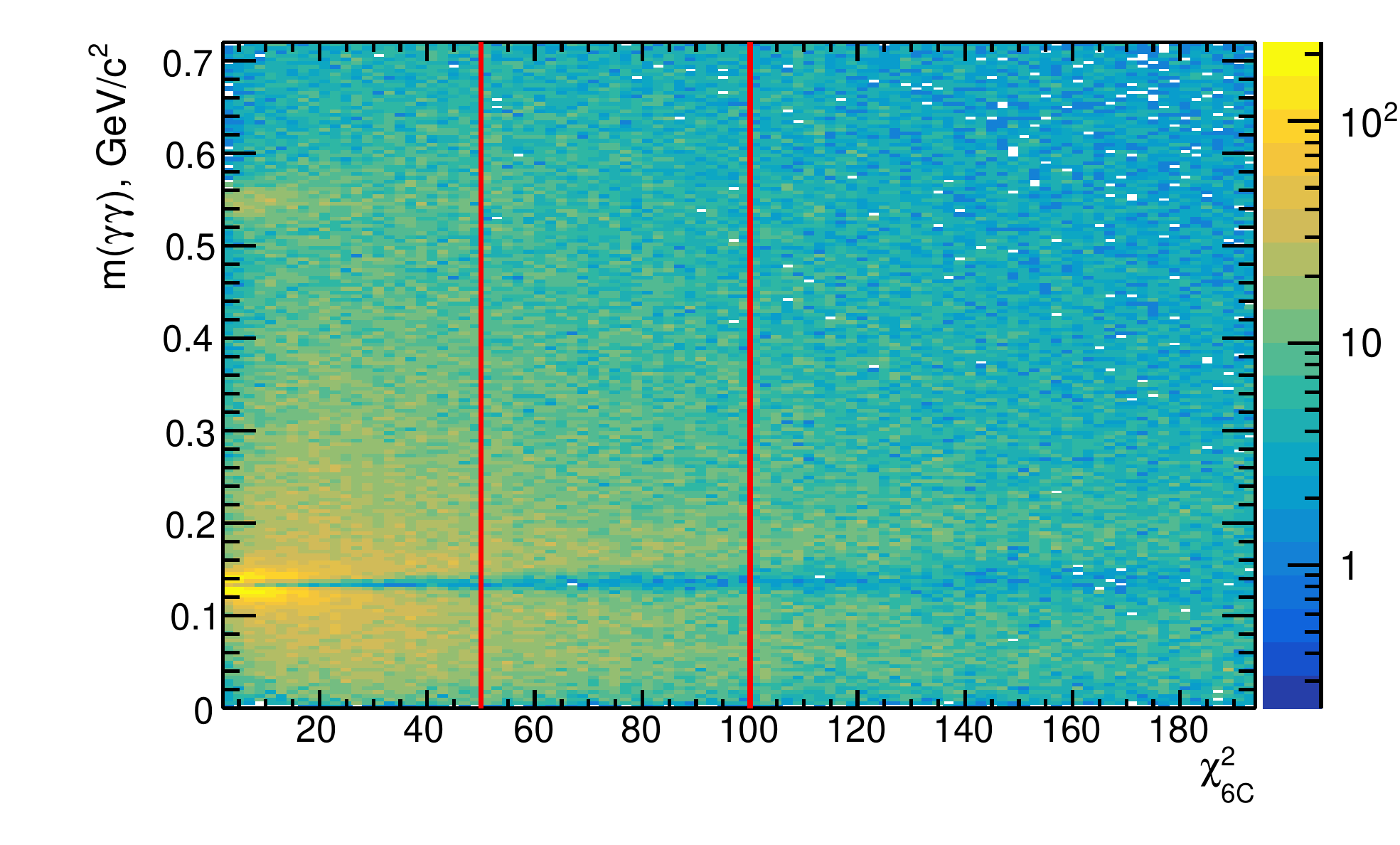}
\put(-50,100){\makebox(0,0)[lb]{\textcolor{white}{\bf (a)}}}\\
\vspace{-0.3cm}
\includegraphics[width=0.95\linewidth]{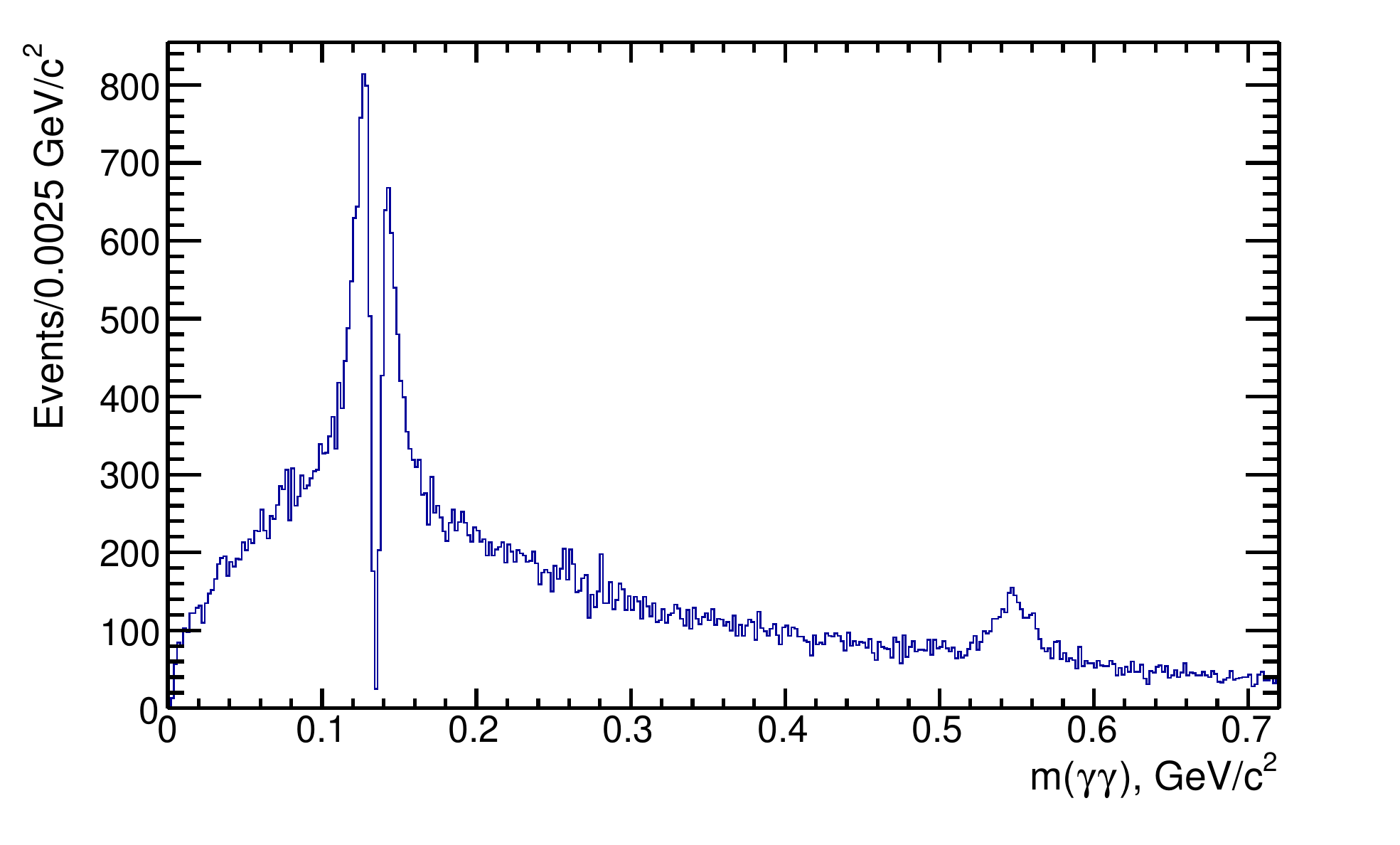}
\put(-50,100){\makebox(0,0)[lb]{\bf(b)}}
\vspace{-0.5cm}
\caption{(a) Distribution of the invariant mass $\mgg$ of the third photon pair  vs
  $\chisq_{4\pi2\piz\gamma\gamma}$. The lines define the boundaries of the
  signal and control regions.
(b) Distribution of $\mgg$  in the signal region 
$\chisq_{4\pi2\piz\gamma\gamma} <50$  with the additional selection
criteria  described in the text. 
}
\label{4pi3pi0_chi2_all}
\end{center}
\end{figure}
\begin{figure}[b]
\begin{center}
\includegraphics[width=0.95\linewidth]{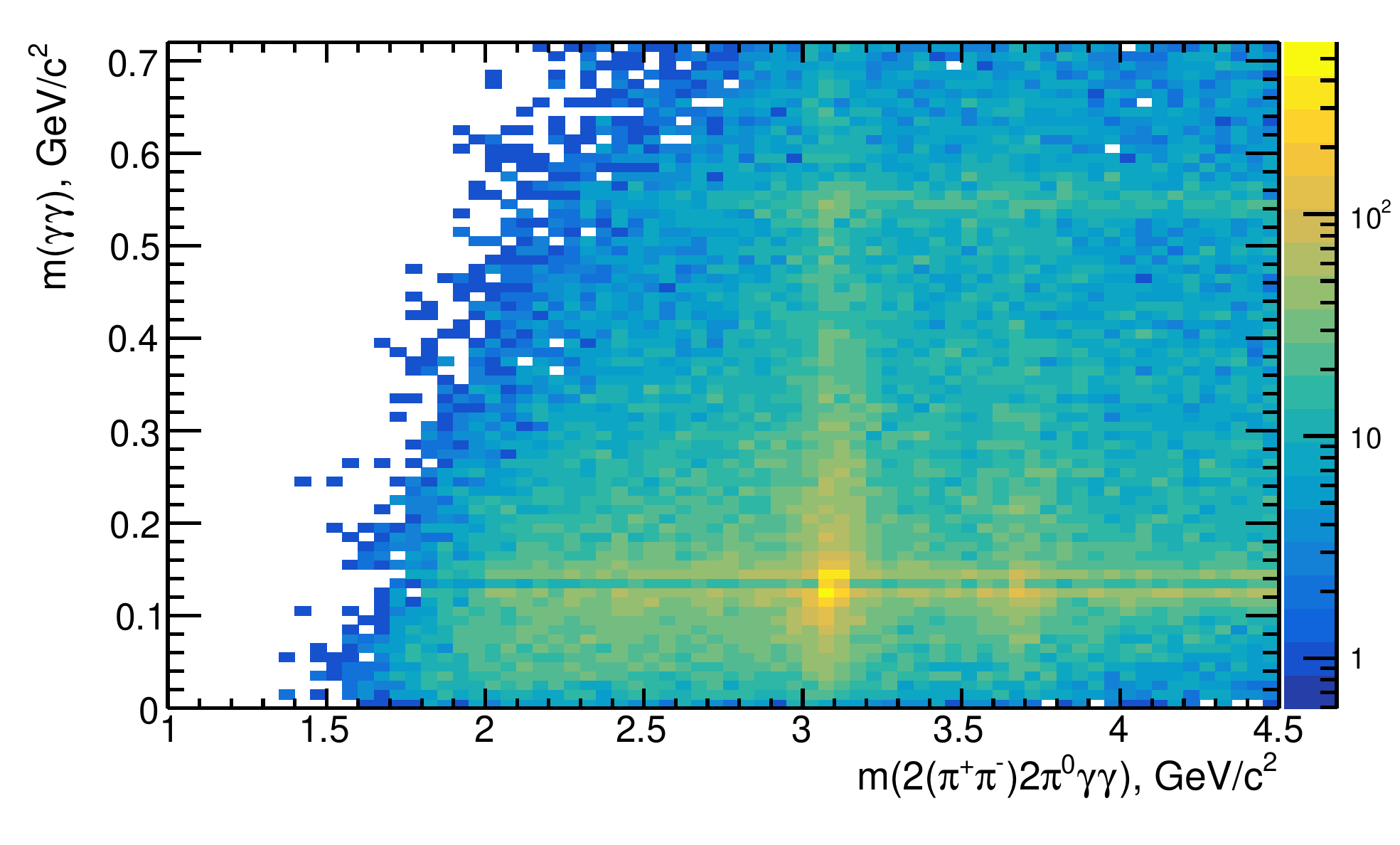}
\put(-185,100){\makebox(0,0)[lb]{\bf(a)}}\\
\vspace{-0.3cm}
\includegraphics[width=0.95\linewidth]{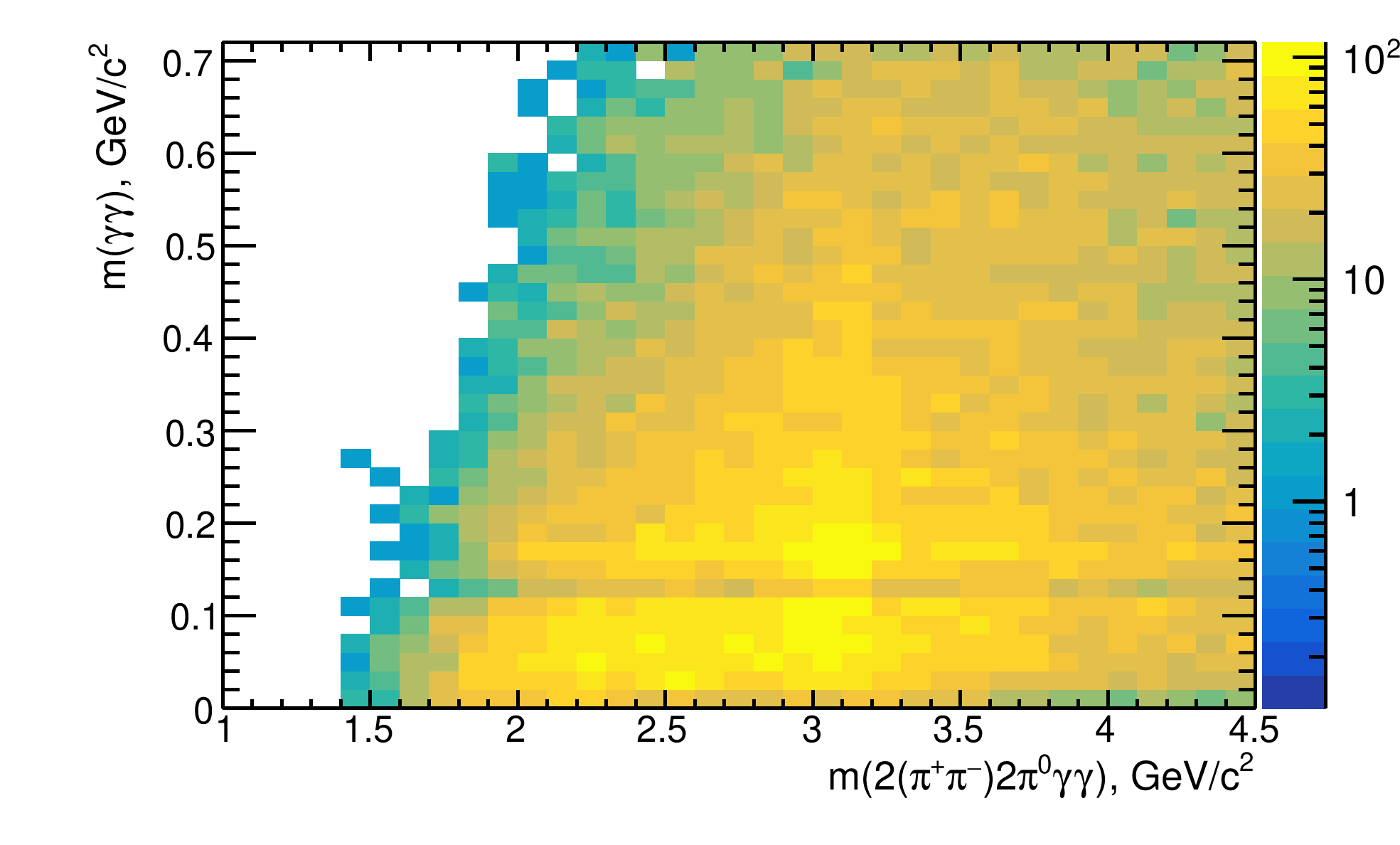}
\put(-185,100){\makebox(0,0)[lb]{\bf(b)}}
\vspace{-0.5cm}
\caption{(a) The third-photon-pair invariant mass $\mgg$ vs
  $m(4\pi2\piz\gamma\gamma)$ for (a) $\chisq_{4\pi2\piz\gamma\gamma}<
  50$ and
(b) $50< \chisq_{2\pi2\piz\gamma\gamma}< 100$.
}
\label{4pi3pi0_mass_all}
\end{center}
\end{figure}

\section{\boldmath Event Selection and Kinematic Fit}
\label{sec:Fits}

Candidates for the
$2(\pipi)3\piz\gamma$ and $2(\pipi)2\piz\eta\gamma$  events are selected
by requiring that there be four well measured  tracks 
 and seven or more detected photons, with an energy above
0.02~\gev  in the EMC. We assume that the photon with the highest energy is  the ISR
photon, and we require its c.m. energy to be larger than 3~\gev.
 
The four tracks must have zero total charge and extrapolate to within
0.25 cm of the beam axis and 3.0 cm 
of the nominal collision point along that axis.
In order to recover a
relatively small fraction of signal events that contain a
background track from secondary decay or interaction,
we allow for the presence of a fifth track in the event, 
which however must not fulfill the above condition.  
The four tracks that satisfy the extrapolation criteria are
fit to a vertex to determine the collision point, which is used 
in the calculation of the photon directions.

We subject each candidate event to a set of constrained 
kinematic fits and use the fit results,
along with charged-particle identification,
to select the final states of interest and evaluate backgrounds
from other processes.
 The kinematic fits use  the
 four-momenta and covariance matrices
of the colliding electrons and selected tracks and photons.
The fitted three-momenta of each track and photon are then used 
in further kinematic calculations.

We exclude the photon with the highest c.m.\  energy,
which is assumed to arise from ISR, and consider each independent set
of six other photons, and combine them  into three pairs.
For each set of six photons,
there are 15 independent combinations of photon pairs.
 We retain those combinations in which the diphoton mass
 of at least two pairs lies within 35~\mevcc of the
\piz mass $m_{\piz}$.
The selected combinations are subjected to a fit
 in which the diphoton masses of the two pairs with
$|m(\gamma\gamma)-m_{\pi^0}|<35~\mevcc$
 are constrained to $m_{\piz}$.
In combination with the constraints
due to four-momentum conservation, there are thus six
 constraints (6C) in the fit.  The photons in the
 remaining (``third'') pair are treated as being independent.
If all three photon pairs in the combination
satisfy $|m(\gamma\gamma)-m_{\pi^0}|<35~\mevcc$,
 then we test all possible combinations,
  allowing each of the three diphoton pairs in turn
 to be the third pair, i.e., the pair without the
 $m_{\piz}$ constraint.

 The above procedure allows us not only to search 
for events with  $\piz\to\gamma\gamma$  in the third photon pair, but
also for events with $\eta\to\gamma\gamma$.

The 6C fit is performed under the signal hypothesis
$\epem\to2(\pipi)\ppz\gamma\gamma\gamma_{ISR}$.
The combination with the smallest \chisq is retained, along with the obtained
$\chisq_{4\pi2\piz\gamma\gamma}$ value and the fitted three-momenta of each
track and photon. 
Each selected event is also subjected to a 6C fit under the
$\epem\to2(\pipi)\ppz\gamma_{ISR}$ background hypothesis, and the
 $\chisq_{4\pi2\piz}$ value is retained.  
The $2(\pipi)\ppz$ process has a larger
cross section than the $2(\pipi)3\piz$ signal process and can
contribute to the background when two background photons are present.


\section{Additional selection criteria}

We require the tracks to lie within the fiducial
region of the DCH (0.45--2.40 radians) and to be
inconsistent with being a kaon or muon.
The photon candidates are required to lie within the
fiducial region of the EMC (0.35--2.40 radians) and to have
an energy larger than 0.035 GeV.
The angular distance between the ISR photon and the closest track must be 
greater than 1 radian; this requirement significantly suppresses the non-ISR background,
in particular reducing the background from $\epem\to\tau^+\tau^-$ to a negligible level.
A requirement that any extra photons in an event must
have an energy below 0.7 GeV reduces the multi-photon
background by 10-20\%. 
Finally, the  background from the ISR process $\epem\to 2(\pipi)2\piz\gamma$ is reduced  
from 30\% to about 1-2\%, with a loss of only 5\% of signal events,
by requiring $\chi_{4\pi2\piz}^2 >30$.

\begin{figure}[tbh]
\begin{center}
\includegraphics[width=0.95\linewidth]{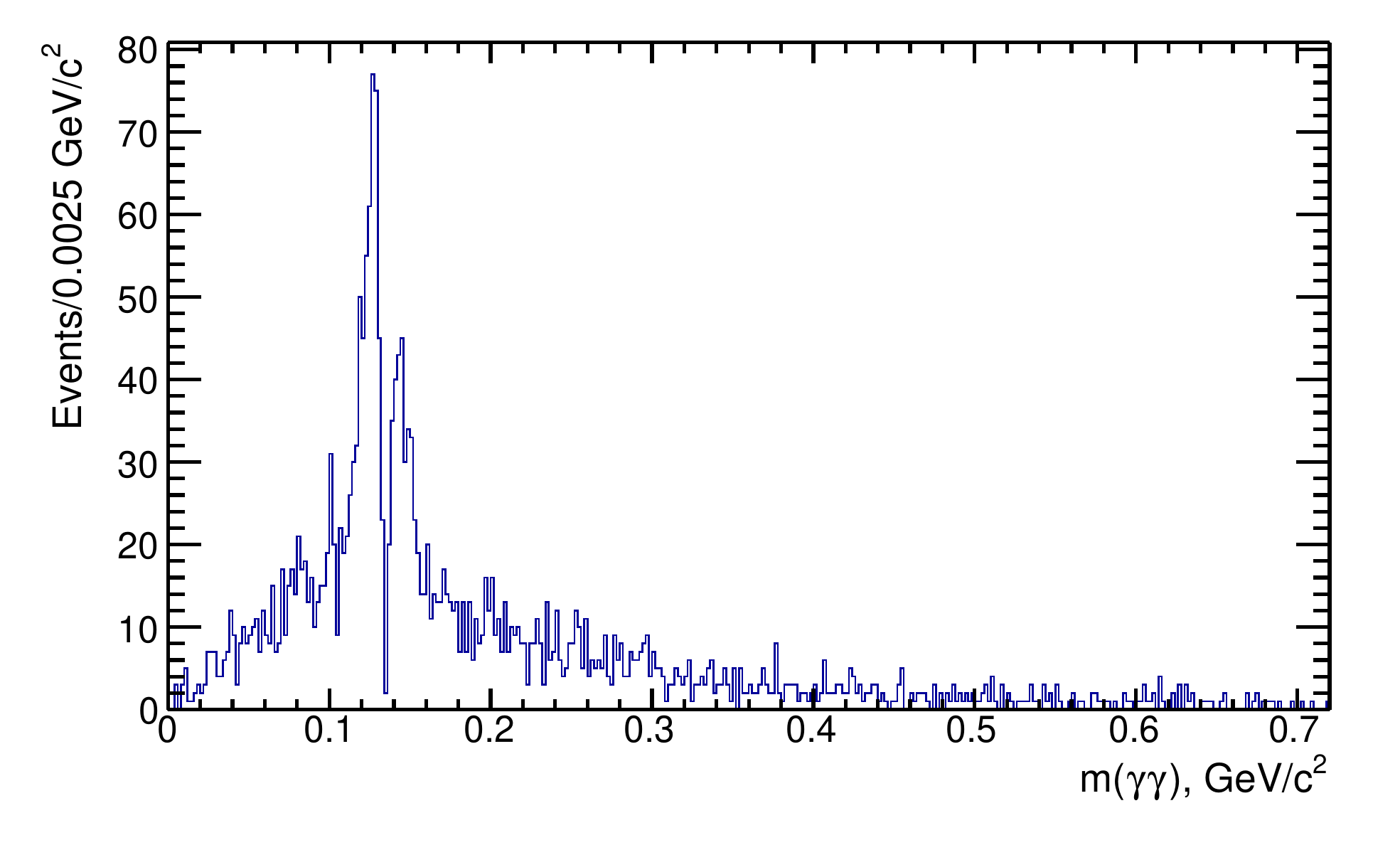}
\put(-50,100){\makebox(0,0)[lb]{\bf(a)}}\\
\vspace{-0.3cm}
\includegraphics[width=0.95\linewidth]{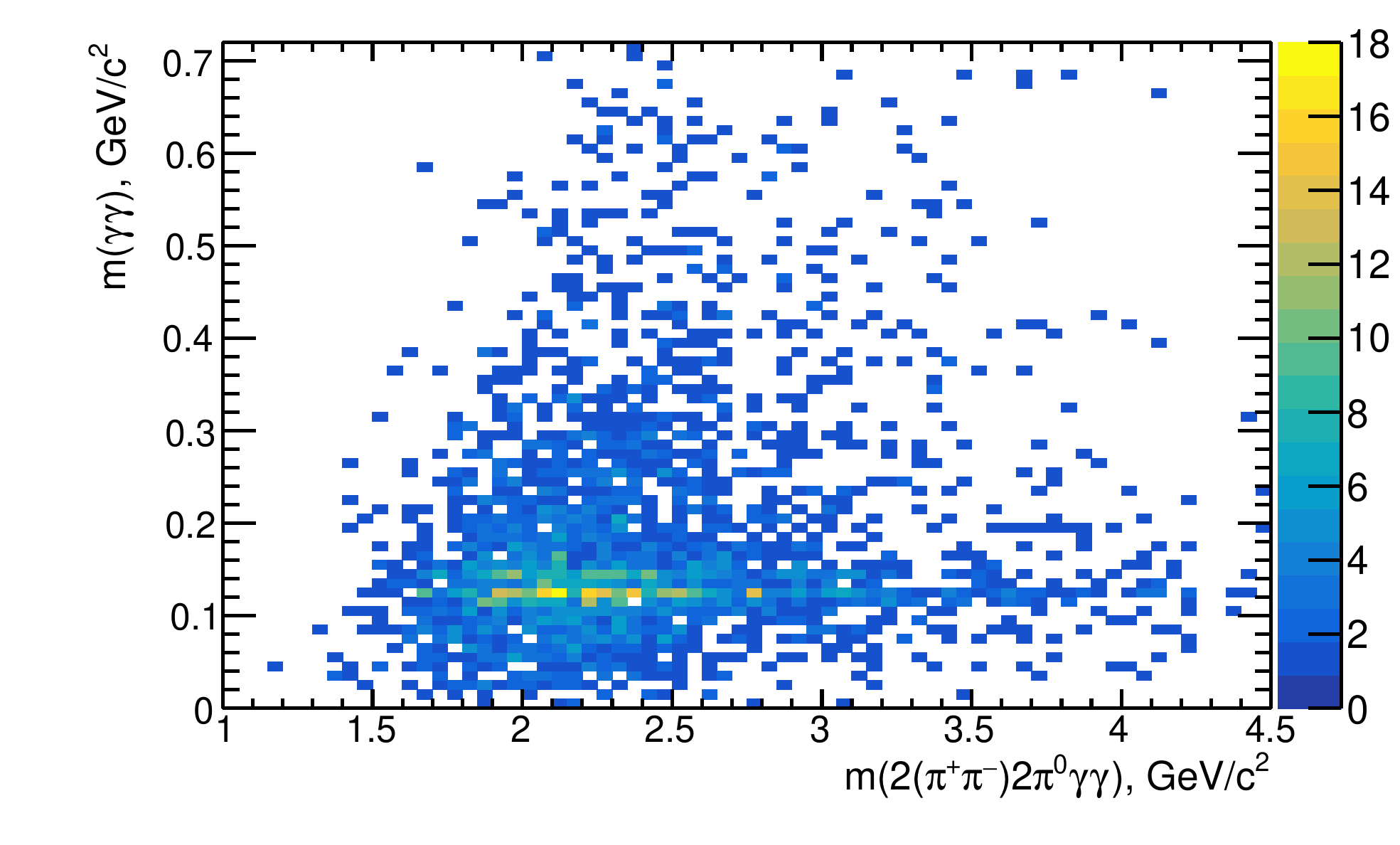}
\put(-50,100){\makebox(0,0)[lb]{\bf(b)}}
\vspace{-0.5cm}
\caption{
The  MC-simulated distribution for $\epem\to\eta\pipi\ppz$ events of (a)
the third-photon-pair invariant mass $\mgg$, and (b) $\mgg$ vs $m(2(\pipi)2\piz\gamma\gamma)$.
}
\label{mgg_eta2pi}
\end{center}
\end{figure}
\begin{figure}[tbh]
\begin{center}
\includegraphics[width=0.95\linewidth]{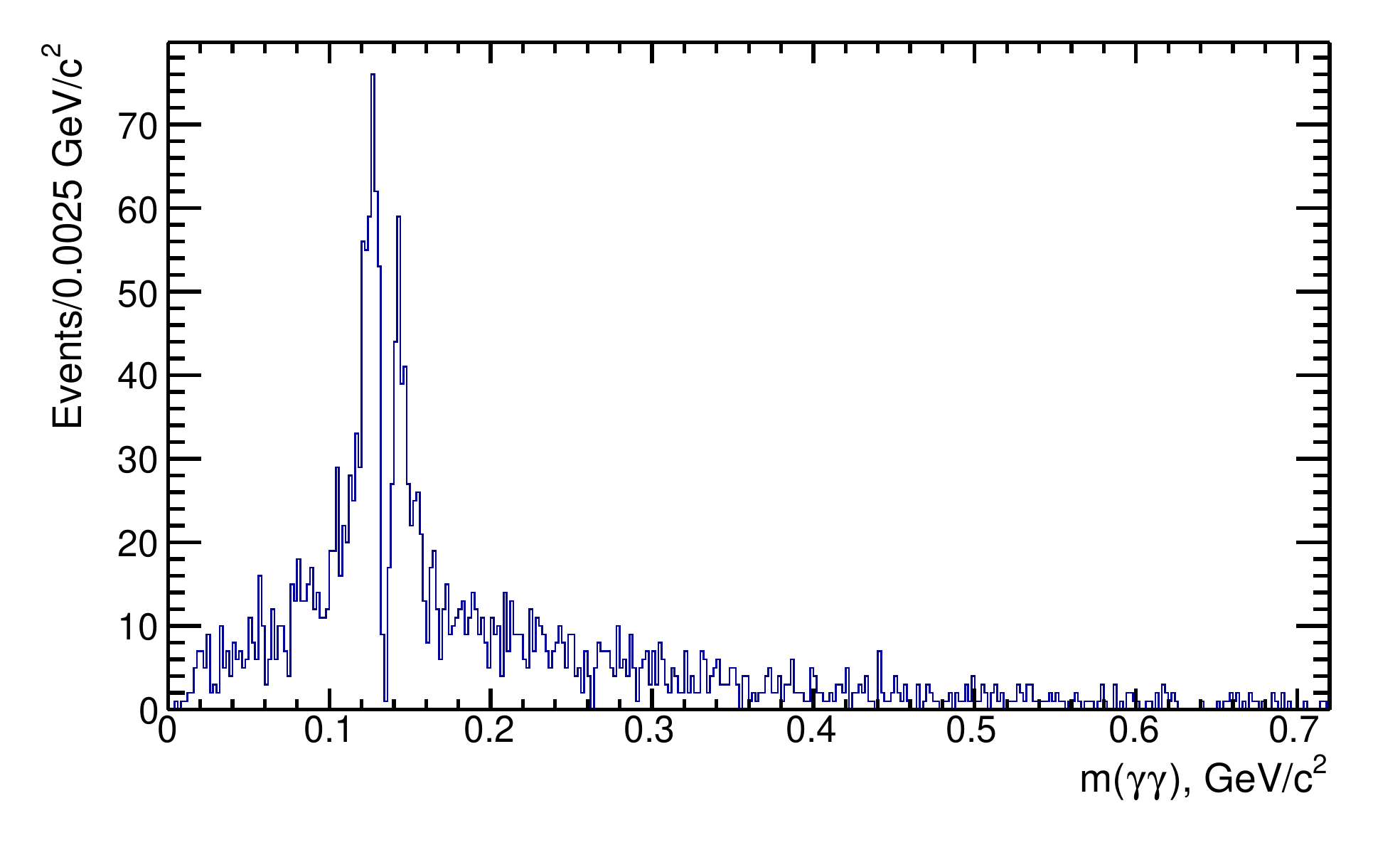}
\put(-50,100){\makebox(0,0)[lb]{\bf(a)}}\\
\vspace{-0.3cm}
\includegraphics[width=0.95\linewidth]{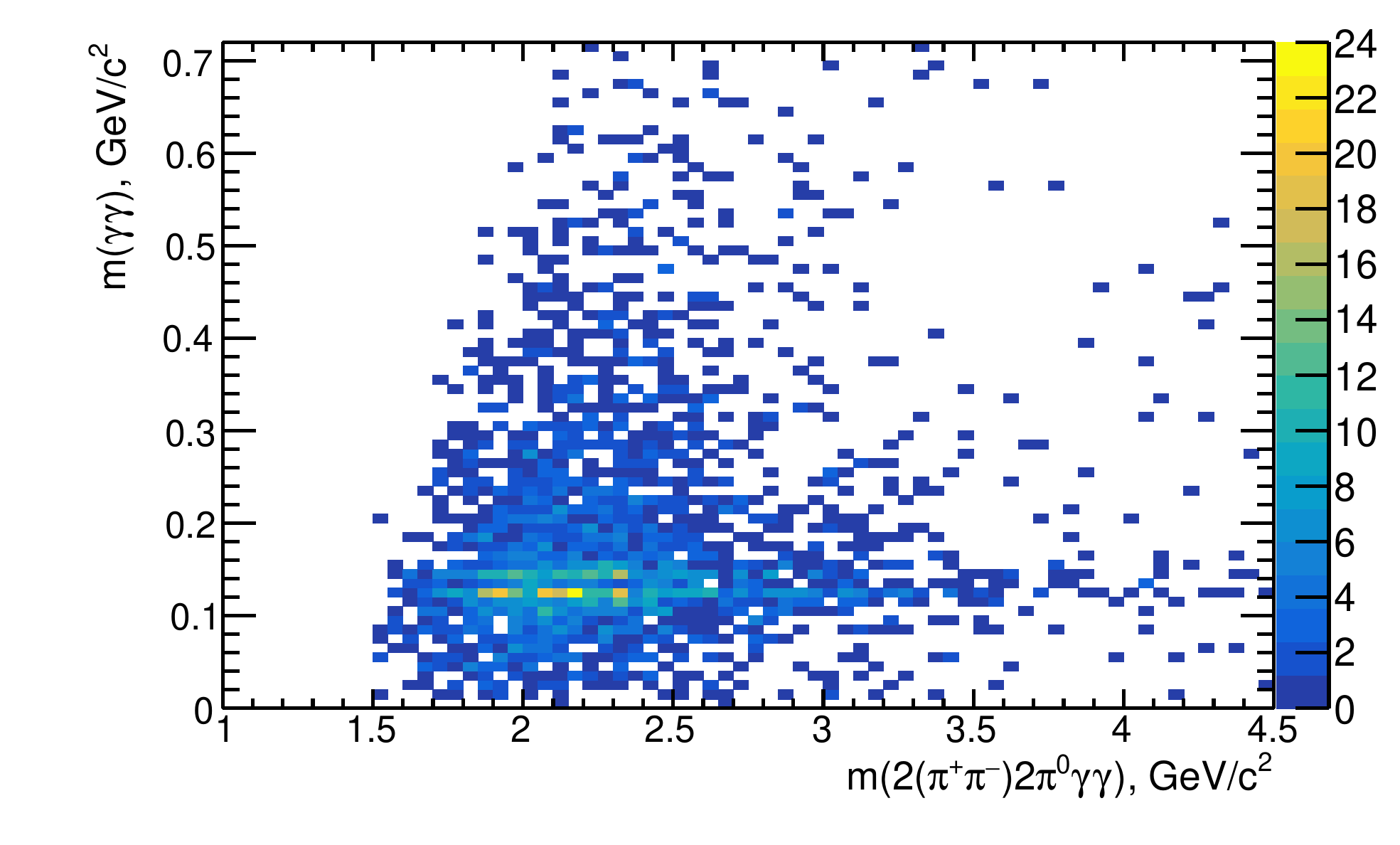}
\put(-50,100){\makebox(0,0)[lb]{\bf(b)}}
\vspace{-0.5cm}
\caption{
(a) Distribution of the third-photon-pair invariant mass $\mgg$ and of (b) $\mgg$ vs $m(2(\pipi)2\piz\gamma\gamma)$ 
for MC-simulated $\epem\to\omega\piz\eta$ events.
}
\label{mgg_omega2pi}
\end{center}
\end{figure}

Figure~\ref{4pi3pi0_chi2_all} (a) shows the invariant mass $\mgg$ of the third
photon pair vs $\chisq_{4\pi2\piz\gamma\gamma}$ after the above requirements. Clear $\piz$
and $\eta$ peaks are visible at small  \chisq values.
The two vertical lines define the signal and  control regions, corresponding to  $\chisq_{4\pi2\piz\gamma\gamma}< 50$ and
$50 < \chisq_{4\pi2\piz\gamma\gamma}< 100$, respectively.

Figure~\ref{4pi3pi0_chi2_all} (b)
shows the $\mgg$ distribution for events in the signal region after the
above requirements have been applied. 
The dip in this distribution at the $\piz$ mass value is a
consequence of the kinematic fit constraint of the best
two photon pairs to the $\piz$ mass, so the third photon pair is
always formed from photon candidates that are less
well measured.

Figure~\ref{4pi3pi0_mass_all} shows the $\mgg$ 
distribution vs the invariant mass $m(2(\pipi)2\piz\gamma\gamma)$ 
 for events in the signal (a) 
and control (b) region. 
Events from the
$\epem\to2(\pipi)3\piz$ and $2(\pipi)2\piz\eta$ processes are
clearly seen in the signal region, as well as $J/\psi$ decays to these final states. In the
control region no significant structures are seen; we use these
events to evaluate background.

Our strategy to extract the signals for the $\epem\to2(\pipi)3\piz$
and $2(\pipi)2\piz\eta$ processes 
is to perform a fit for the $\piz$ and $\eta$ yields
in intervals of 0.05~\gevcc in the distribution of
the $2(\pipi)2\piz\gamma\gamma$ invariant mass.
This mass interval is about three times wider than
the experimental resolution.

\section{Detection efficiency}\label{sec:efficiency}
\subsection{Number of signal events in simulation}
As mentioned in Sec.~\ref{sec:babar}, the model used in the MC simulation 
assumes that the seven-pion final state results  from 
$\omega\piz\eta$ and $\eta\pipi\ppz$ production,  with $\omega$
decays to three pions and $\eta$ decays to all modes.  As shown below, these  two final
states dominate the observed cross section.
Also, the ISR photon is
simulated to wider angles than the EMC acceptance, reducing
the nominal efficiency.
For each mode we have
200,000 simulated events from the  primary generator. 

The selection procedure applied to the data is also applied to the         
MC-simulated events. Figures~\ref{mgg_eta2pi}
and \ref{mgg_omega2pi} show
(a) the  $\mgg$ distribution and (b) the distribution of
$\mgg$ vs $m(2(\pipi)2\piz\gamma\gamma)$ for the simulated  $\eta\pipi\ppz$
and $\omega\piz\eta$ events, respectively. 
The $\piz$ peak is not Gaussian in either reaction.
Background photons are included in the simulation.
Therefore, the simulation accounts for the combinatorial  
background that arises when background
photons are combined with photons from
the signal reactions.

\begin{figure}[t]
\begin{center}
\includegraphics[width=0.95\linewidth]{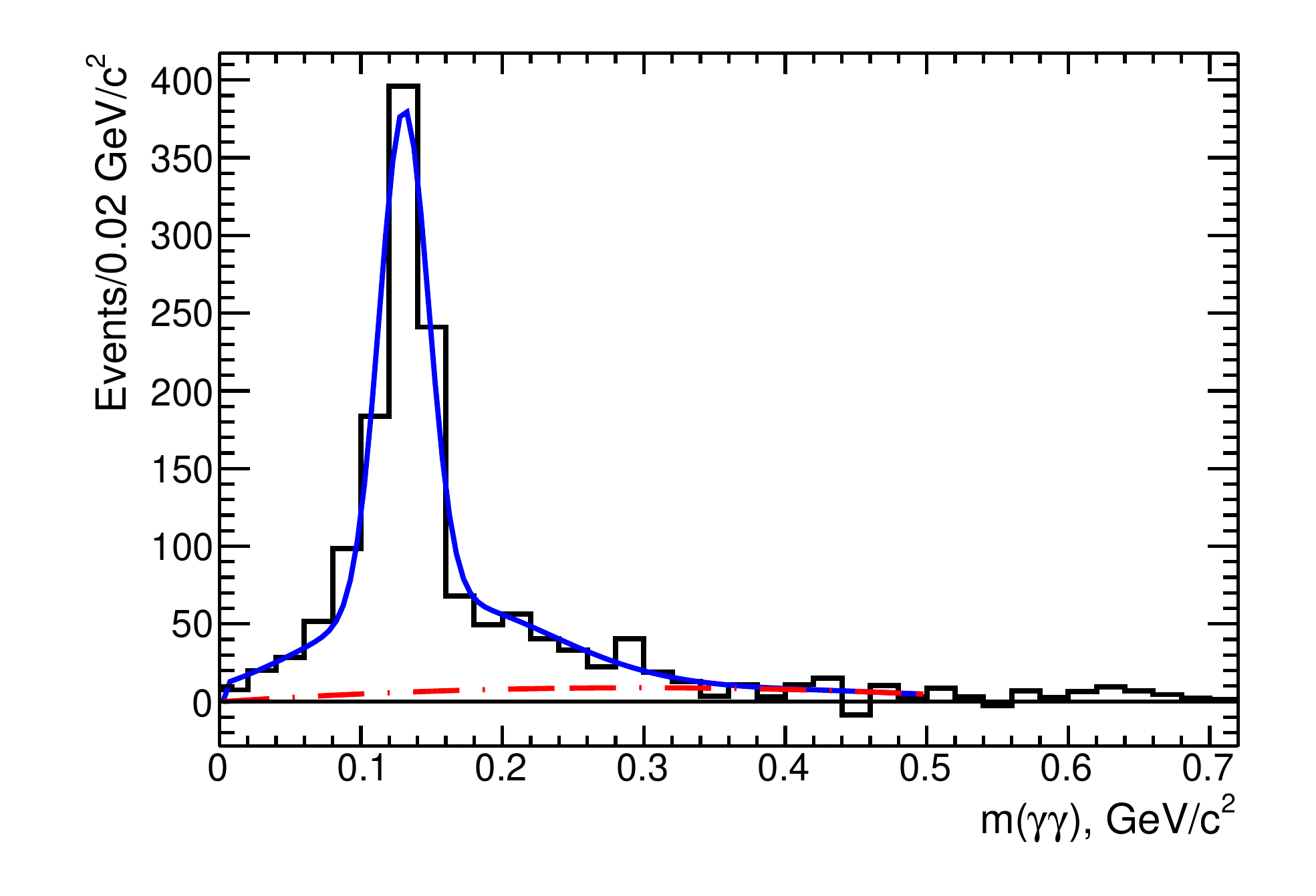}
\put(-50,100){\makebox(0,0)[lb]{\bf(a)}}\\
\vspace{-0.4cm}
\includegraphics[width=0.95\linewidth]{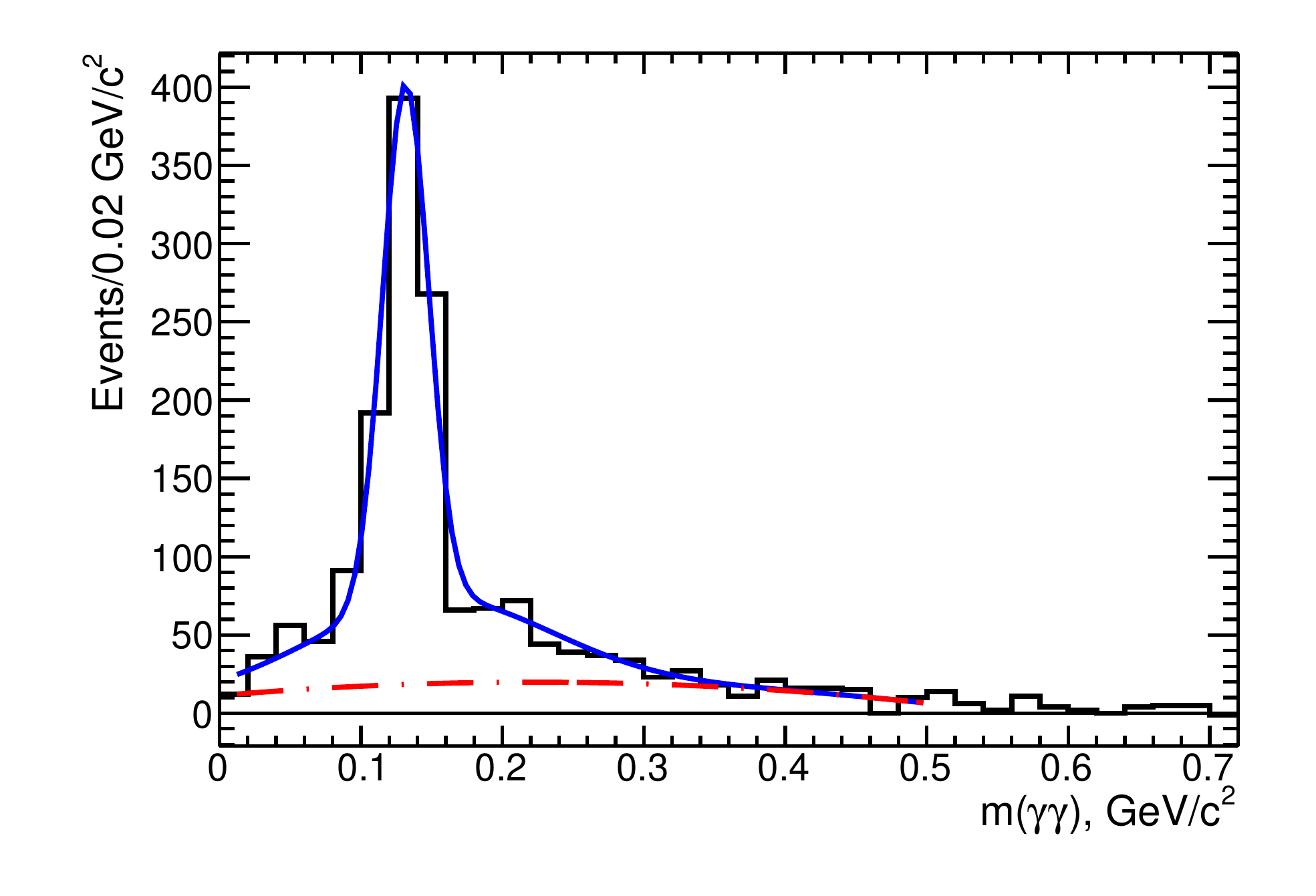}
\put(-50,100){\makebox(0,0)[lb]{\bf(b)}}
\vspace{-0.5cm}
\caption{
Background subtracted $\mgg$ distribution for MC-simulated 
(a) $\epem\to\eta\pipi\ppz$  and (b)  $\epem\to\omega\piz\eta$
events.
The fit function is described in the text.
The dashed line shows a fit of the remaining contribution
from the \chisq control region. 
}
\label{mgg_eta2pi_fit}
\end{center}
\end{figure}

\begin{figure}[b]
\begin{center}
\includegraphics[width=0.95\linewidth]{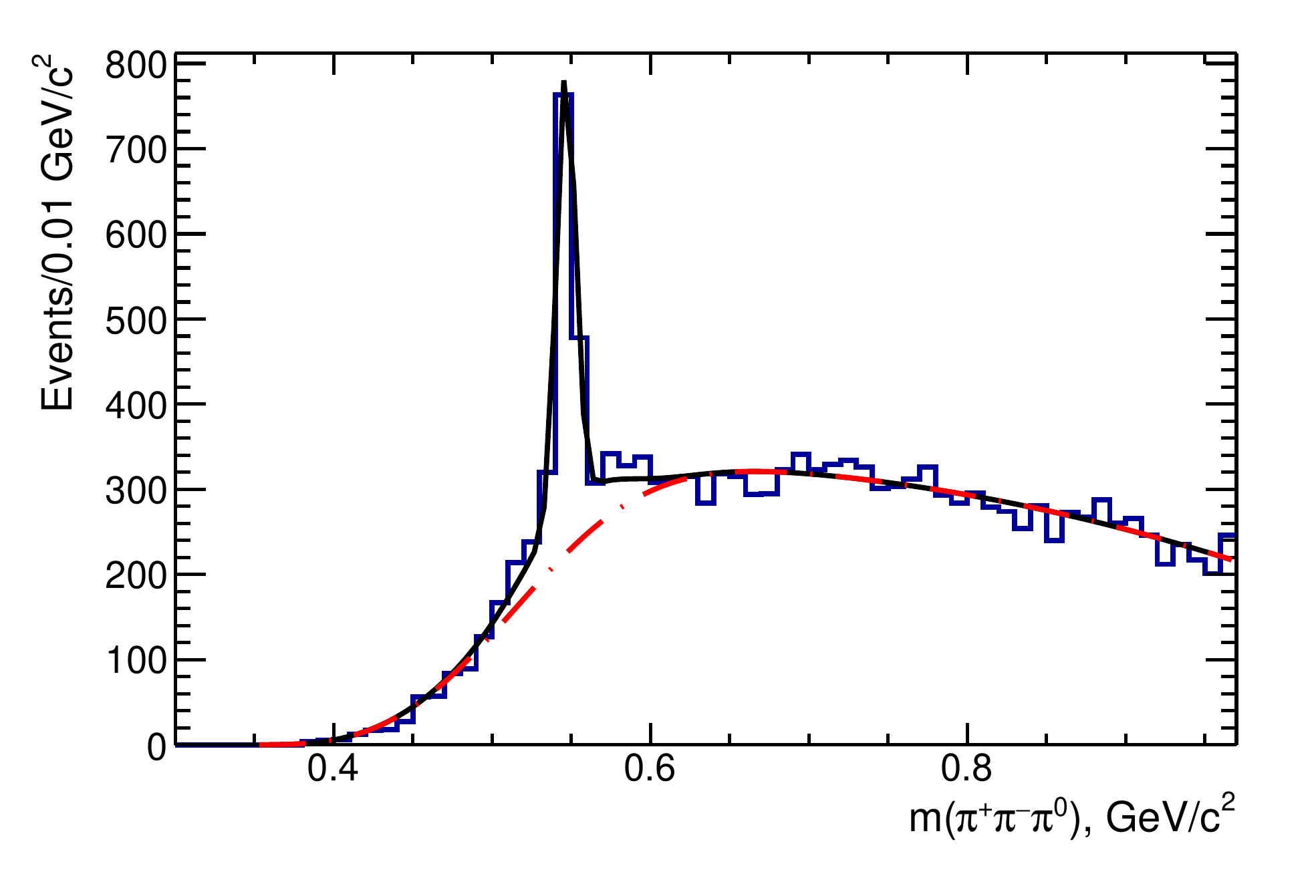}
\put(-50,100){\makebox(0,0)[lb]{\bf(a)}}\\
\vspace{-0.4cm}
\includegraphics[width=0.95\linewidth]{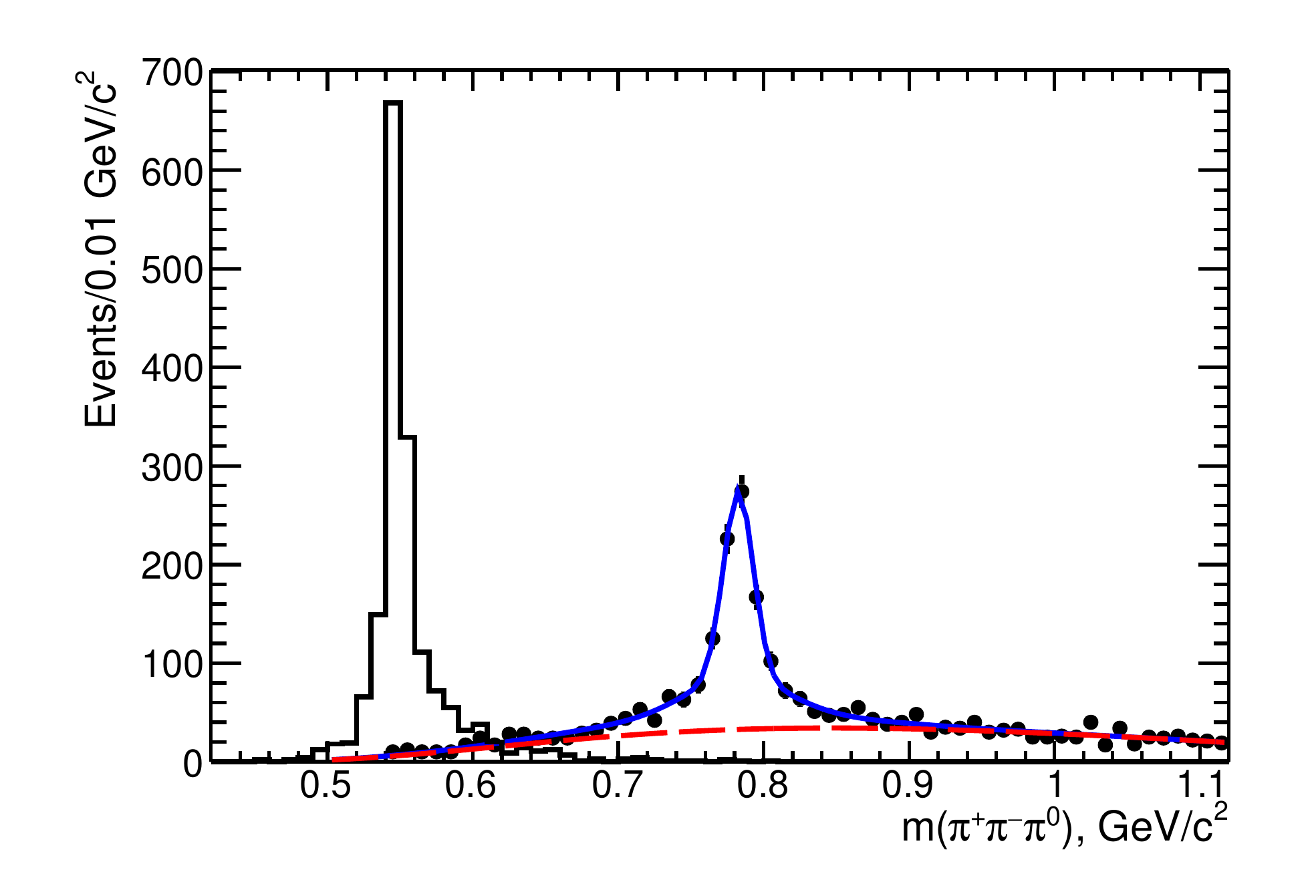}
\put(-50,100){\makebox(0,0)[lb]{\bf(b)}}
\vspace{-0.5cm}
\caption{(a) The 
$\pipi\piz$ invariant mass distribution for the MC-simulated $\epem\to\eta\pipi\ppz$ events. 
The dashed curve is for  the combinatorial background.
(b) The $\pipi\piz$  invariant masses for the
  MC-simulated $\epem\to\omega\piz\eta$ events. The histogram shows the  $\pipi\piz$
  combination closest to the $\eta$ mass, while the two
  remaining combinations (dots) exhibit  the $\omega$ meson. 
 The curves show the fit functions used to obtain the number of signal (solid)
  events, and the combinatorial background (dashed) contribution.
}
\label{m3pi_omega_eta_mc}
\end{center}
\end{figure}

The combinatorial background  is subtracted using the data
 from the \chisq control region. We do not know how large  the
 combinatorial background is in the signal region, and we use a scale factor varying from 1.0
 to 1.5 for the subtraction to estimate the uncertainty in the number of
 signal events.
 The method is illustrated using simulation in Fig.~\ref{mgg_eta2pi_fit},
which shows the $\mgg$ distribution with a bin width of 0.02~\gevcc.
The solid histograms show the simulated results from the
signal region after subtraction of the simulated
combinatorial background with the scale factor 1.5.
The sum of three Gaussian functions
is used to describe the $\piz$ signal shape. A third-order polynomial function is
used to describe the shape of the remaining combinatorial background. 
  The fitted  function is shown by the smooth solid curve, 
  while the dashed curve is for the
  contribution of the remaining combinatorial background.
The remaining combinatorial background contribution is almost negligible for the
scale factor value 1.5. We obtaine 1122$\pm$46 and 1161$\pm$55
simulated signal events for each mode, respectively.
If the scale factor 1.0 is used, the
remaining background is well described by the polynomial function and
the signal yield does not change by more than 3\%.     

Alternatively, for the $\eta\pipi\ppz$ events, we determine the number of
events by fitting the
$\eta$ signal from the $\eta\to\pipi\piz$ decay: the simulated
distribution is shown  in Fig.~\ref{m3pi_omega_eta_mc}(a) (twelve
entries per event). The fit functions are again the sum of three
Gaussian functions and a polynomial for the combinatorial background.
In total we obtain 1183$\pm$49 events.
A similar fit of the $\eta$ signal is performed for the $\omega\piz\eta$
final state simulation with  1110$\pm$54 selected events.

Similarly, as an alternative for the $\omega\piz\eta$ events,
the $\omega$ mass peak can be used. To reduce the number of combinatorial
entries, we require one $\pipi\piz$ combination to have invariant mass
close to the $\eta$ mass, and fit the remaining two combinations to extract
the numbers of signal events with an $\omega$, as shown in
Fig.~\ref{m3pi_omega_eta_mc}(b).
In total 1104$\pm$71 signal events are found.
A Breit-Wigner (BW) function,
convolved with a Gaussian distribution to account
for the detector resolution, is used to describe the $\omega$ signal.
A second-order polynomial is used to describe the background.

\begin{figure}[t]
\begin{center}
\includegraphics[width=0.99\linewidth]{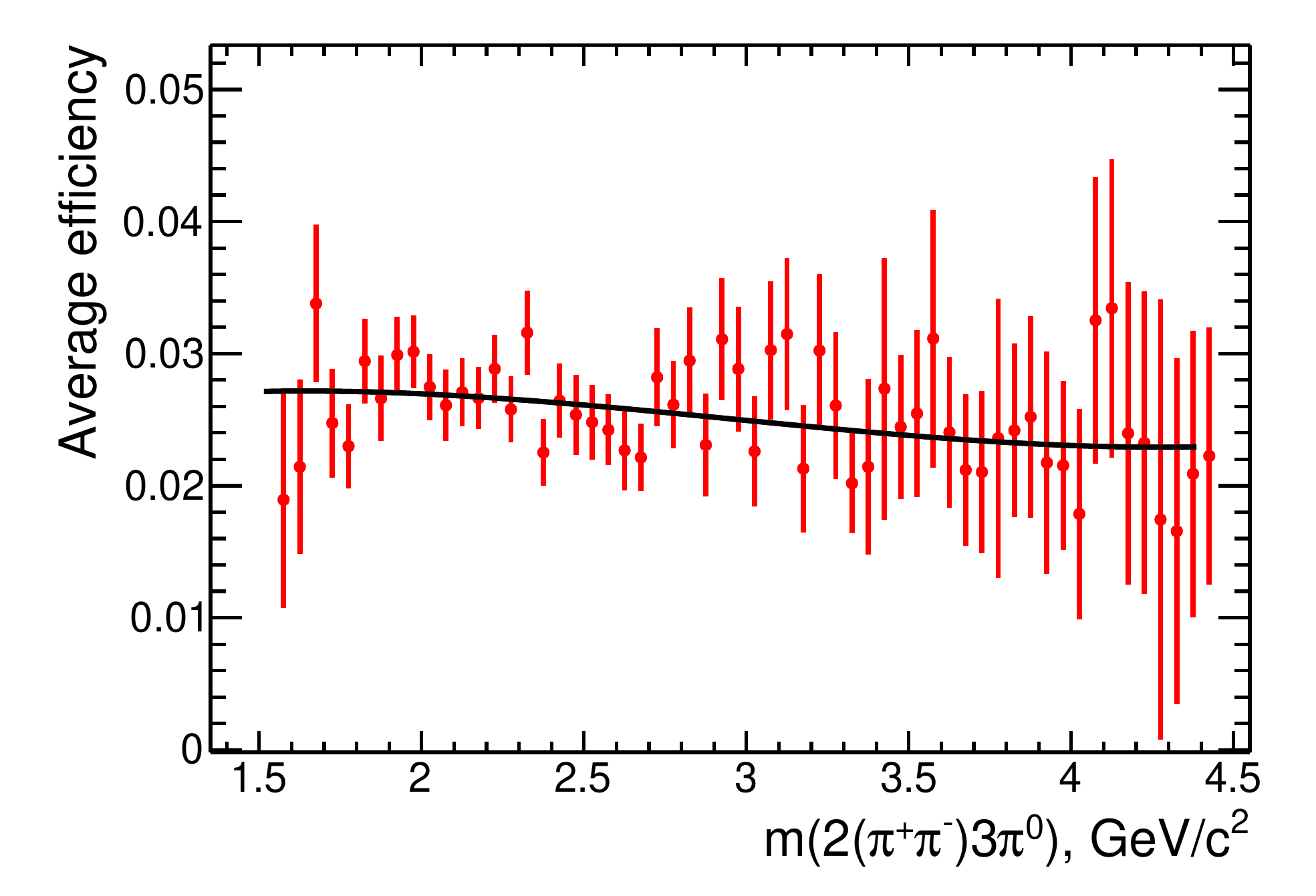}
\vspace{-0.5cm}
\caption{ The energy-dependent reconstruction efficiency for 
  $\epem\to2(\pipi)3\piz$ events, determined using
  five different methods: see text. 
The curve shows the results of a fit to the
average values, which is used in the cross
section calculation.
}
\label{mc_acc}
\end{center}
\end{figure}
\begin{figure}[tbh]
\begin{center}
\includegraphics[width=0.99\linewidth]{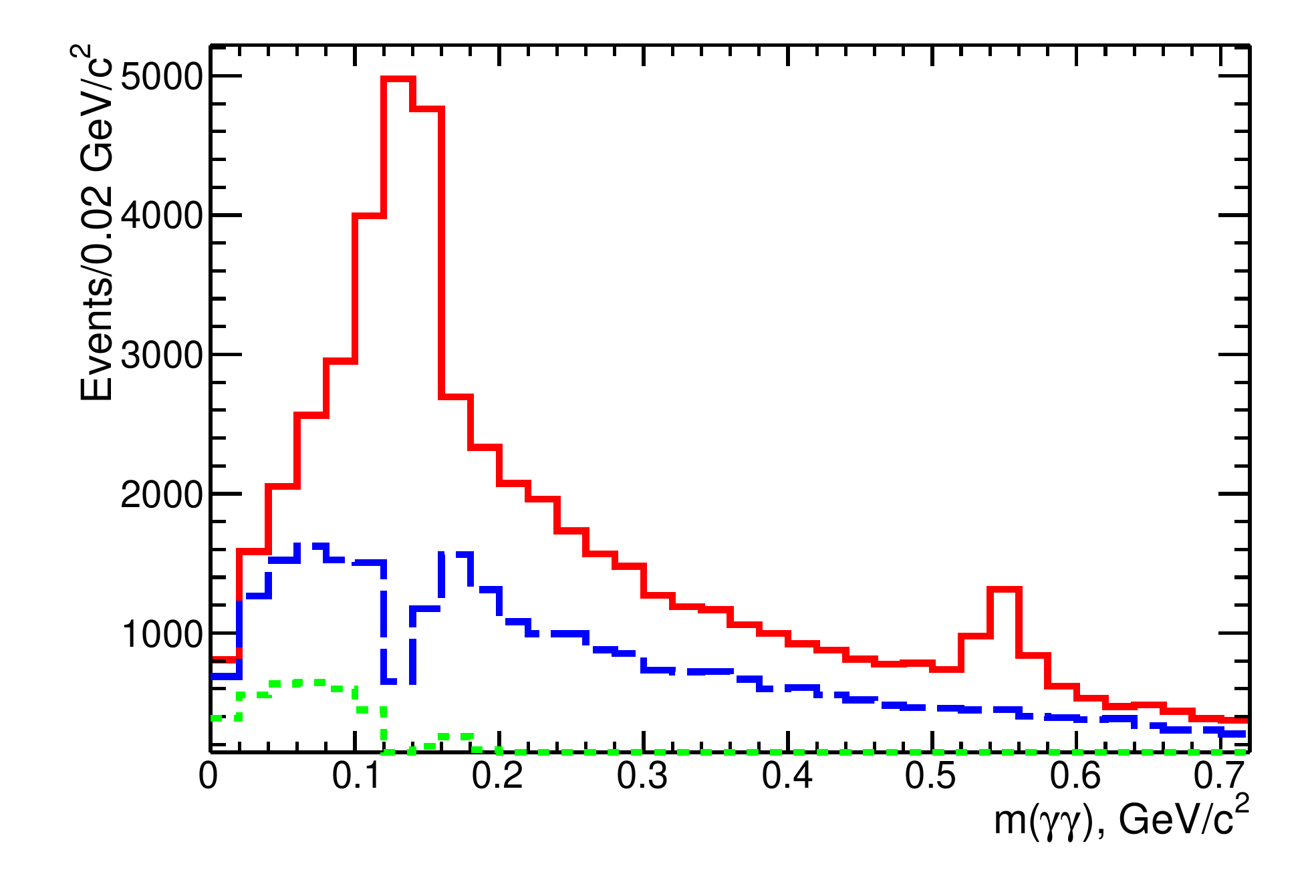}
\put(-50,90){\makebox(0,0)[lb]{\bf(a)}}\\
\vspace{-0.35cm}
\includegraphics[width=0.99\linewidth]{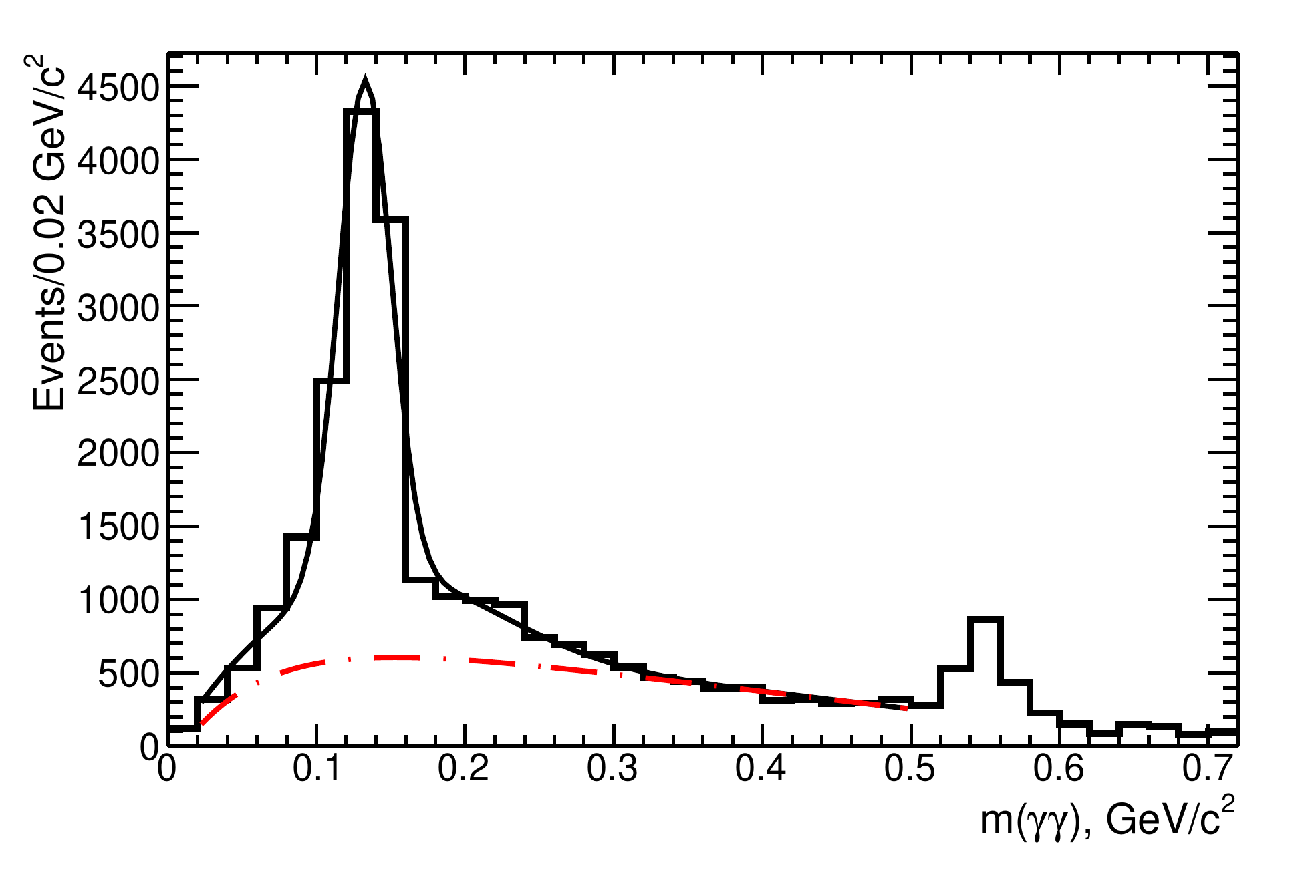}
\put(-50,90){\makebox(0,0)[lb]{\bf(b)}}
\vspace{-0.5cm}
\caption{
  (a) Invariant mass $\mgg$ for data in the \chisq 
signal (solid) and control (dashed) regions.
The dotted histogram shows the estimated background from
$\epem\to2(\pipi)\ppz$.
(b) The $\mgg$ invariant mass for data after the background
subtraction. The curves are the fit results as described in the text.
}
\label{2pi2pi0_bkg}
\end{center}
\end{figure}

\subsection{Efficiency evaluation}
The mass-dependent  detection
efficiency is obtained by dividing the number of fitted MC
events in each 0.05~\gevcc mass interval of the hadronic system by the number generated in          
the same interval.
The number of signal events in the simulation, obtained by
fitting the $\piz$, $\eta$, or $\omega$ signals,   is consistent
within uncertainties not only in total, but also in every mass
interval.
We do not see any significant difference in mass-dependent efficiency
between the different methods.
The uncertainty in the value of the efficiency in each mass bin
is dominated by the fluctuation of the combinatorial background.  We
average the five efficiencies in each 0.05~\gevcc mass interval and
fit the result with a third-order polynomial function, shown in
Fig.~\ref{mc_acc}. The result of this fit is used for the cross section calculation. 

Although the signal simulation accounts for all
$\eta$ decay modes, the efficiency calculation
considers the signal $\eta\to\pipi\piz$ decay mode only.
This efficiency estimate takes into account the geometrical acceptance of the detector 
for the final-state photons and the charged pions, the inefficiency of 
the  detector subsystems, and  the event loss due to additional
soft-photon  emission from the initial and final states.
Corrections that account for data-MC differences are discussed below.

From Fig.~\ref{mc_acc} it is seen that the reconstruction
efficiency is about 2.7\%, roughly independent of mass.
By comparing the results of the five different
methods used to evaluate the efficiency, 
we conclude that the relative overall efficiency does not change by more than 
5\% because of variations of the functions used to extract the number
of events or the use of different models. This
value is taken as an estimate of the systematic uncertainty in the acceptance
associated with the simulation model used and with the fit
procedure. 

We do not simulate the $2(\pipi)\eta$  and $2(\pipi)2\piz\eta$
intermediate states, which are observed in data (see below) in the
$\eta\to 3\piz$ and $\eta\to\gamma\gamma$ decays.
But  our previous studies \cite{isr2pi3pi0} have 
demonstrated that, for these and similar decays, the variations in efficiency due to
model dependence do not exceed 5\%. 
In combination with the selections
 above,  we assign 7\% as a sistematic uncertainty to the detection efficiency. 

\begin{figure}[t]
\begin{center}
  \includegraphics[width=0.99\linewidth]{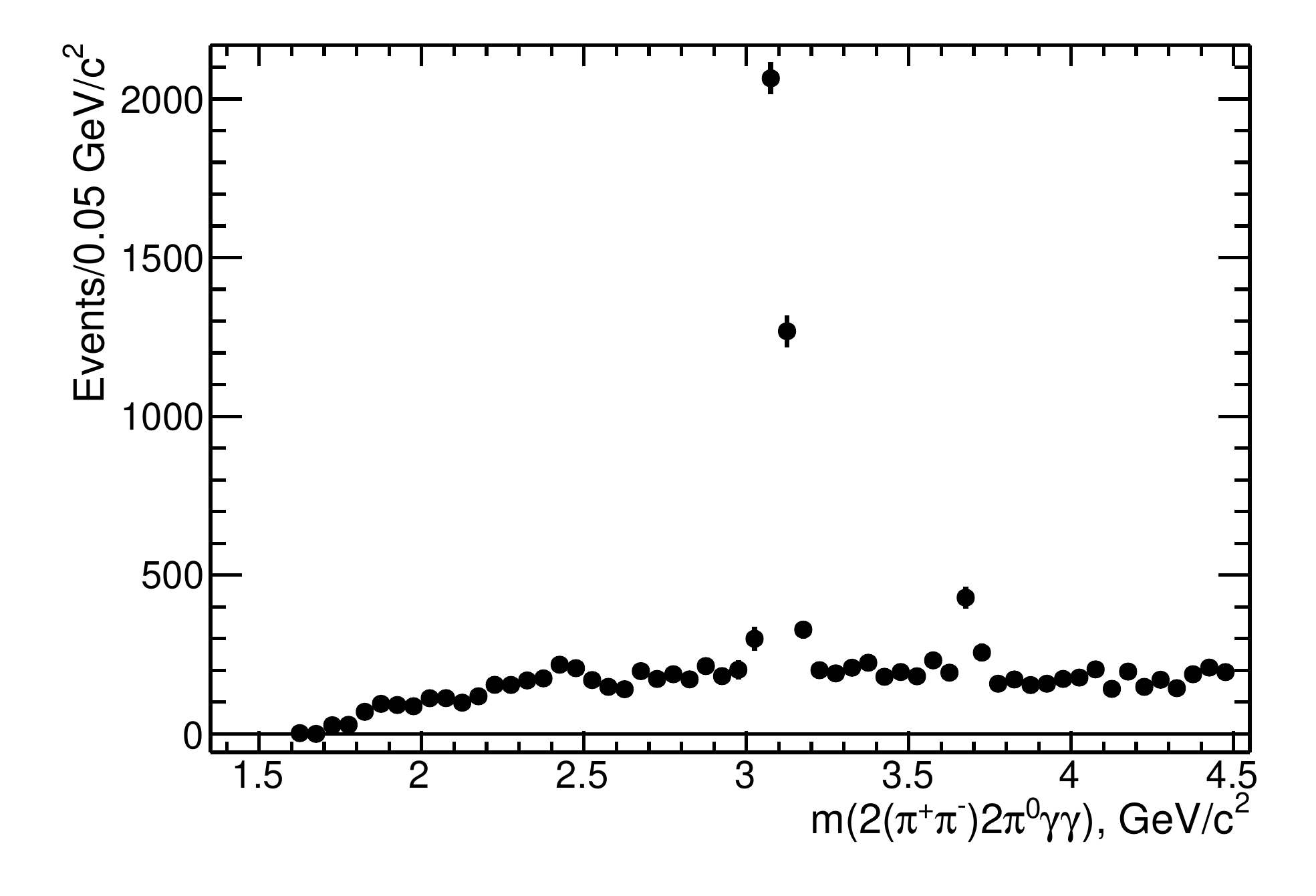}
  \put(-50,90){\makebox(0,0)[lb]{\bf(a)}}\\
\includegraphics[width=0.99\linewidth]{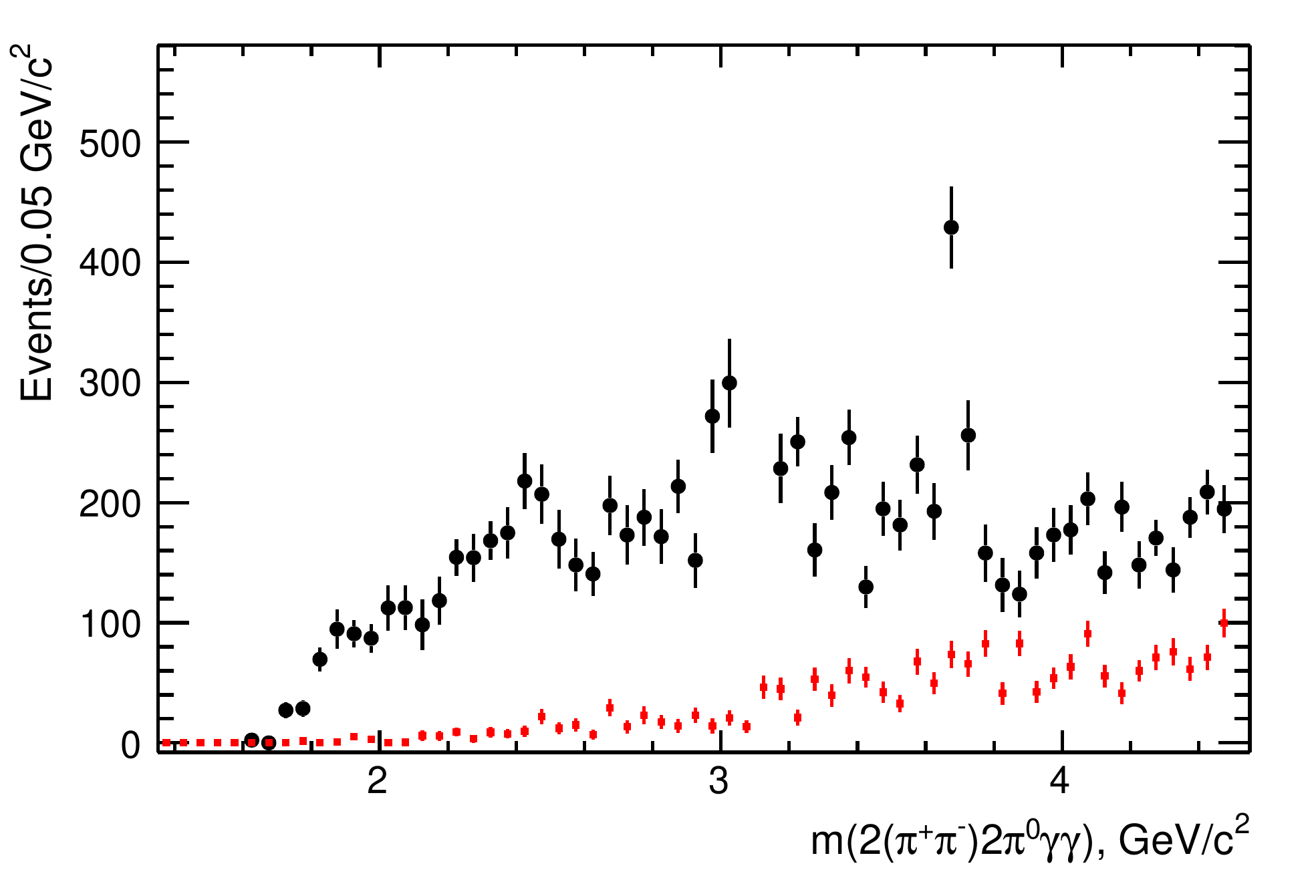}
\put(-50,90){\makebox(0,0)[lb]{\bf(b)}}
\vspace{-0.2cm}
\caption{(a) The invariant-mass distribution of 
  $\pipi3\piz$ events, obtained from the fit to the $\piz$ mass peak.
  (b) Expanded view of (a) to show
the contribution from non-ISR $uds$ background, shown by squares.
}
\label{nev_4pi3pi0_data}
\end{center}
\end{figure}

\begin{figure}[tbh]
\begin{center}
\includegraphics[width=0.98\linewidth]{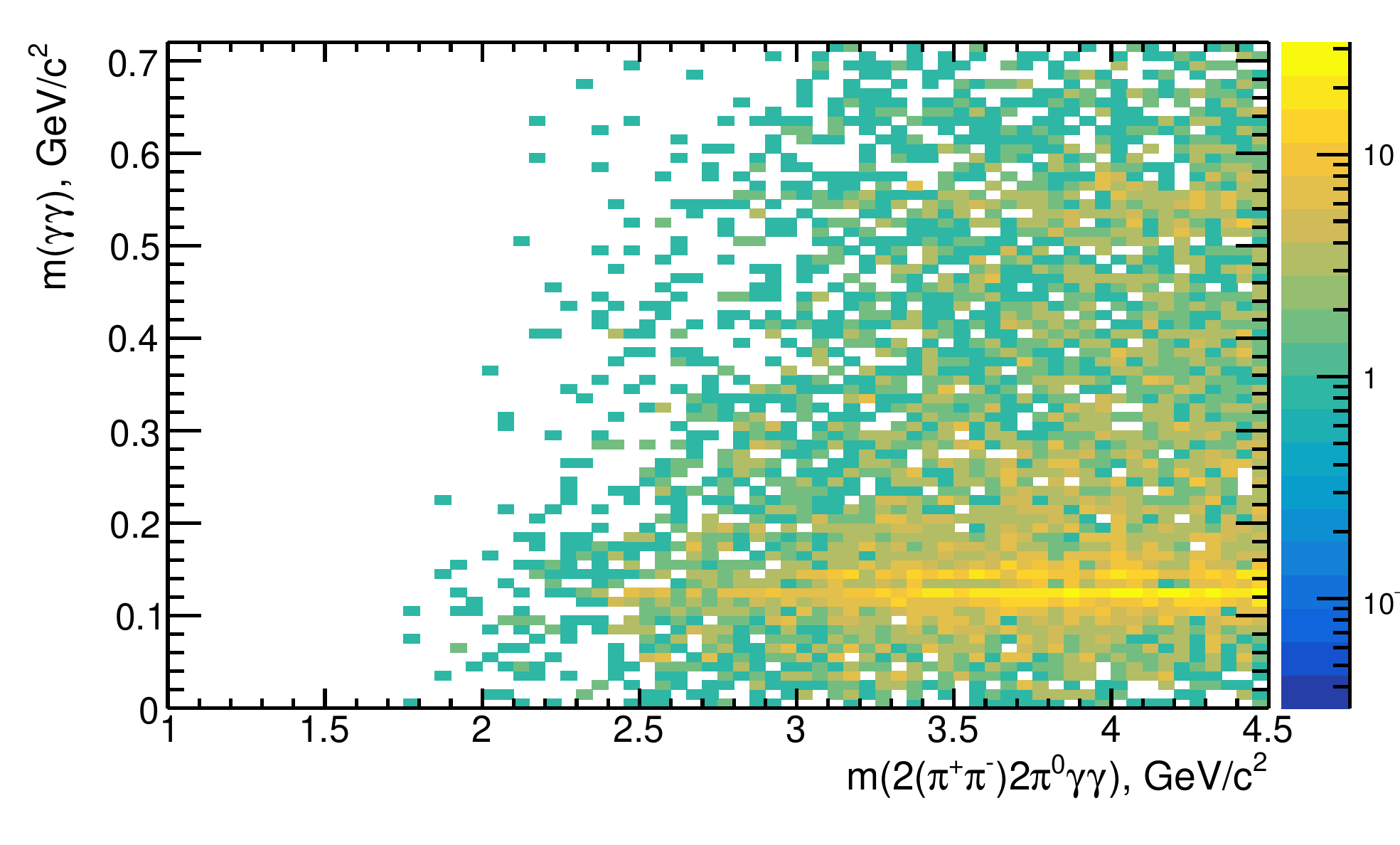}
\put(-180,100){\makebox(0,0)[lb]{\bf(a)}}\\
\includegraphics[width=0.98\linewidth]{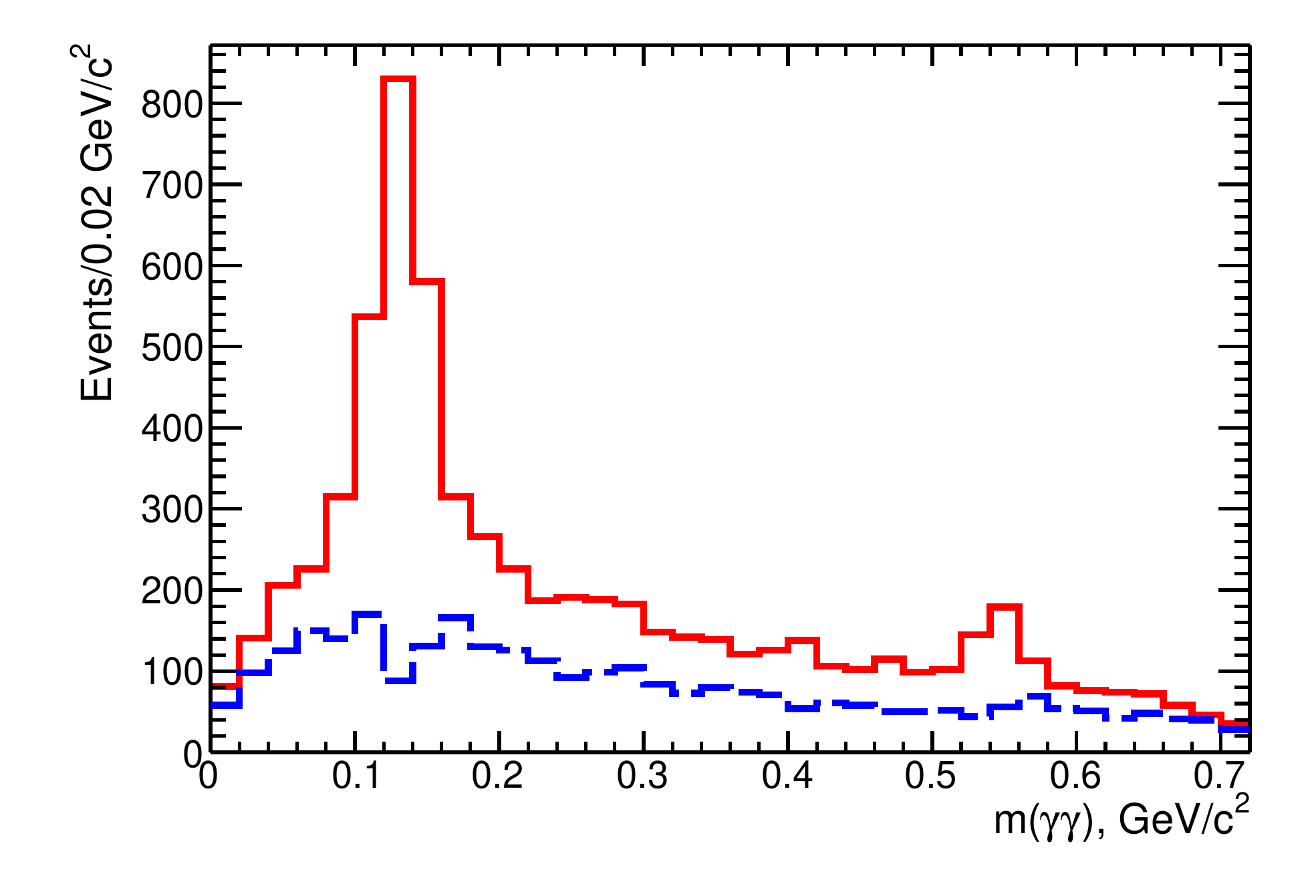}
\put(-100,100){\makebox(0,0)[lb]{\bf(b)}}
\vspace{-0.2cm}
\caption{(a) The third-photon-pair invariant mass vs
  $m(2(\pipi)\ppz\gamma\gamma)$ for the $uds$ simulation.
(b) Projected events from  (a) for the signal region
$\chisq_{4\pi2\piz\gamma\gamma} <50$ (solid histogram), and
the control region $50<\chisq_{4\pi2\piz\gamma\gamma} <100$
(dashed histogram). 
}
\label{udsbkg}
\end{center}
\end{figure}

\section{The {\boldmath $2(\pipi)3\piz$} final state}
\subsection{\boldmath Number of $2(\pipi)3\piz$ events}\label{sec:signal}

The solid histogram in Fig.~\ref{2pi2pi0_bkg} (a) shows the same $\mgg$
distribution of Fig.~\ref{4pi3pi0_chi2_all} (b) binned in mass intervals of 0.02~\gevcc.
The dashed histogram corresponds instead to the distribution of data from the 
\chisq control region, and the dotted
histogram is the estimated
remaining background from $\epem\to2(\pipi)\ppz$ events produced via ISR. 
No evidence for a peaking background is seen in either
of the two background distributions.
We subtract the background evaluated using the \chisq control region
with the scale factor 1.0, and vary it to 1.5 to check the stability
of the result.
The resulting $\mgg$ distribution is shown in Fig.~\ref{2pi2pi0_bkg} (b).

We fit the data of Fig.~\ref{2pi2pi0_bkg} (b) with a combination of a
signal function, taken from simulation, and a
background function, taken to be a third-order polynomial.
The fit is performed in the $\mgg$ mass range from
0.0 to 0.5~\gevcc.
The result of the fit is shown by the solid and
dashed curves. A total of $12\,559\pm174$
events is obtained.
Note that this number includes a relatively small
peaking background component, due to $\qqbar$ events,
which is discussed in Sect.~\ref{sec:udsbkg}.
The same fit is applied to the corresponding \mgg
distribution in each 0.05~\gevcc interval in the
$2(\pipi)2\piz\gamma\gamma$  invariant mass.
The resulting number of $2(\pipi)3\piz$ event candidates
as a function of $m(2(\pipi)2\piz\gamma\gamma)$, including the
peaking $\qqbar$ background, is reported 
 in Fig.~\ref{nev_4pi3pi0_data} (a).
\input{xs4pi3pi0_table}

\begin{figure}[tbh]
\begin{center}
\includegraphics[width=1.05\linewidth]{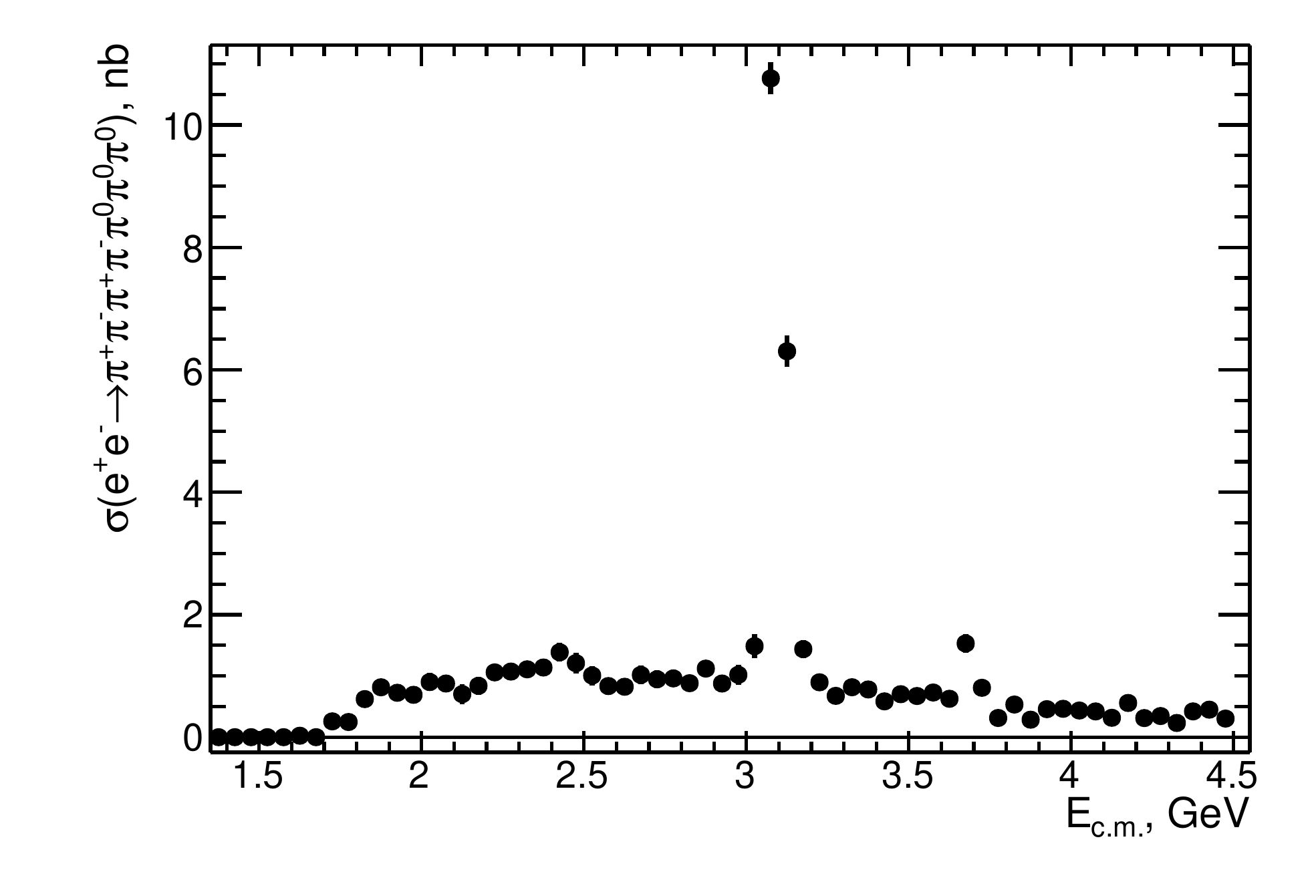}
\vspace{-1.0cm}
\caption{
The  measured $\epem\to2(\pipi)\ppz\piz$ cross section.
The uncertainties are statistical only.
}
\label{4pi3pi0_ee_babar}
\end{center}
\end{figure}

%
%
\begin{figure*}[tbh]
\begin{center}
\includegraphics[width=0.315\linewidth]{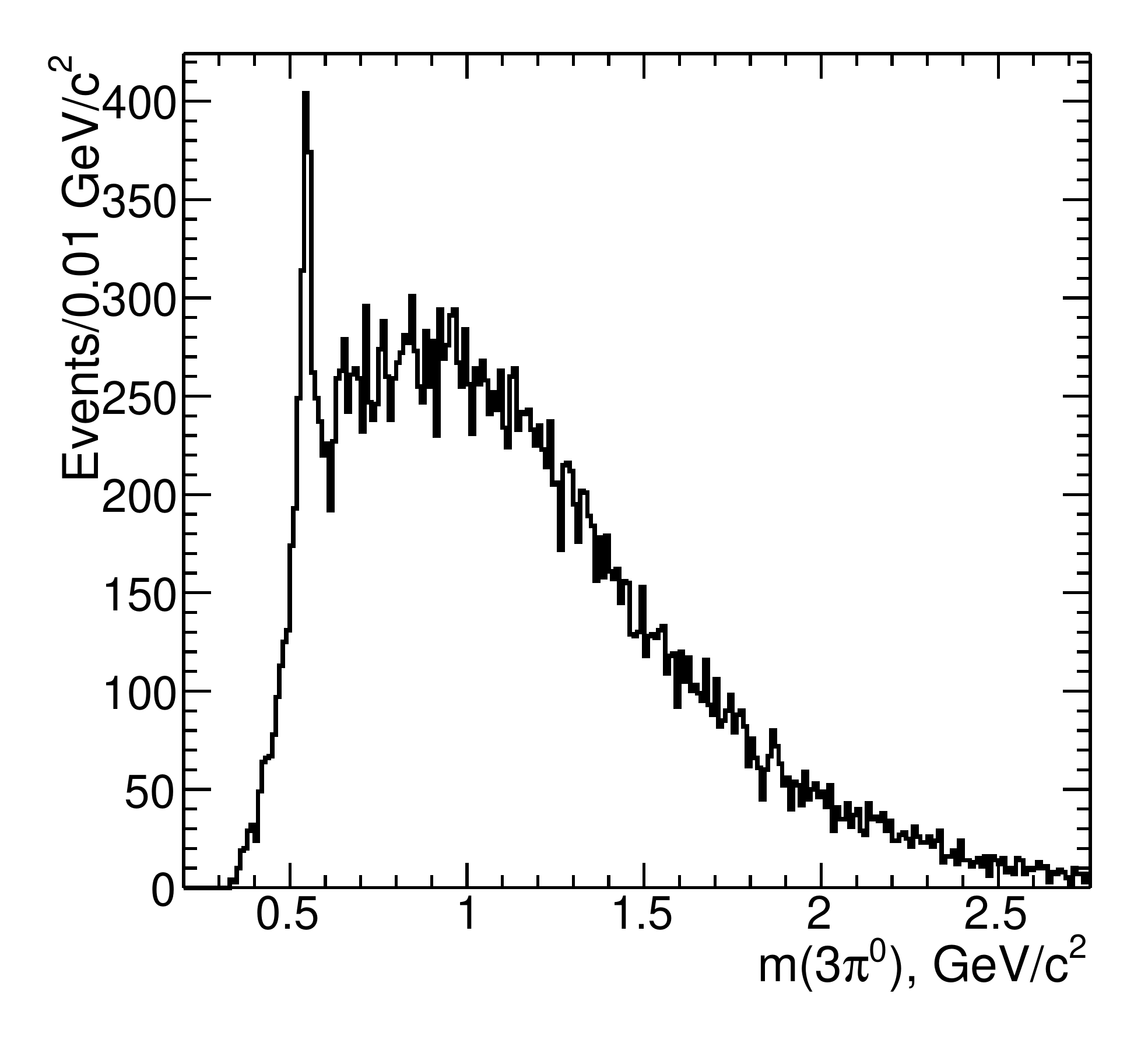}
\put(-50,120){\makebox(0,0)[lb]{\bf(a)}}
\includegraphics[width=0.315\linewidth]{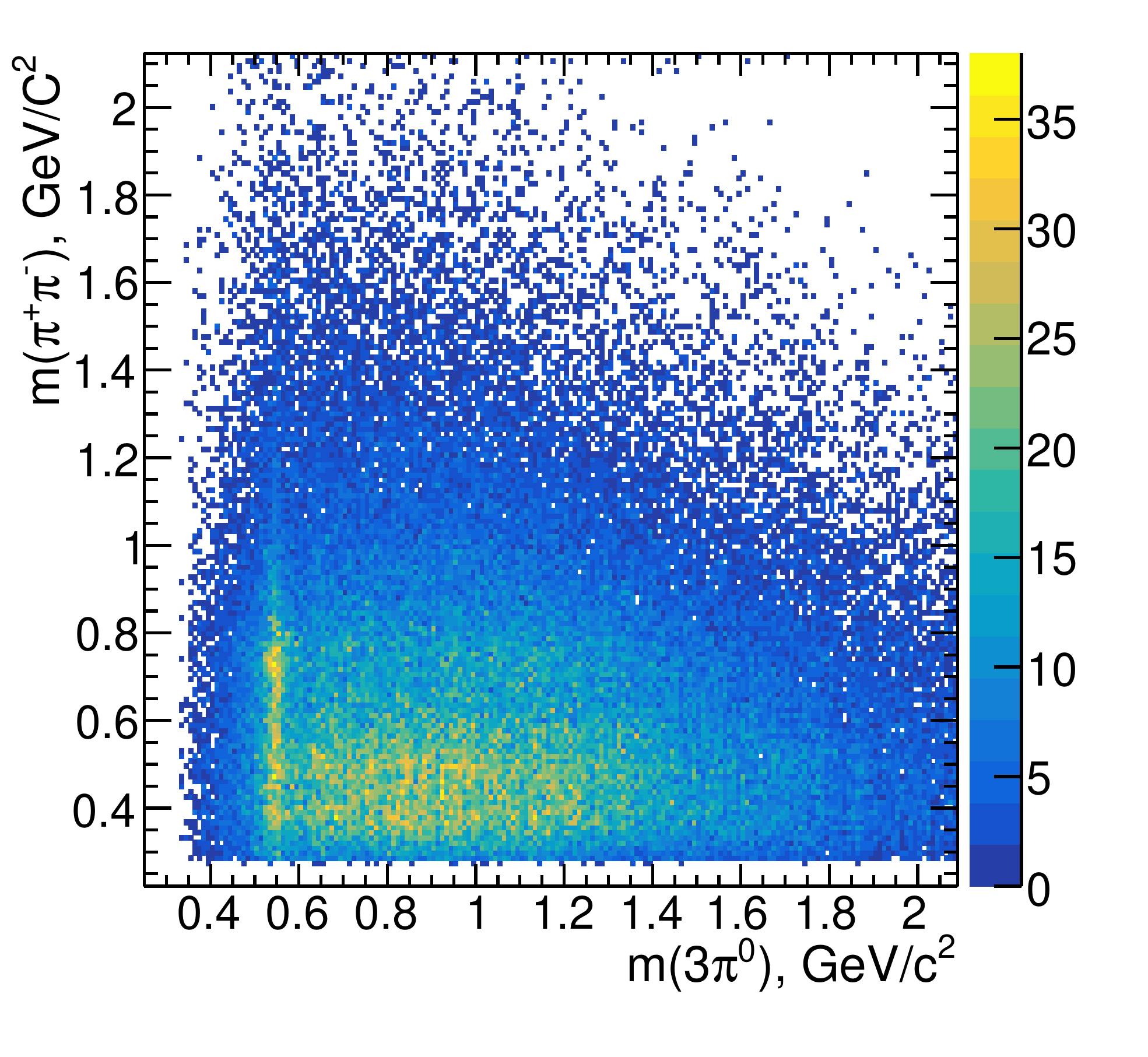}
\put(-50,120){\makebox(0,0)[lb]{\bf(b)}}
\includegraphics[width=0.315\linewidth]{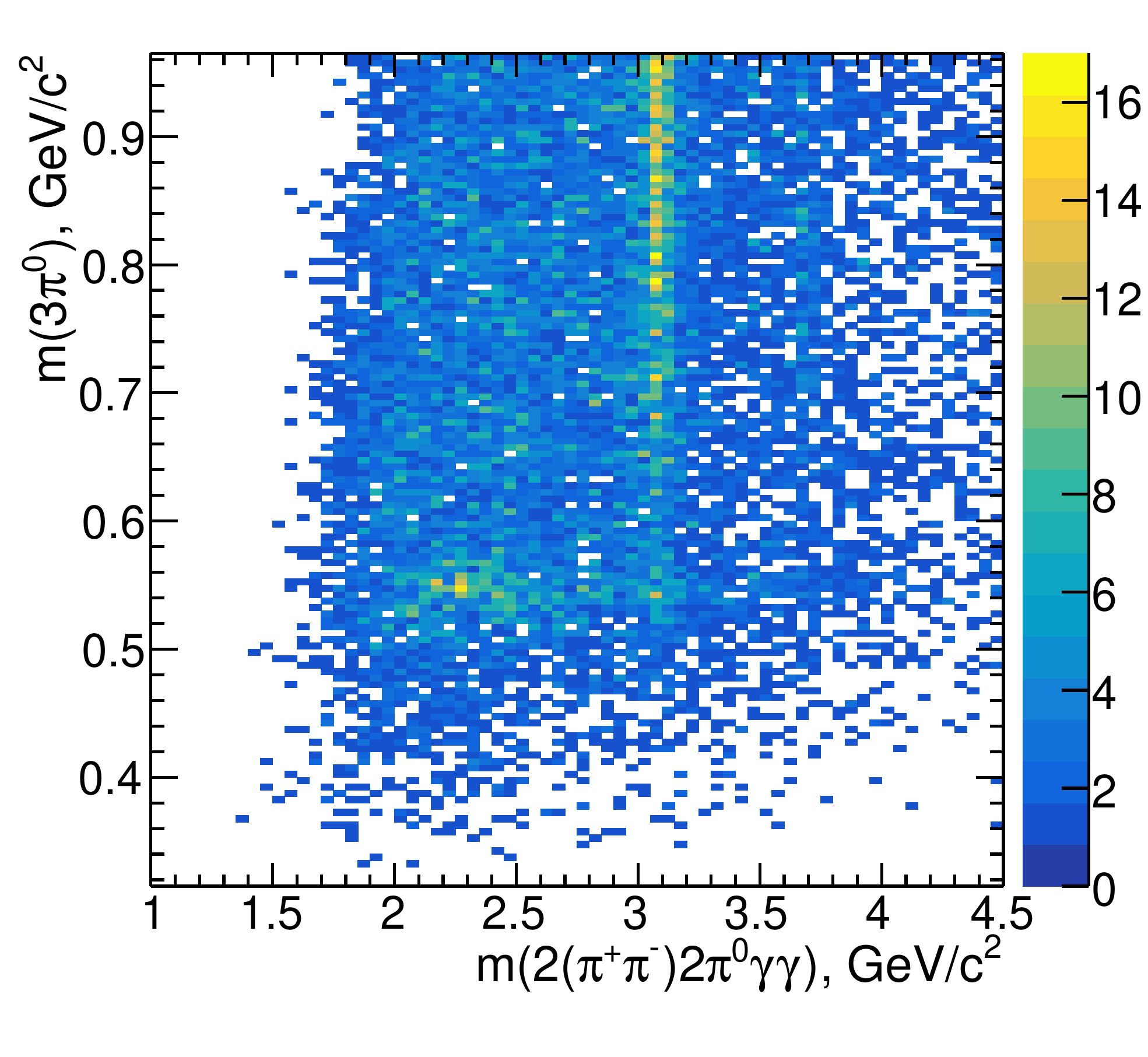}
\put(-135,120){\makebox(0,0)[lb]{\bf(c)}}
\vspace{-0.5cm}
\caption{
(a) The $\ppz\pi^0$  invariant mass.
(b) The $\pipi$ vs the $\ppz\pi^0$ invariant mass.
(c)  The $\ppz\pi^0$  invariant mass vs the seven-pion invariant mass.
}
\label{3pi0vs7pi}  
\end{center}
%
\begin{center}
\includegraphics[width=0.315\linewidth]{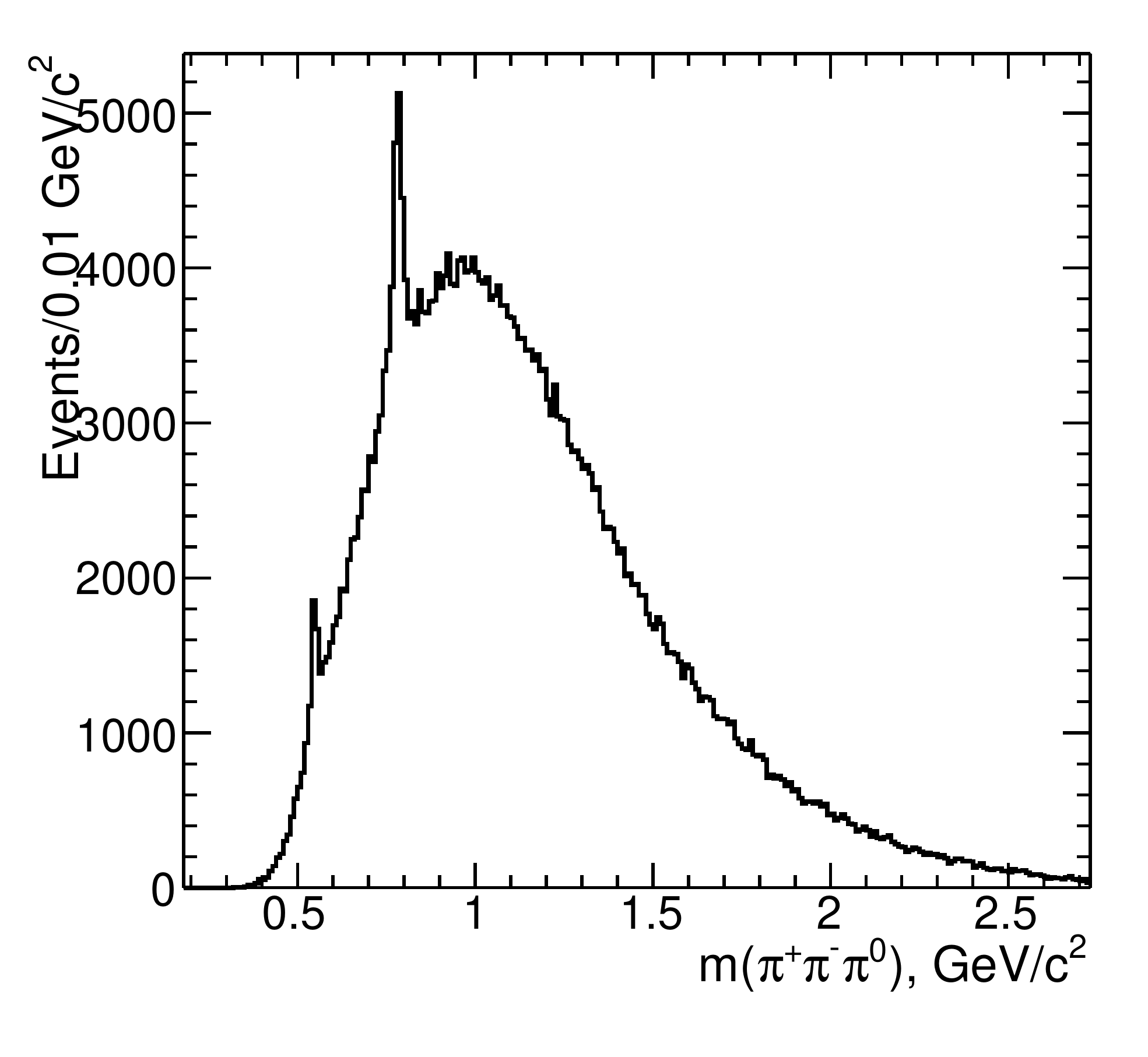}
\put(-50,120){\makebox(0,0)[lb]{\bf(a)}}
\includegraphics[width=0.315\linewidth]{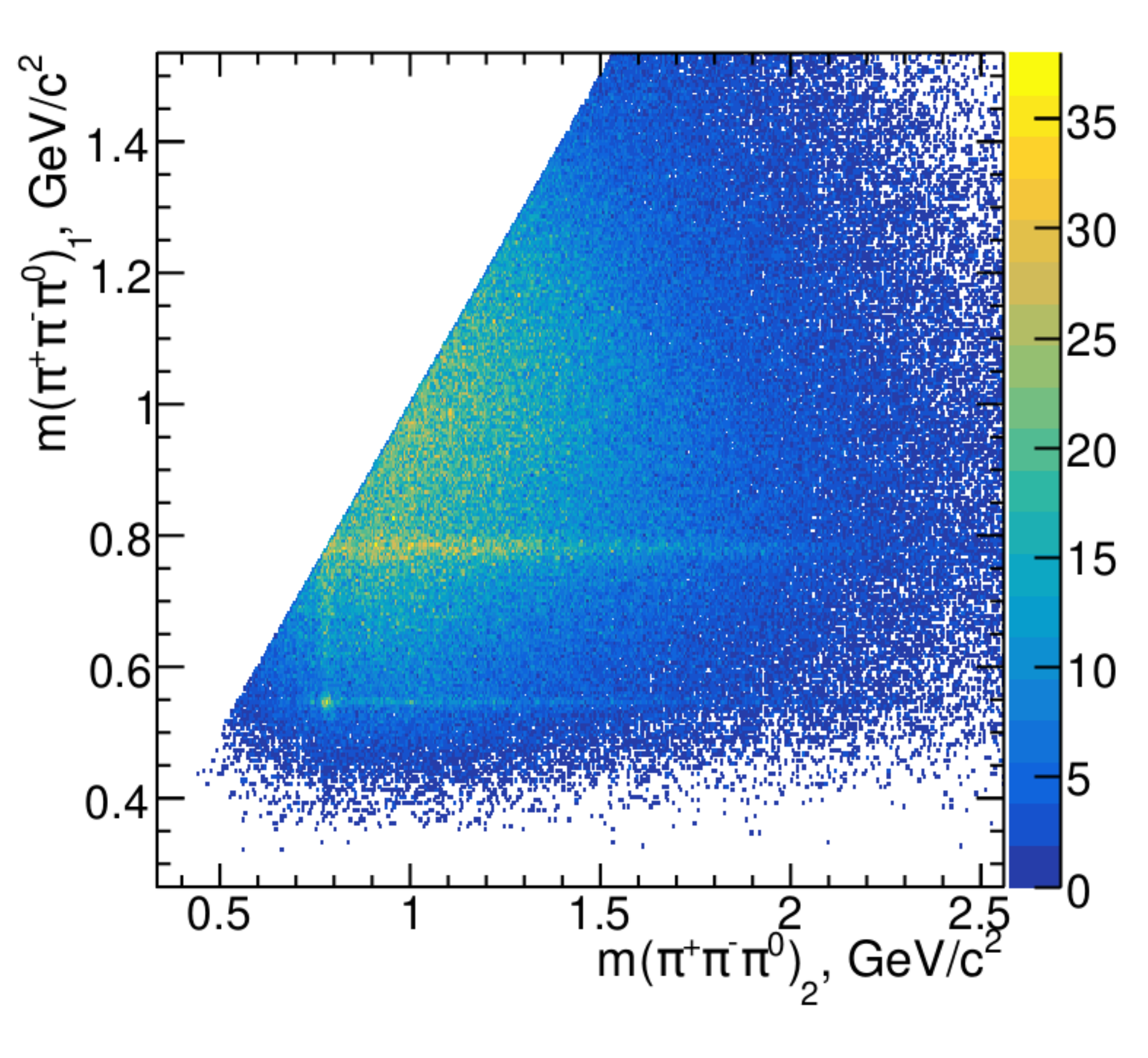}
\put(-130,120){\makebox(0,0)[lb]{\bf(b)}}
\includegraphics[width=0.315\linewidth]{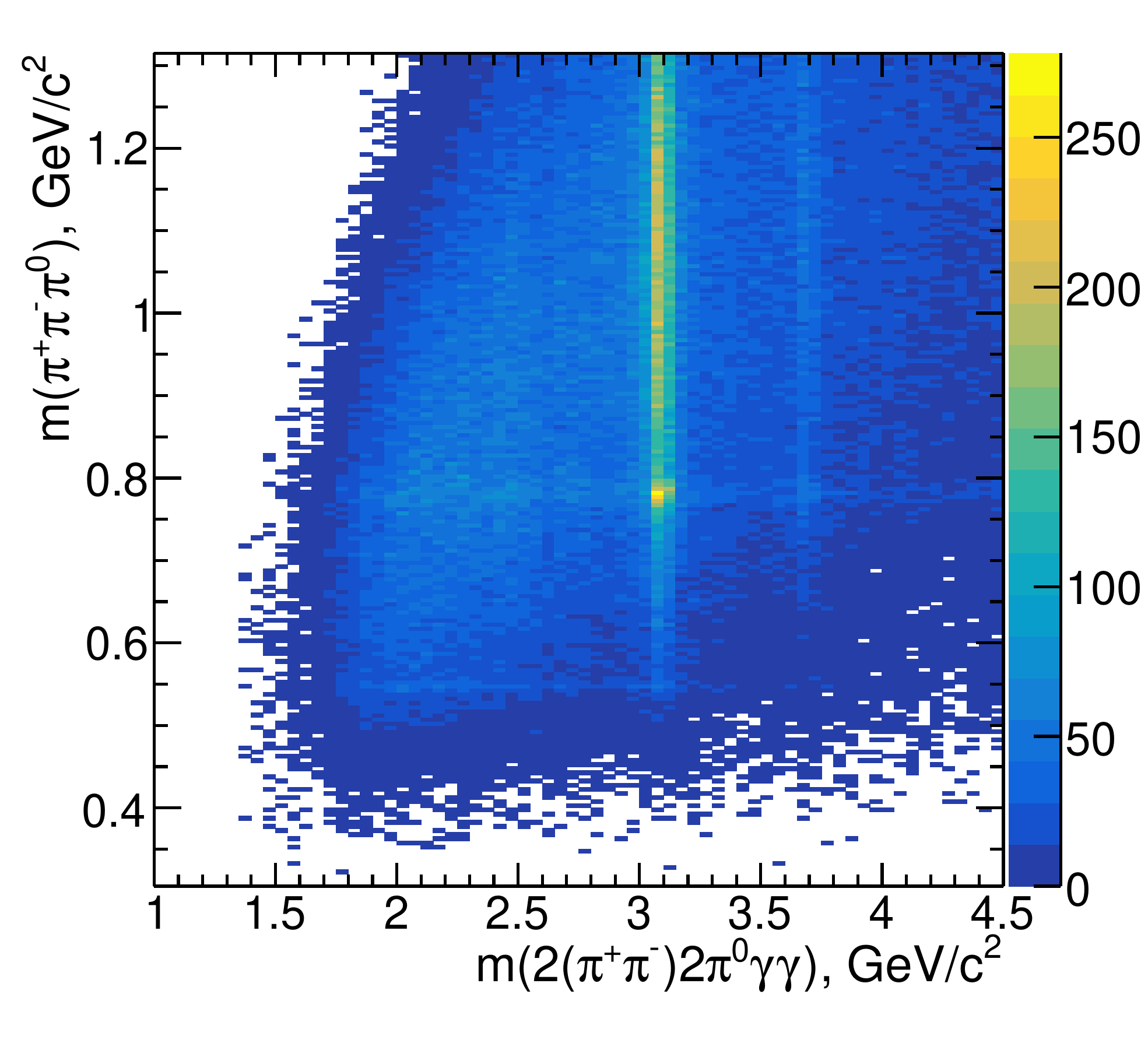}
\put(-135,120){\makebox(0,0)[lb]{\bf(c)}}
\vspace{-0.5cm}
\caption{
(a) The $\pipi\piz$  invariant mass (twelve combinations per event).
(b) The $\pipi\piz$ vs the $\pipi\pi^0$ invariant mass.
(c)  The $\pipi\piz$  invariant mass vs the seven-pion invariant mass.
}
\label{3pivs7pi}  
\end{center}
%
\begin{center}
\includegraphics[width=0.315\linewidth]{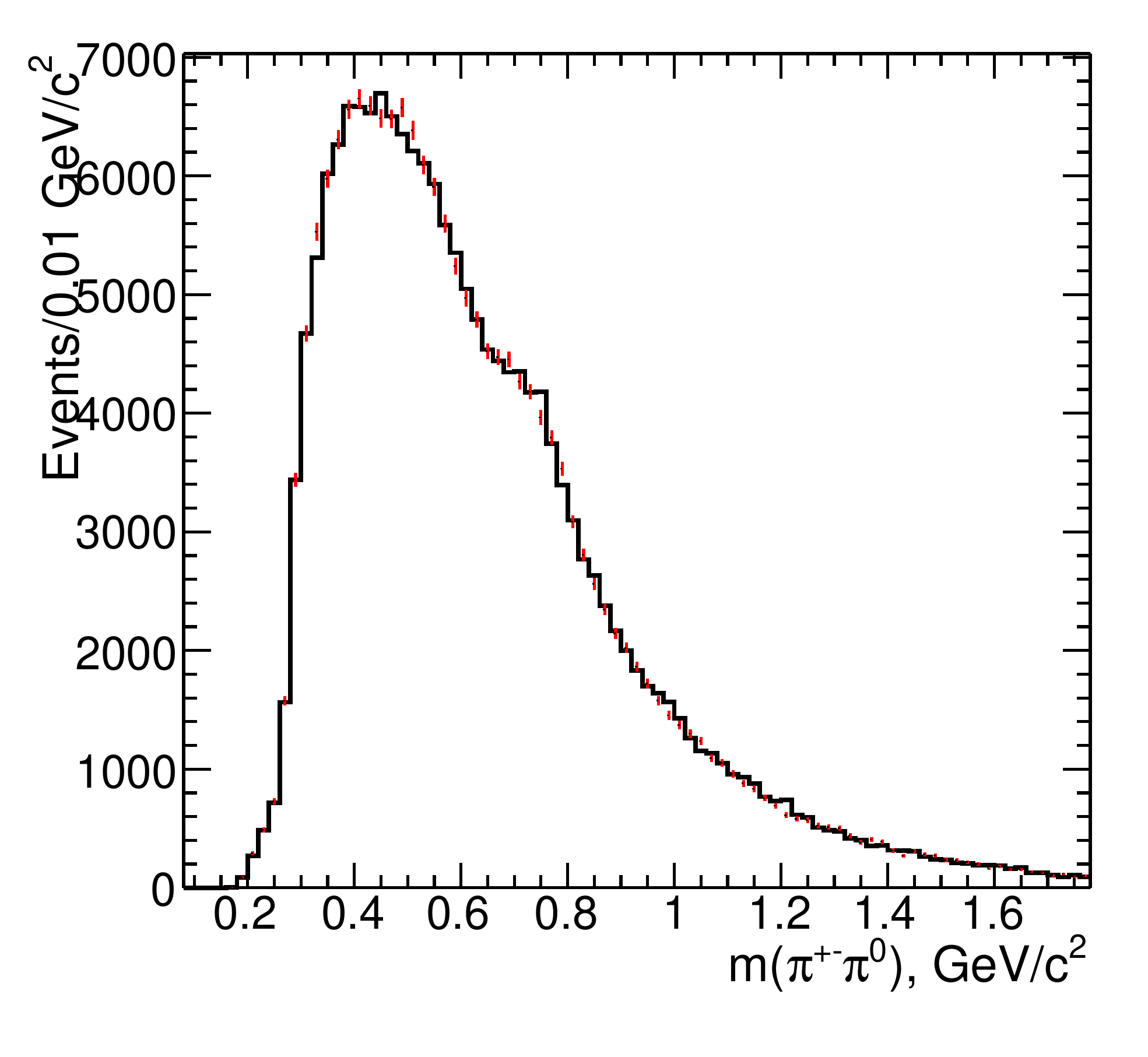}
\put(-50,120){\makebox(0,0)[lb]{\bf(a)}}
\includegraphics[width=0.315\linewidth]{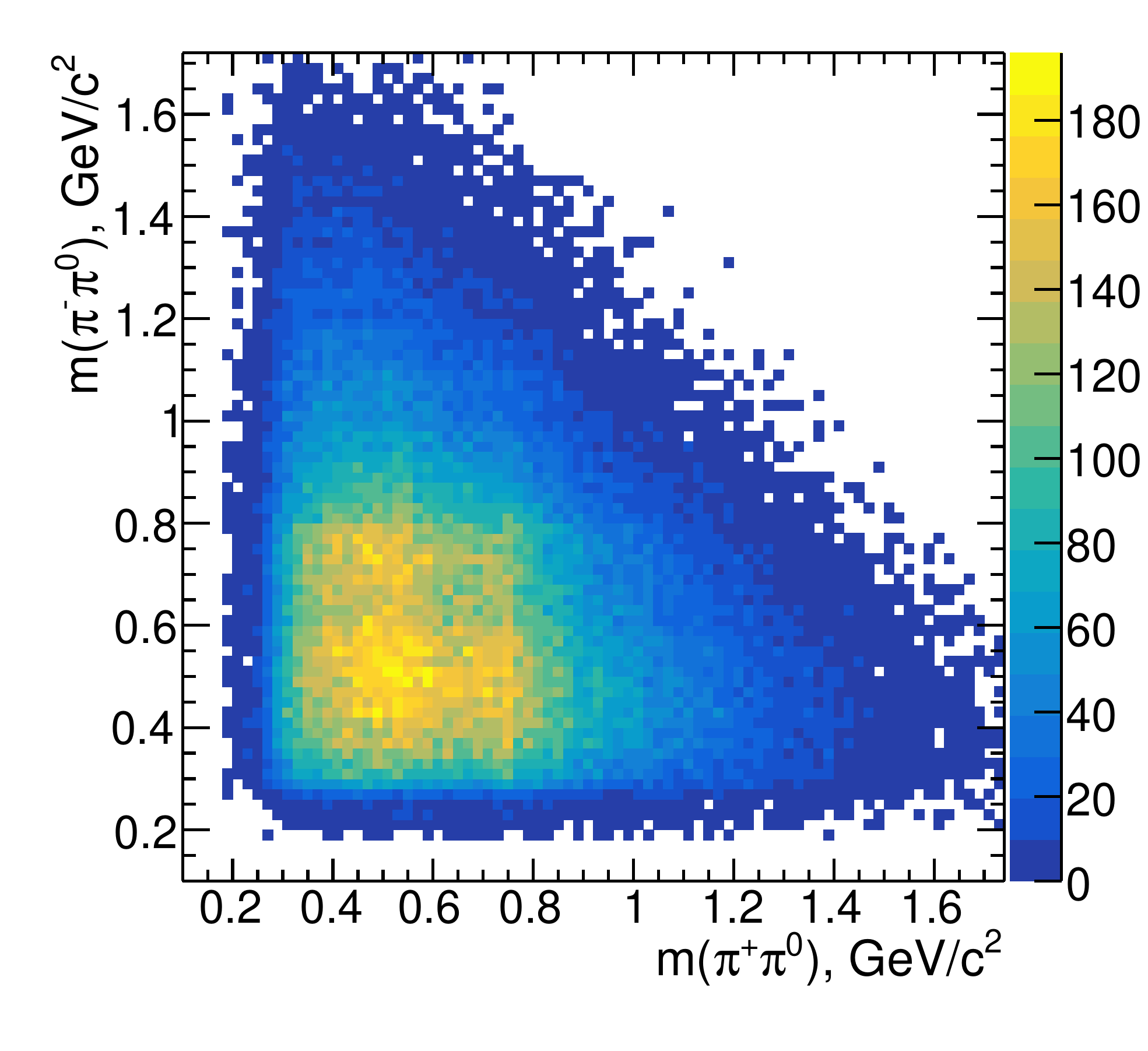}
\put(-50,120){\makebox(0,0)[lb]{\bf(b)}}
\includegraphics[width=0.315\linewidth]{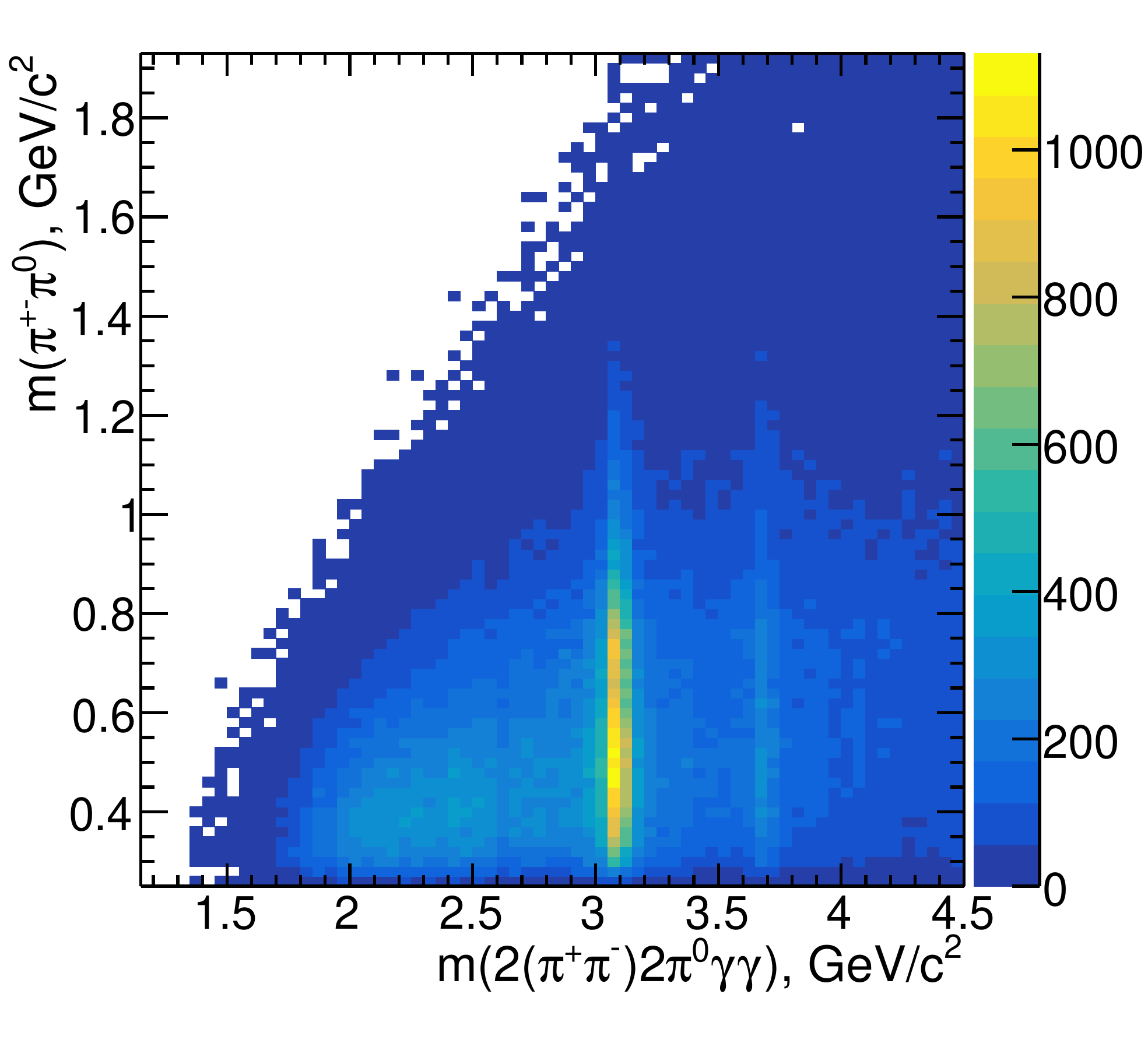}
\put(-130,120){\makebox(0,0)[lb]{\bf(c)}}
\vspace{-0.5cm}
\caption{
(a) The $\pi^+\piz$ (solid) and $\pi^-\piz$ (dashed)  invariant masses
(twelve combinations per event). 
(b) The $\pi^-\piz$ vs the $\pi^+\piz$ invariant mass.
(c)  The $\pi^{\pm}\piz$  invariant mass vs the seven-pion invariant mass.
}
\label{pipi0vs7pi}  
\end{center}
\end{figure*}

\subsection{Peaking background}\label{sec:udsbkg}
The major background producing a $\piz$ peak
following application of the
selection criteria of Sect. IV.A is
from non-ISR \qqbar events, the most important
channel being $\epem\to2(\pipi)\ppz\ppz$
in which one of the
neutral pions decays asymmetrically, 
yielding a high energy photon
that mimics an ISR photon.
Figure~\ref{udsbkg} (a) shows the third-photon-pair invariant
mass vs $m(2(\pipi)\ppz\gamma\gamma)$ for the non-ISR light quark
$\qqbar$ ($uds$) simulation:
clear signals from $\piz$ and 
$\eta$ are seen. Figure~\ref{udsbkg}(b) shows the $m(\gamma\gamma)$ projection
for  $\chisq_{4\pi2\piz\gamma\gamma} <50$ and 
$50<\chisq_{4\pi2\piz\gamma\gamma} <100$. 

To normalize the $uds$ simulation, we calculate the diphoton invariant mass
distribution of the ISR candidate with all the remaining candidate photon in
the event.  A $\piz$ peak is observed, with approximately the same
number of events in data and simulation, leading
to a normalization factor of $1.0\pm0.1$.
The resulting $uds$ background is shown by the squares
in Fig.~\ref{nev_4pi3pi0_data} (b), the $uds$ background is negligible
below 2~\gevcc, but accounts for more
than half of the total event yield around 
4~\gevcc and above.

\begin{table}[b]
\caption{
Summary of the systematic uncertainties in the $\epem\to
\pipi\ppz\pi^0$ cross section measurement.
}
\label{error_tab}
\begin{tabular}{l c c} 
Source & Correction & Uncertainty\\
\hline
Luminosity  &  --  &  $1\%$ \\
MC-data difference in ISR\\ photon efficiency & +1.5\%  & $1\%$\\ 
\chisq cut uncertainty & -- & $3\%$\\
MC-data difference in track losses & $+4\%$ & $2\%$ \\
MC-data difference in $\pi^0$ losses & $+9\%$ & $3\%$ \\
Radiative corrections accuracy & -- & $1\%$ \\
Efficiency from MC \\(model-dependence) & -- & $5\%$  \\
\hline
Total (assuming no correlations)    &  $+14.5\%$   & $10\%$  \\
\end{tabular}
\end{table}
\subsection{\boldmath Cross section for $\epem\to 2(\pipi)\ppz\piz$}
\label{4pi3pi0}

The $\epem\to2(\pipi)\ppz\piz$ Born cross section is  determined from
\begin{equation}
  \sigma(4\pi3\piz)(\Ecm)
  = \frac{dN_{7\pi\gamma}(\Ecm)}
         {d{\cal L}(E_{\rm c.m.})\epsilon_{7\pi}^{\rm corr}
          \epsilon_{7\pi}^{\rm MC}(E_{\rm c.m.})(1+\delta_R)}\ ,
\label{xseq}
\end{equation}
where $\Ecm$ is the invariant mass of
the seven-pion system, $dN_{7\pi\gamma}$ is the background-subtracted number of selected
events in the interval $dE_{\rm c.m.}$,  and
$\epsilon_{7\pi}^{\rm MC}(E_{\rm c.m.})$ is the corresponding detection
efficiency from simulation. The factor $\epsilon_{7\pi}^{\rm corr}$ 
accounts for the difference between data and
simulation: the MC efficiency is larger by 
(1.0$\pm$1.0)\%/per charged track~\cite{isr4pi} and 
by (3.0$\pm$1.0)\% per $\piz$~\cite{isr2pi2pi0}.  
The ISR differential luminosity~\cite{isr3pi}, $d{\cal L}$, is calculated using the 
total integrated \babar~ luminosity of 469 fb$^{-1}$~\cite{lumi}.
The initial- and final-state soft-photon emission is accounted for
by the radiative correction factor $(1+\delta_R)$, which is
close to unity for our selection criteria.
The cross section results contain the effect of
 vacuum polarization because this effect is not accounted for in
the luminosity calculation.

Our results for the $\epem\to2(\pipi)\ppz\piz$ cross section
are shown in Fig.~\ref{4pi3pi0_ee_babar}.  The cross section does not
exhibit any clear structures except signals from the $J/\psi$ and
$\psi(2S)$ resonances.
Because we present our data in bins of width 0.050~\gevcc, compatible   
with the experimental resolution, we do not apply an unfolding procedure to the data.
Numerical values for the cross section are presented in Table~\ref{4pi3pi0_tab}.
The $J/\psi$ region is discussed below.

\subsection{\boldmath Summary of  systematic uncertaintes}
\label{sec:Systematics}
The systematic 
uncertainties, presented in the previous sections, are summarized in
Table~\ref{error_tab},  along 
with the corrections that are applied to the measurements.

The three corrections applied to the cross sections sum
up to 14.5\%.
The systematic uncertainties are considered to be uncorrelated and are
added in  quadrature, summing to 10\%.
The largest systematic uncertainty arises from the fitting
and background subtraction procedures.
It is estimated by varying the background levels and the parameters of the functions used.

\subsection{Overview of the intermediate structures}
The $\epem\to2(\pipi)\ppz\piz$ process has a rich internal
substructure. 
To study this substructure, we restrict events to $\mgg <
0.35$~\gevcc, eliminating the region populated
by $\epem\to2(\pipi)\ppz\eta$.  We then assume that
the $2(\pipi)2\piz\gamma\gamma$ invariant mass can be taken to
represent $m(2(\pipi)3\piz)$.

Figure~\ref{3pi0vs7pi}(a) shows the distribution of the $\ppz\piz$
invariant mass.  The distribution is seen to exhibit a  
prominent $\eta$ peak, which is due to the
$\epem\to\eta2(\pipi)$ reaction.
Figure~\ref{3pi0vs7pi}(b) presents a scatter plot of the $\pipi$ (four
entries per event) vs the $3\piz$ invariant mass.
From this plot, the $\rho(770)\eta$ correlation in the intermediate state is seen.
Figure~\ref{3pi0vs7pi}(c)  presents a scatter plot of the $3\piz$
invariant mass versus $m(2(\pipi)\ppz\gamma\gamma)$.

The distribution of the $\pipi\piz$ invariant mass (twelve entries per event)
is shown in ~\ref{3pivs7pi}(a).  Prominent $\eta$ and $\omega$ peaks are seen.
The scatter plot in Fig.~\ref{3pivs7pi}(b) shows one $\pipi\piz$  vs
another $\pipi\piz$ invariant mass for the same event. Correlated
$\eta$ and $\omega$ production  from
$\epem\to\omega\piz\eta$ is seen. 
A scatter plot of the $\pipi\piz$ vs the seven-pion 
 mass is shown in Fig.~\ref{3pivs7pi}(c).
A clear signal for a $J/\psi$ peak is also observed.

Figure~\ref{pipi0vs7pi}(a) shows the
$\pi^+\piz$(dotted) and $\pi^-\piz$(solid) invariant masses (twelve entries per
event). A prominent $\rho(770)$ peak, corresponding to  
$\epem\to5\pi\rho$, is visible.
The scatter plot in Fig.~\ref{pipi0vs7pi}(b) shows the $\pi^-\piz$  vs
the $\pi^+\piz$ invariant mass.  An indication of the $\rho^+\rho^-\pipi\piz$
intermediate state is visible. Figure~\ref{pipi0vs7pi}(c) shows
the $\pi^{\pm}\piz$ invariant mass  vs the seven-pion invariant mass: a clear signal for
the $J/\psi$ and an indication of the $\psi(2S)$  are seen. 

\begin{figure}[tbh]
  \begin{center}
    \vspace{-0.3cm}
  \includegraphics[width=0.9\linewidth]{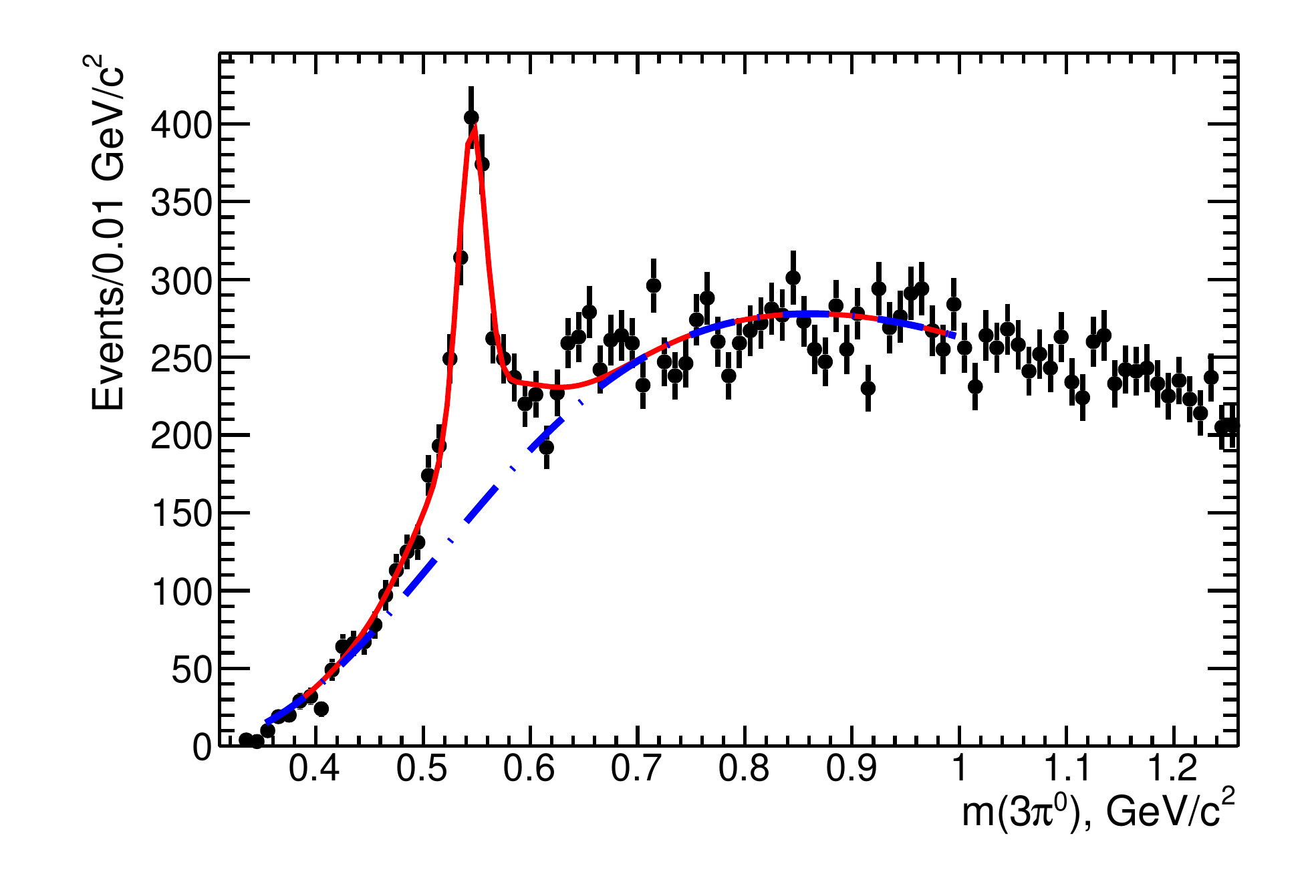}
  \vspace{-0.5cm}
\caption{(a) The $3\piz$ invariant mass for data. 
  The curves show the fit functions.  The solid curve shows 
the $\eta$ peak (based on MC simulation) plus the non-$\eta$ continuum
background (dashed).
}
\label{3pi0slices}
\end{center}
\end{figure}
\begin{figure}[tbh]
  \begin{center}
    \vspace{-0.5cm}
  \includegraphics[width=0.9\linewidth]{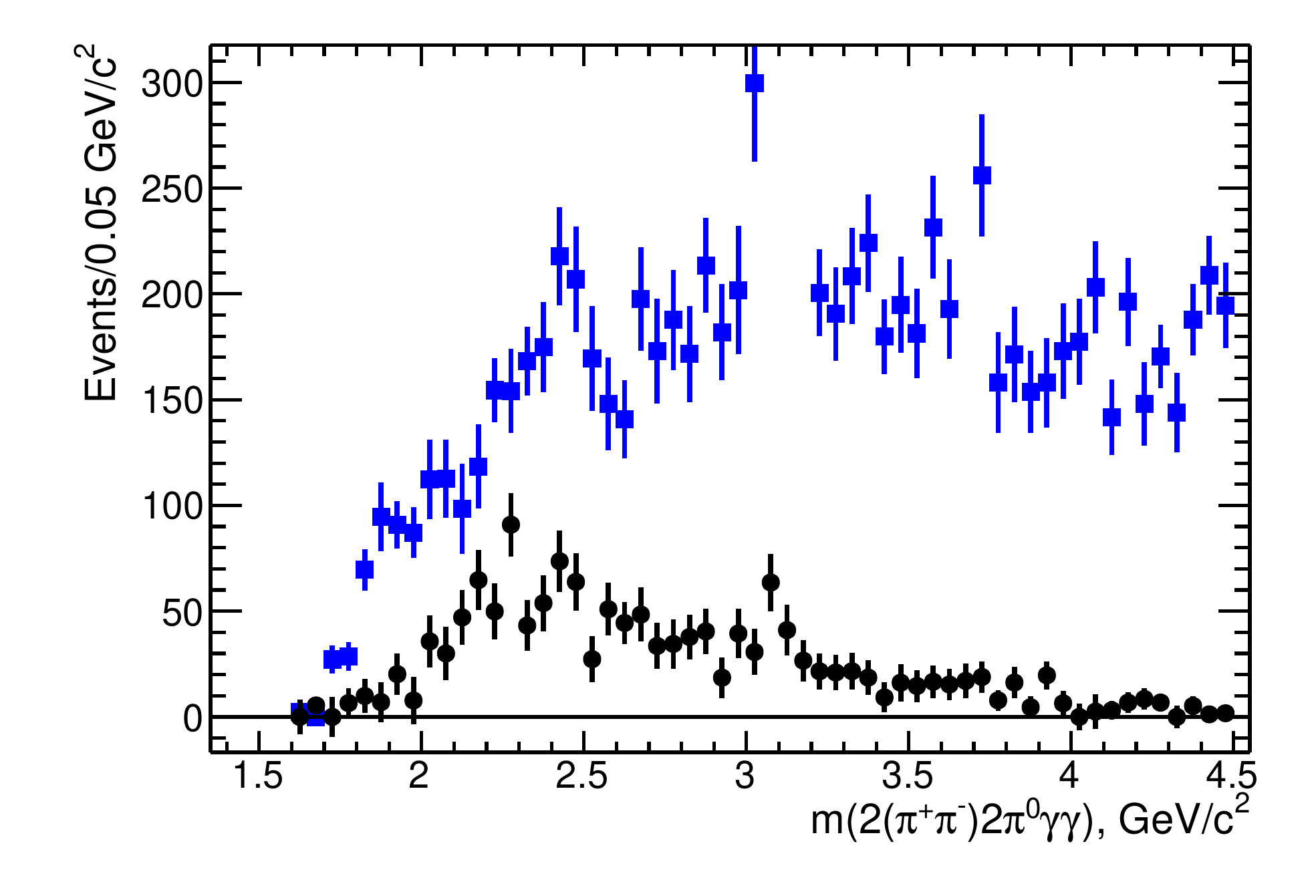}
  \vspace{-0.5cm}
\caption{ The $m(2(\pipi)3\piz)$ invariant mass dependence of the selected data events
for $\epem\to\eta2(\pipi), \eta\to 3\piz$ (dots) in comparison with
all seven-pion events (squares). The $J/\psi$ signal is off-scale.
}
\label{neveta2pi}
\end{center}
\end{figure}

\begin{figure}[tbh]
\begin{center}
  \includegraphics[width=0.95\linewidth]{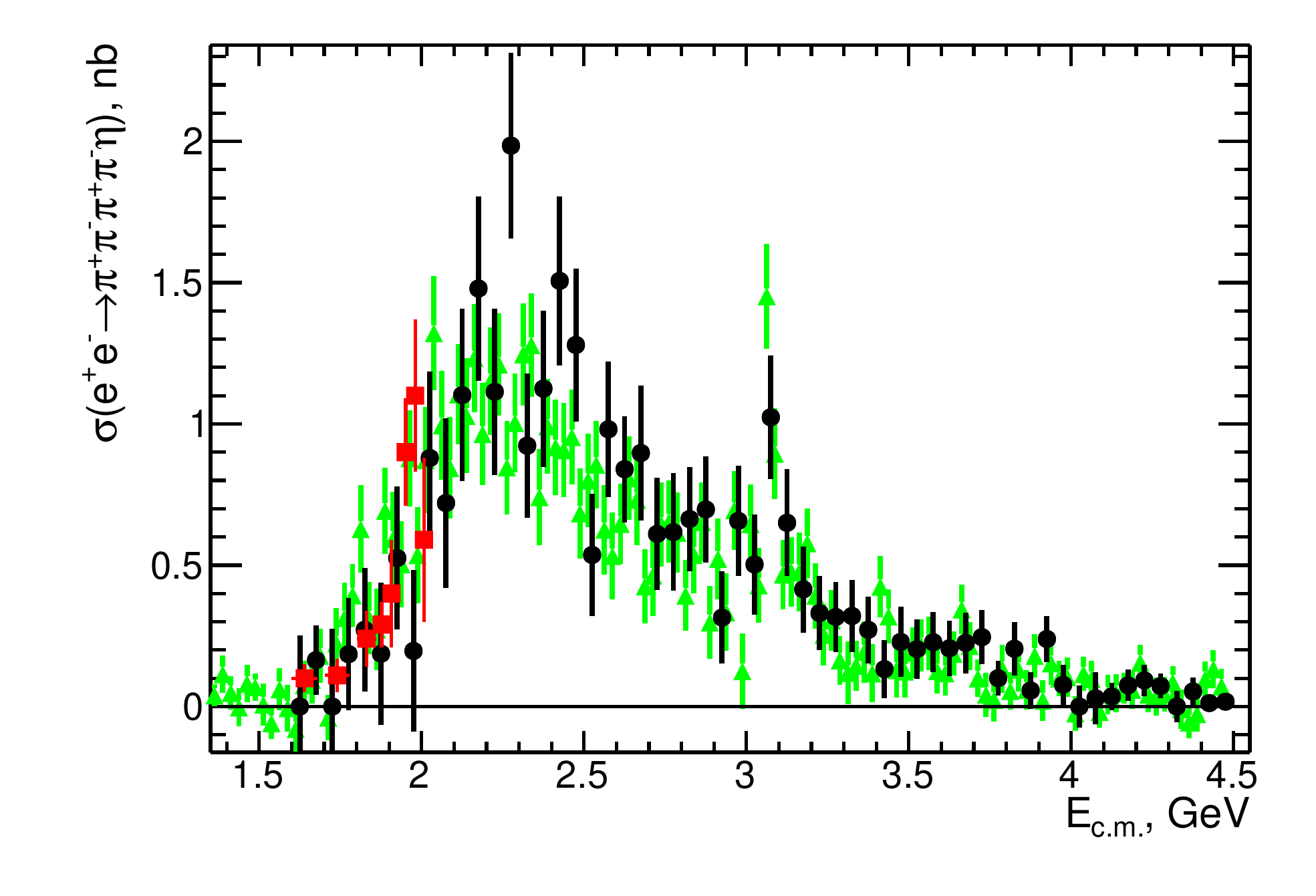}
  \vspace{-0.5cm}
\caption{Comparison of the present results (dots) with previous measurements of
the $\epem\to2(\pipi)\eta$ cross section
from \babar~ in  $\eta\to\gamma\gamma$ (triangles)~\cite{isr5pi} and
from CMD-3  (squares)~\cite{cmd7pi} in  $\eta\to\pipi\piz$.
}
\label{xs_2pieta}
\end{center}
\end{figure}

\input{xs4pieta_table}[b]

\begin{figure}[t]
\begin{center}
\includegraphics[width=0.85\linewidth]{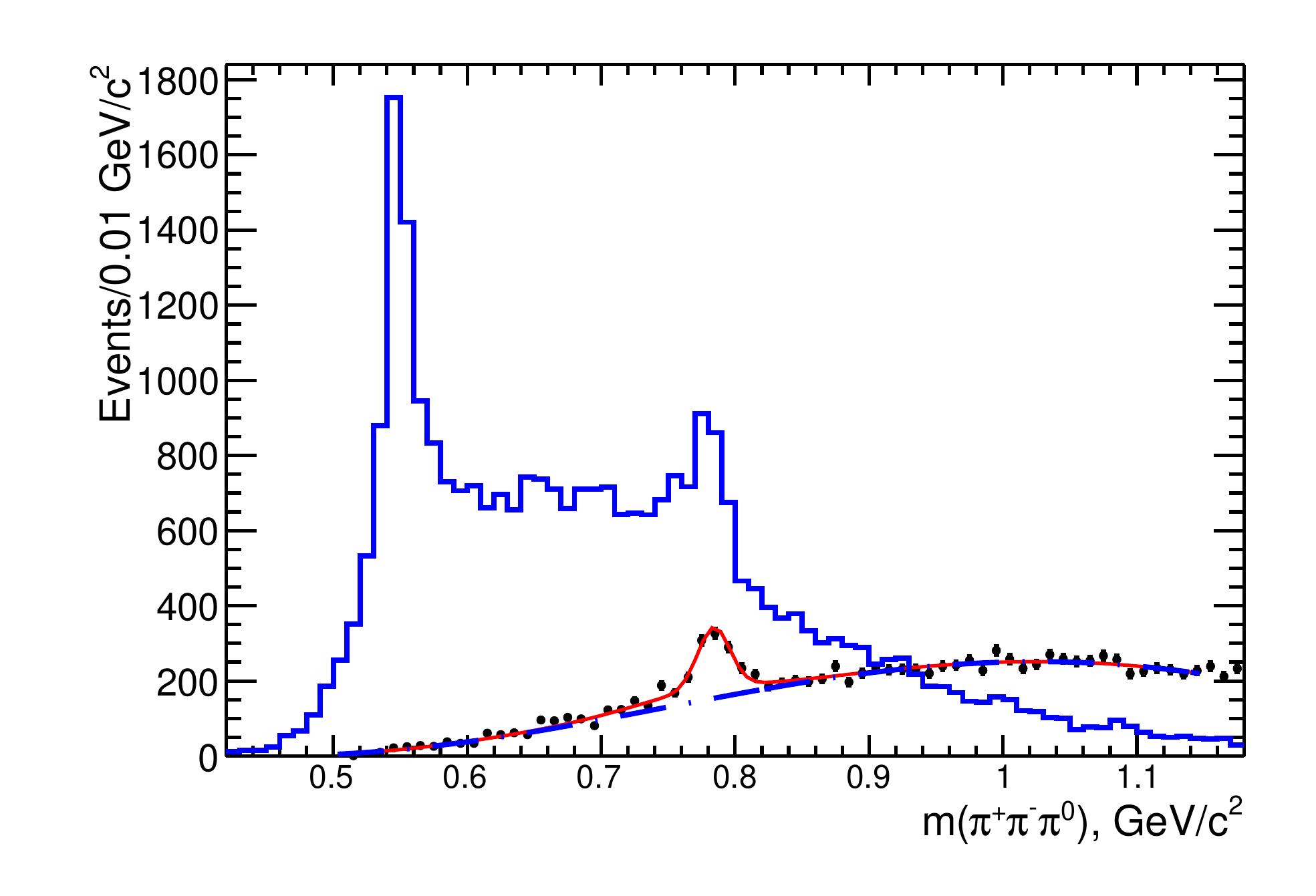}
\put(-60,110){\makebox(0,0)[lb]{\bf(a)}}\\
\vspace{-0.35cm}
\includegraphics[width=0.85\linewidth]{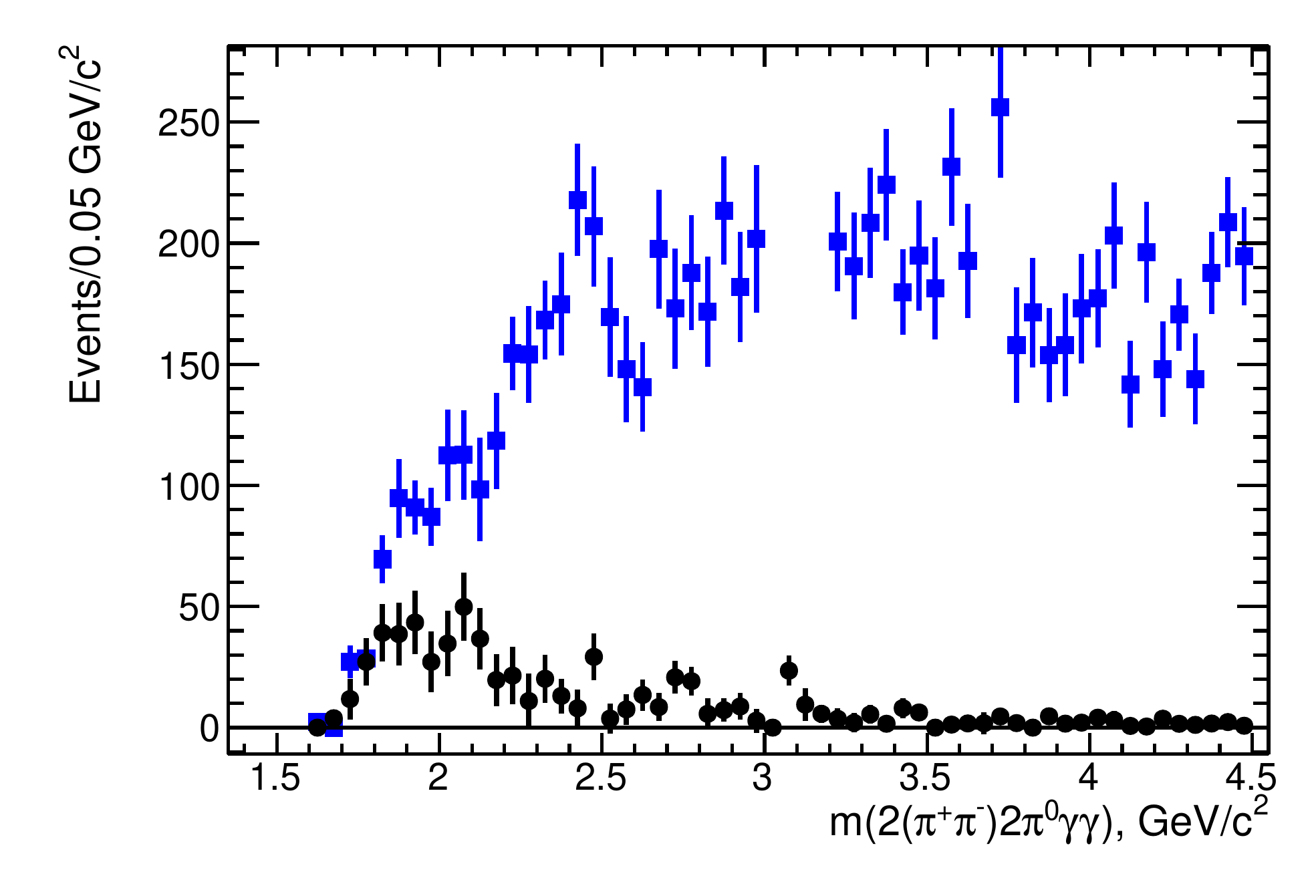}
\put(-150,100){\makebox(0,0)[lb]{\bf(b)}}
\vspace{-0.5cm}
\caption{Mass plots for the $\omega\piz\eta$ intermediate state:
  (a) The solid histogram
  is for the $\pipi\piz$ invariant mass closest to the $\eta$ mass, while the dots are for the two
  remaining combinations of  $\pipi\piz$. The solid curve
  shows the fit function for the $\omega$ signal plus
 the combinatorial background (dashed curve).
(b) The mass distribution of the $2(\pipi)3\piz$ events in the 
$\omega$ peak (circles) correlated with $\eta$ production in
comparison with all $2(\pipi)3\piz$ events(squares).
}
\label{nevomegapi0eta}
\end{center}
\end{figure}

\subsubsection{\bf\boldmath The $\eta2(\pipi)$ intermediate state}
\label{Sec:etapipi}
To determine the contribution of the $\eta2(\pipi)$
 intermediate state, we fit the events of Fig.~\ref{3pi0vs7pi}(a)
using a triple-Gaussian function to describe the
signal peak, as in Fig.~\ref{m3pi_omega_eta_mc}(a), and a polynomial to
describe the background.
The result of the fit is shown in Fig.~\ref{3pi0slices}(a).
We obtain $1410\pm58$ $\eta2(\pipi)$ events. 
The number of $\eta2(\pipi)$ events as a function of the seven-pion invariant mass
is determined by performing an analogous fit 
in each 0.05~\gevcc interval of $m(2(\pipi)3\piz)$.  
The resulting distribution is shown in Fig.~\ref{neveta2pi}.

The very rich intermediate structures in the $\eta2(\pipi)$ mode were carefully studied in our
previous paper~\cite{isr5pi} with significantly larger statistical precision.

Using Eq.~(\ref{xseq}), we determine the cross section for the
$\epem\to\eta2(\pipi)$ process. Our simulation takes into account all
$\eta$ decays, so the cross section results, shown in Fig.~\ref{xs_2pieta}
and listed in Table~\ref{2pieta_table}, correspond to all $\eta$ decays.
Systematic uncertainties in this measurement are the same as those listed in Table~\ref{error_tab}.
Figure~\ref{xs_2pieta} shows our measurement
in comparison to our previous result~\cite{isr5pi}  and to
those from the CMD-3 experiment~\cite{cmd7pi}.
These previous results are based on different
$\eta$ decay modes than those considered here.
The different results are seen to agree within the
uncertainties.  Including the results of the present study,
we have thus now measured the  $\epem\to\eta2(\pipi)$ cross
section in three different $\eta$ decay modes.

\begin{figure}[t]
\begin{center}
  \includegraphics[width=1.0\linewidth]{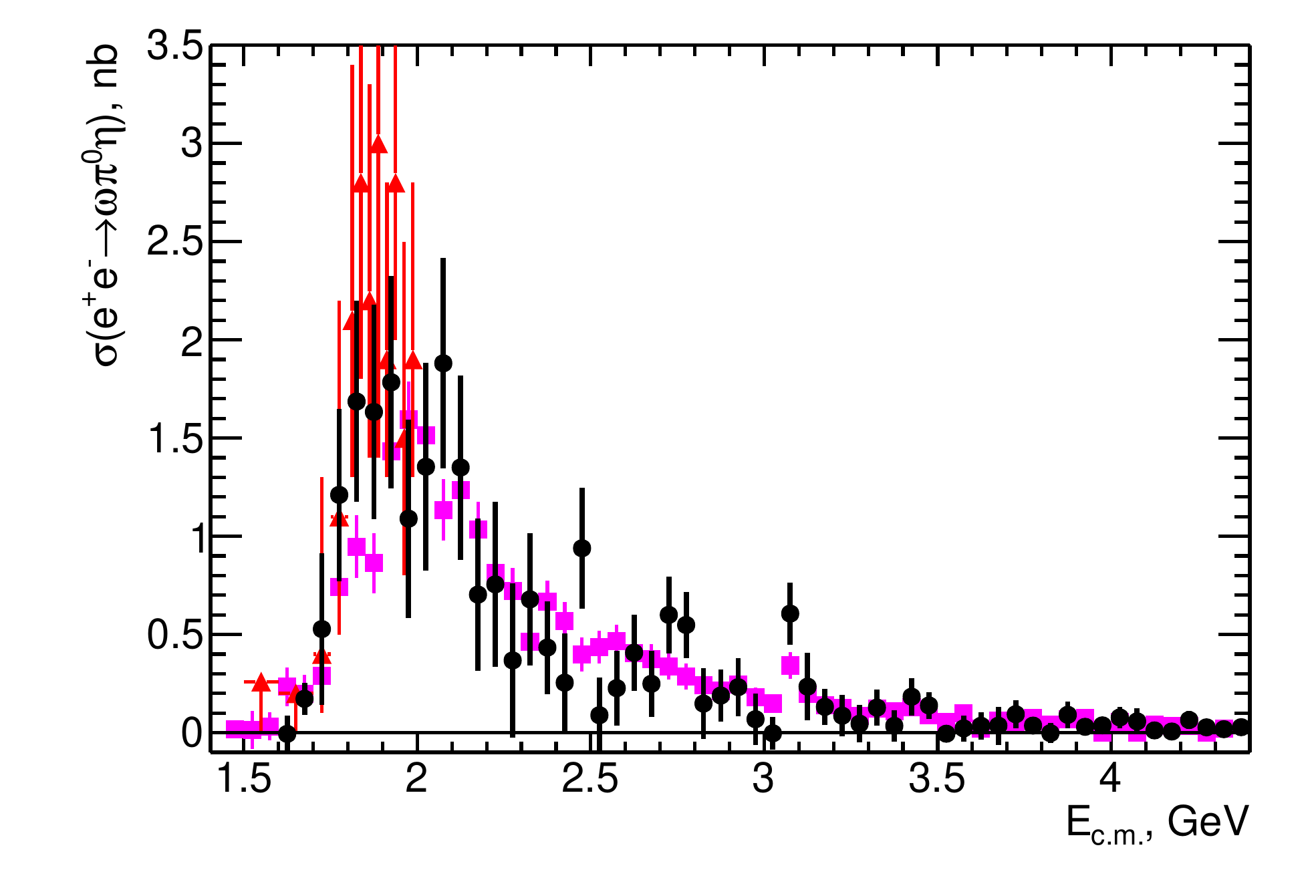}
  \vspace{-0.9cm}
\caption{Comparison of the present results (dots) with previous measurements of
the $\epem\to\omega\piz\eta$ cross section
from \babar~ in  $\eta\to\gamma\gamma$ (squares)~\cite{isr2pi3pi0} and
from SND  (triangles)~\cite{SNDompi0eta} in  $\eta\to3\piz$.
}
\label{xsompi0eta}
\end{center}
\end{figure}

\begin{figure}[t]
  \begin{center}
    \vspace{-0.3cm}
\includegraphics[width=0.9\linewidth]{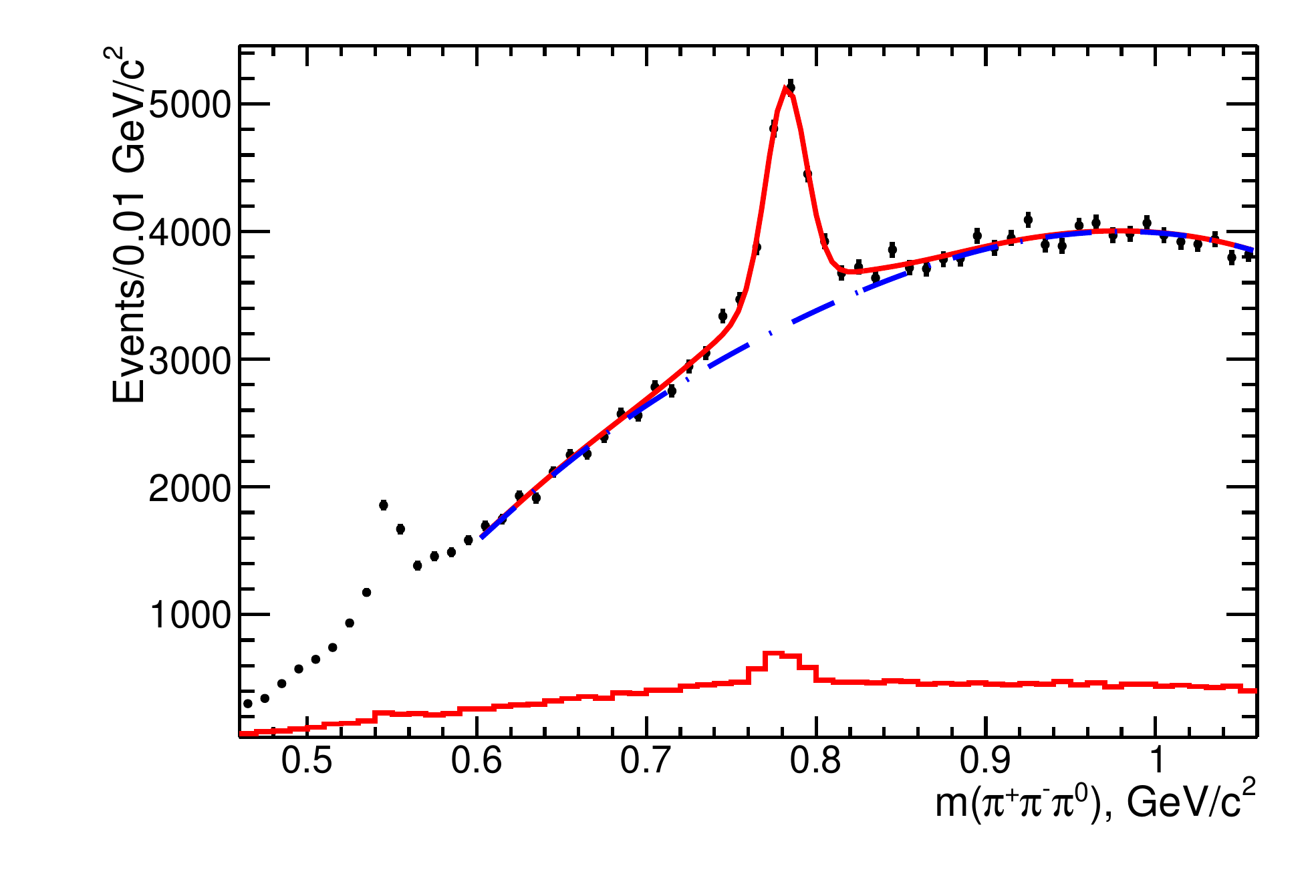}
\put(-60,130){\makebox(0,0)[lb]{\bf(a)}}\\
\vspace{-0.35cm}
\includegraphics[width=0.9\linewidth]{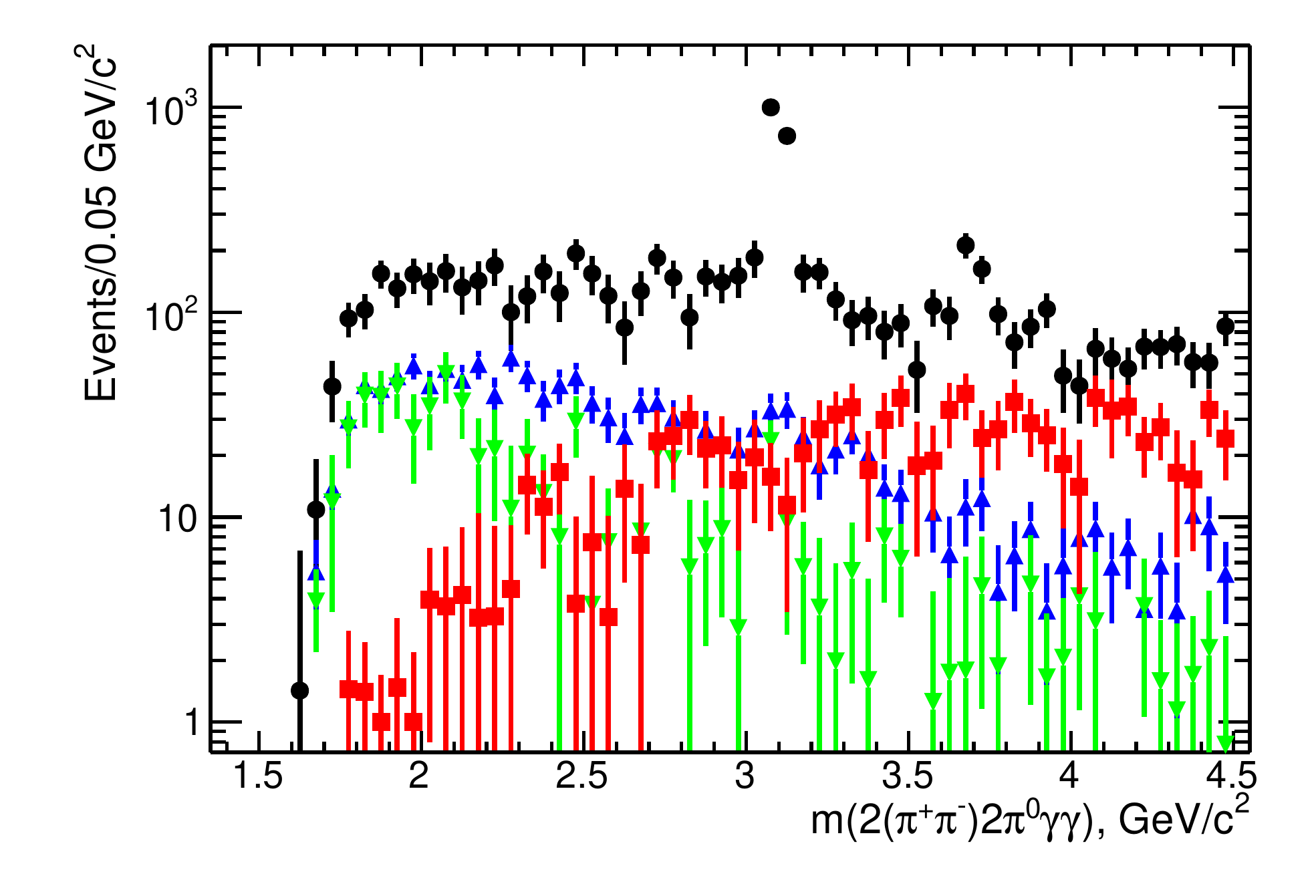}
\put(-50,130){\makebox(0,0)[lb]{\bf(b)}}
\vspace{-0.5cm}
\caption{(a) The $\pipi\piz$ invariant mass for data with the fit
  function for the $\omega$ signal (solid)  plus
 the combinatorial background (dashed curve). The solid histogram shows
 peaking background from the simulated $\epem\to\omega\eta$ ISR events.
(b) The mass distribution of the $2(\pipi)3\piz$ events in the 
$\omega$ peak (circles) and estimated contribution from the
$\omega\eta$ background (triangles), from $\omega\piz\eta$ (up-down
triangles), and from $uds$ (squares).
}
\label{nevomega2pi0}
\end{center}
\end{figure}
\input{xsomegapi0eta_table}

\subsubsection{\bf\boldmath The $\omega\piz\eta$ intermediate state}
\label{Sec:omegapi0eta}

As  demonstrated in  Fig.~\ref{3pivs7pi}(a,b) we can expect
$\eta\pipi\ppz$, $\omega\pipi\ppz$ intermediate final states, or  correlated  $\eta$ and $\omega$
production in the $\omega\piz\eta$ mode.  

The solid histogram in Fig.~\ref{nevomegapi0eta}(a)  shows the mass distribution of
the $\pipi\piz$ combination closest to the nominal $\eta$ mass, while
the dotted histogram reports the invariant mass of the remaining two
combinations of three pions after selecting the first combination
within a window of $\pm 80$~\mevcc from the nominal $\eta$ mass.
A fit to the dotted distribution with a sum of a BW for the $\omega$
signal and a combinatorial background, as shown in Sect.~\ref{sec:efficiency}, allows
the extraction of the  $\omega\eta\piz$ intermediate state signal,
which amounts to 739$\pm$51 events.
The contribution of the $\omega\piz\eta$ intermediate state to
all $2(\pipi)3\piz$ events is shown in Fig.~\ref{nevomegapi0eta}(b).

Using Eq.~(\ref{xseq}), we determine the cross section for the
$\epem\to\omega\piz\eta$ process. The energy dependence of the cross
section is shown in Fig.~\ref{xsompi0eta} by the dots: we are in agreement with our previous
measurement~\cite{isr2pi3pi0} and still slightly below  the SND
result~\cite{SNDompi0eta}. The numerical values of the  cross
section are listed in Table~\ref{ompi0eta_table}.
Again, we have the measurements of this reaction in three different
decay modes of $\eta$.

\begin{figure}[tbh]
\begin{center}
\includegraphics[width=0.95\linewidth]{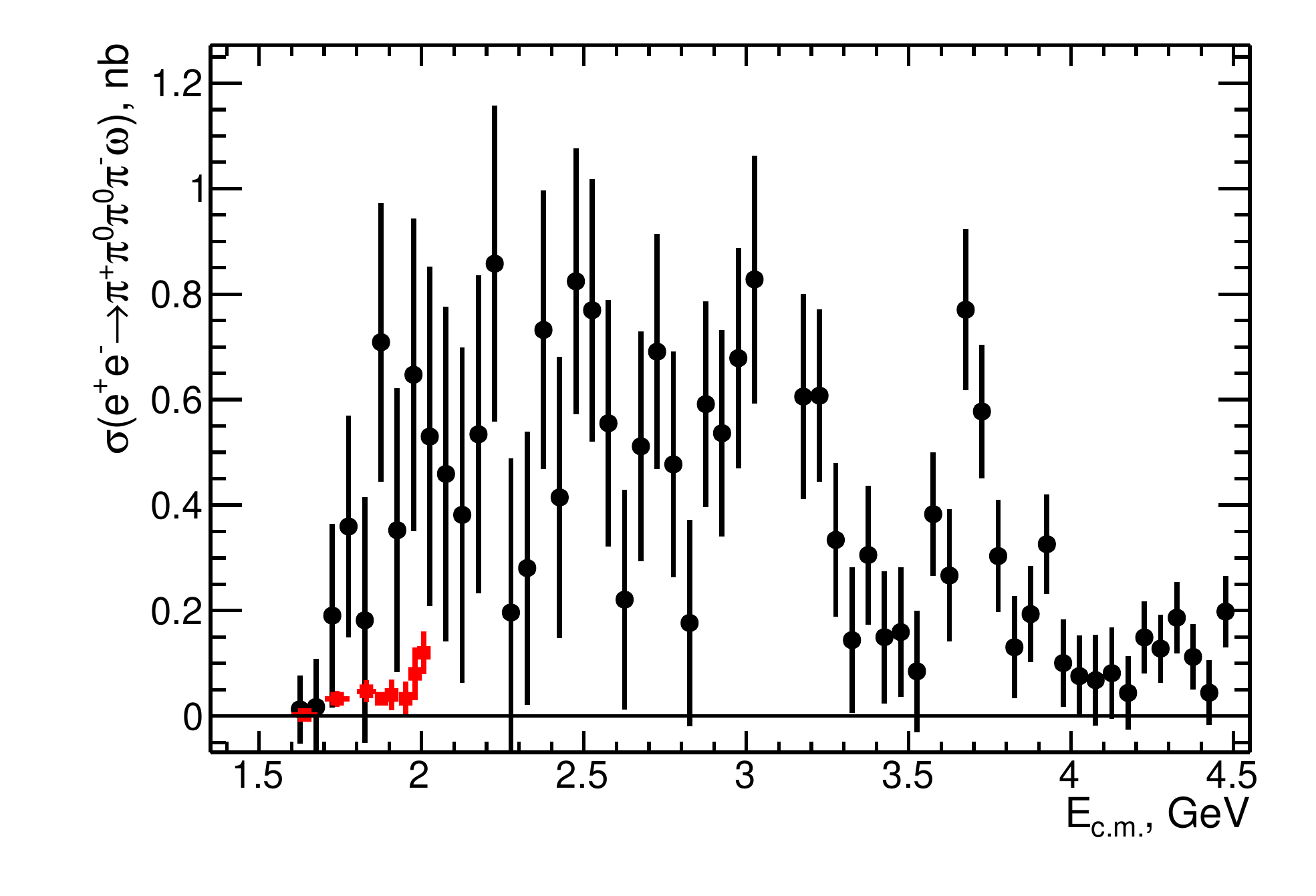}
\vspace{-0.5cm}
\caption{The energy dependent $\epem\to\omega\pipi\ppz$ cross
  section  in the $2(\pipi)3\piz$ mode (the $J/\psi$ signal is
  off-scale). The result of  CMD-3 for the $\epem\to\omega\pipi\pipi$ cross
  section~\cite{cmd7pi} is shown by squares.
}
\label{xs_2pi0omega}
\end{center}
\end{figure}

\begin{figure}[tbh]
\begin{center}
\includegraphics[width=0.9\linewidth]{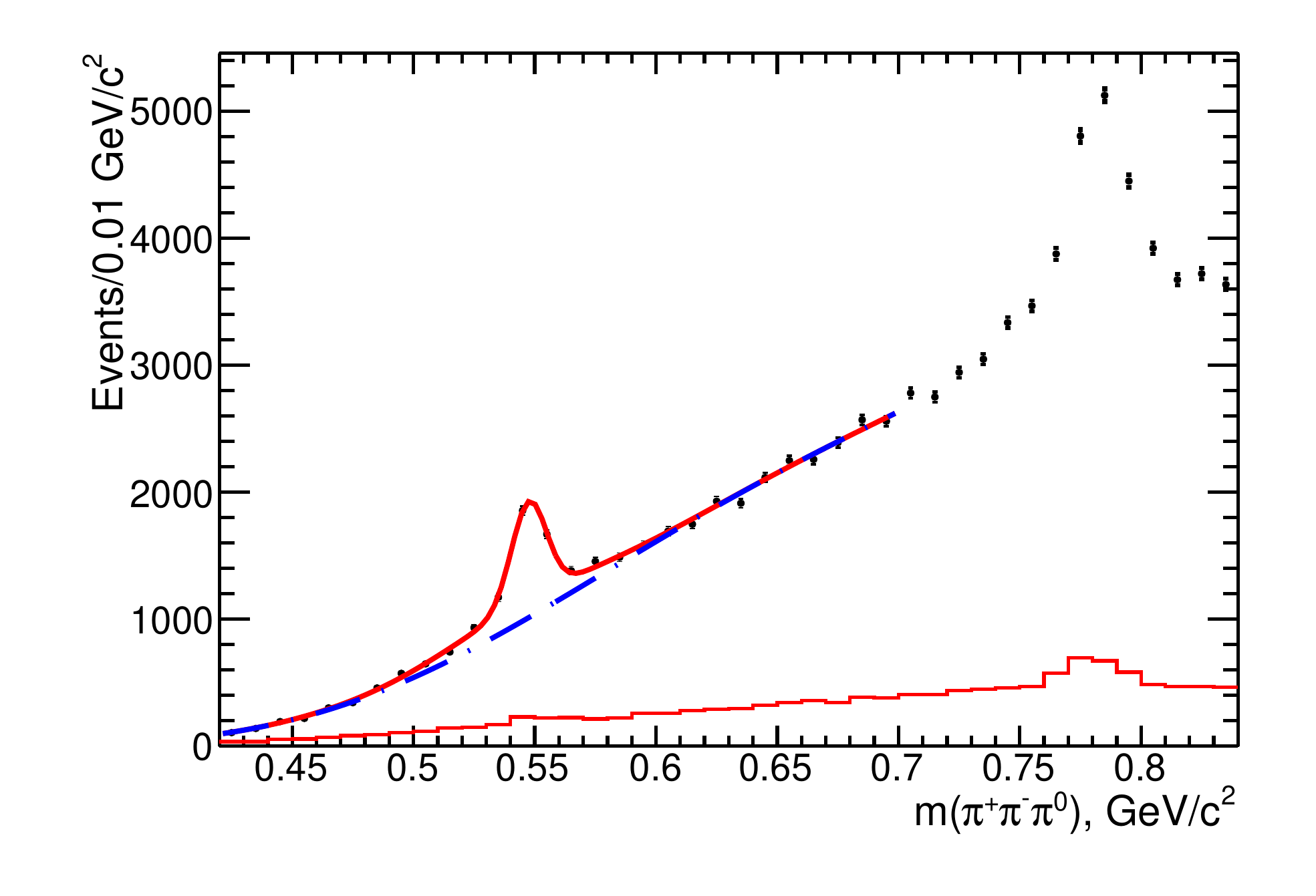}
\put(-60,110){\makebox(0,0)[lb]{\bf(a)}}\\
\vspace{-0.35cm}
\includegraphics[width=0.9\linewidth]{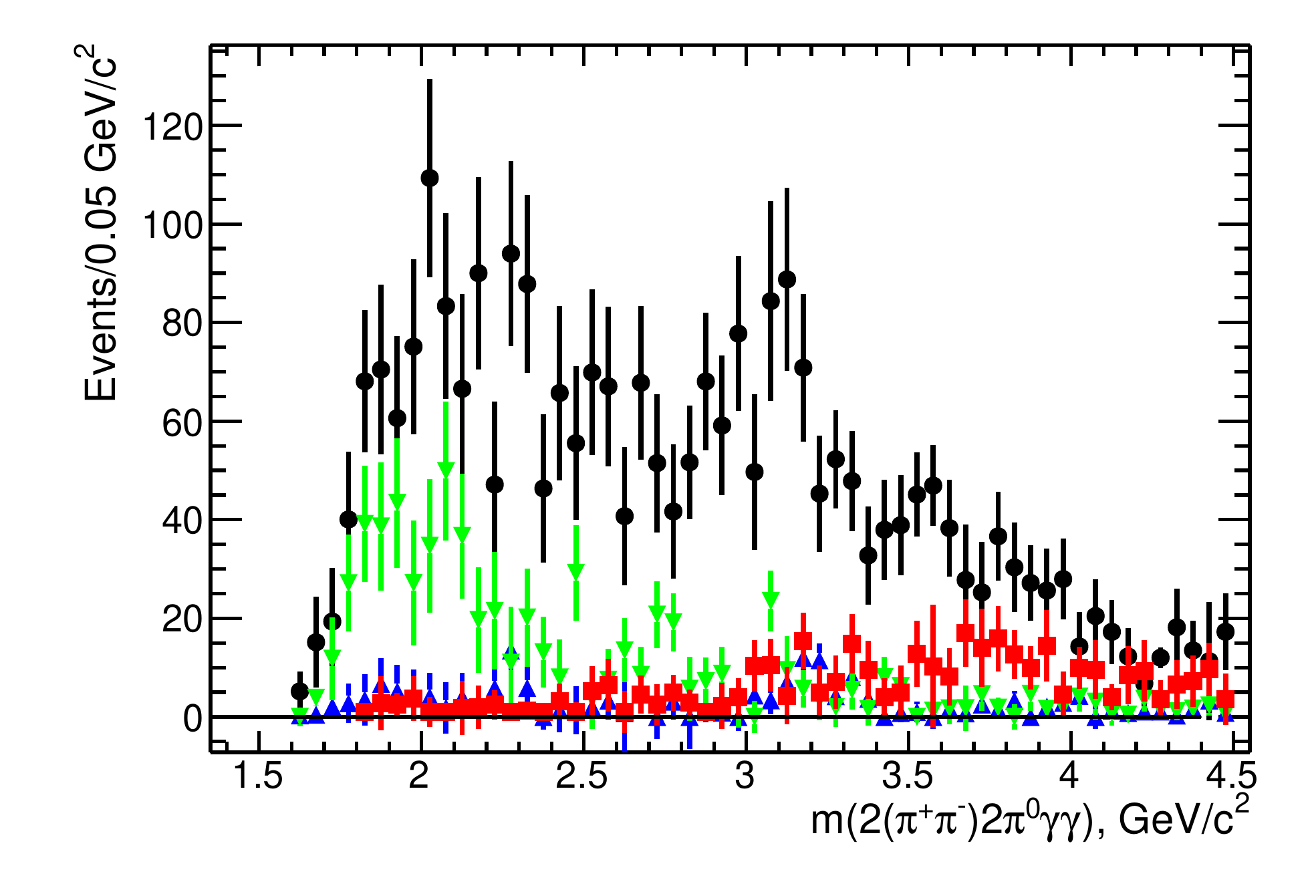}
\put(-50,100){\makebox(0,0)[lb]{\bf(b)}}
\vspace{-0.5cm}
\caption{Mass distributions for the $\eta\pipi\ppz$ intermediate state:
  (a) The curves
  show the fit function for the $\eta$ signal in the $\pipi\piz$ invariant
  mass (solid)  plus
 the combinatorial background (dashed curve). The solid histogram
 shows estimated contributions from the simulated $\epem\to\omega\eta$ ISR events.
(b) The mass distribution of the $2(\pipi)3\piz$ events in the 
$\eta$ peak (circles) and estimated contribution from the
$\omega\eta$ background (triangles), from $\omega\piz\eta$ (up-down
triangles), and from $uds$ (squares).
}
\label{neveta2pi0}
\end{center}
\end{figure}
\begin{figure}[tbh]
\begin{center}
  \includegraphics[width=0.95\linewidth]{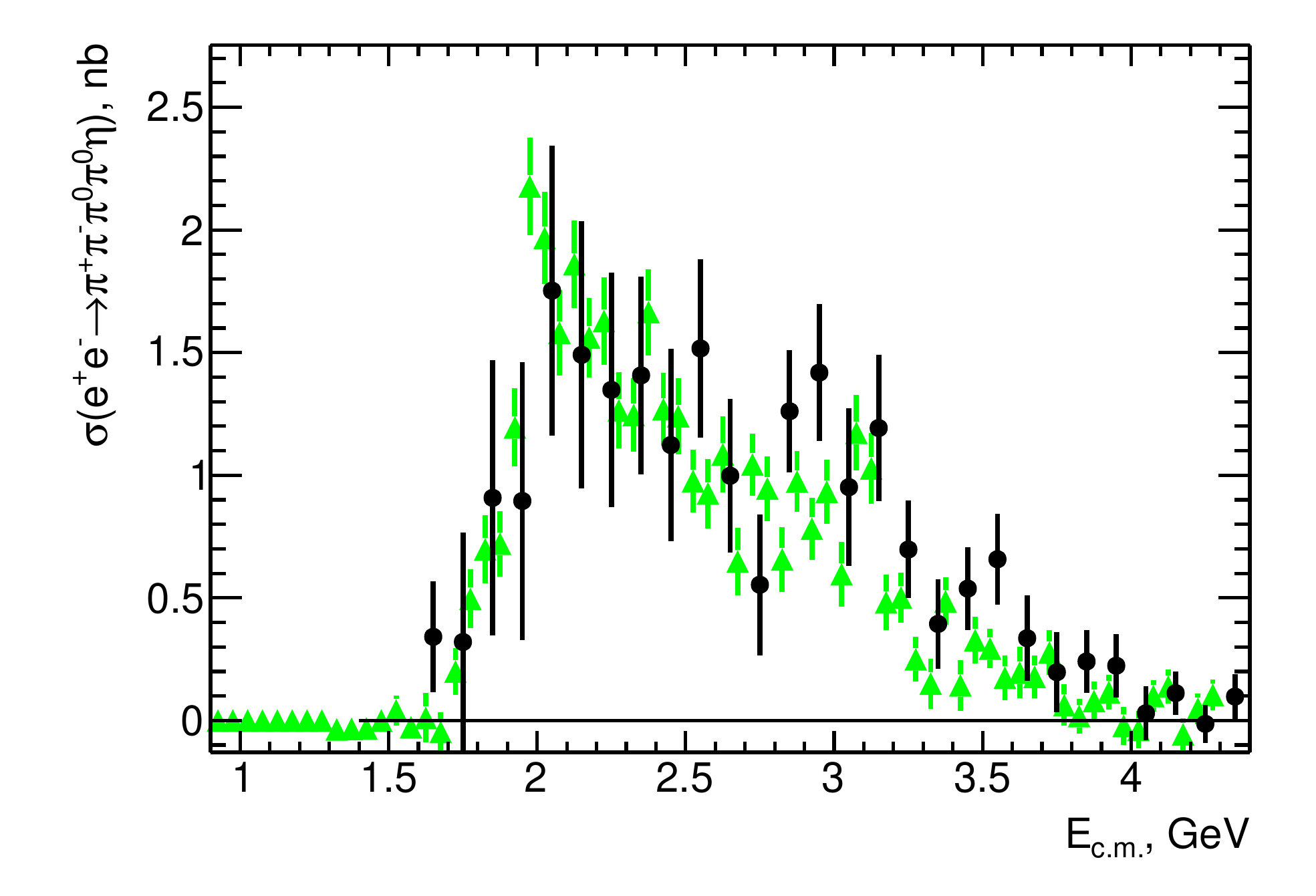}
  \vspace{-0.5cm}
\caption{The result of the energy dependent $\epem\to\eta\pipi\ppz$ cross
  section  in the $\eta\to\pipi\piz$ mode. The result of the \babar~ experiment in the
  $\eta\to\gamma\gamma$ mode~\cite{isr2pi3pi0} is shown by triangles.
}
\label{xs_2pi0eta}
\end{center}
\end{figure}
\input{xs2pi2pi0omega_table}

\subsubsection{\bf\boldmath The $\omega\pipi\ppz$ intermediate state}
\label{Sec:omegapipi}
%
To determine the contribution of the $\omega\pipi\ppz$
 intermediate state, we fit the events of Fig.~\ref{3pivs7pi}(a)
 using a BW function to model the signal and a polynomial to model the background. 
The BW function is convolved with a Gaussian distribution that accounts for the detector
resolution, as described for the fit of Fig.~\ref{m3pi_omega_eta_mc}(b).
The result of the fit is shown in
Fig.~\ref{nevomega2pi0}(a).
We obtain $7808\pm176$ $\omega\pipi\ppz$ events. 
The number of  $\omega\pipi\ppz$ events as a function of the seven-pion invariant mass
is determined by performing an analogous fit 
in each 0.05~\gevcc interval of $m(2(\pipi)3\piz)$.  
The resulting distribution is shown by the circle symbols in
Fig.~\ref{nevomega2pi0}(b).

For the $\epem\to\omega\pipi\ppz$ channel, there is a peaking background from
$\epem\to\omega\eta$ when $\omega$ and $\eta$ decay to $\pipi\piz$. 
A simulation of this reaction
with proper normalization leads to the peaking-background
estimation  shown by the histogram in Fig.~\ref{nevomega2pi0}(a) and by the triangle symbols in
Fig.~\ref{nevomega2pi0}(b).
We also have peaking background from the general $uds$ reactions (also 
shown in Fig.~\ref{nevomega2pi0}(b)).

These background contributions, as well as the events from the correlated $\omega$ and
$\eta$ production from the $\omega\piz\eta$ final state, are subtracted from the 
 $\omega\pipi\ppz$ signal candidate distribution.

The resulting $\epem\to\omega\pipi\ppz$ cross section, corrected
for the $\omega\to\pipi\piz$ branching fraction,
is shown in Fig.~\ref{xs_2pi0omega} and tabulated in Table~\ref{omega2pi0_table}.
The uncertainties are statistical only.
The systematic uncertainties are about 10\%.
No previous measurement exists for this process.
The cross section exhibits a rise at threshold,
a decrease at large \Ecm, and a possibly resonant activity
at around 2.3-2.5 GeV.
The result by CMD-3 for the significantly lower
$\epem\to\omega\pipi\pipi$ cross  section~\cite{cmd7pi}  is shown by squares.
\input{xs2pi2pi0eta_table}

\subsubsection{\bf\boldmath The $\eta\pipi\ppz$ intermediate state}
\label{Sec:etapipi2pi0}
%
A similar approach is used 
to determine the contribution of the $\eta\pipi\ppz$
 intermediate state. We fit the events of Fig.~\ref{3pivs7pi}(a)
 using the three-Gaussian function for the signal and a polynomial to model the background. 
The result of the fit is shown in Fig.~\ref{neveta2pi0}(a).
The fitted $\eta\pipi\ppz$  yield corresponds to $2522\pm91$  events.
The signal distribution 
as a function of the seven-pion invariant mass
is determined by performing an analogous fit 
in each 0.05~\gevcc interval of $m(2(\pipi)3\piz)$, and is shown 
by the circle symbols in Fig.~\ref{neveta2pi0}(b).

Also in this case a peaking background arises from the process 
$\epem\to\omega\eta$ when $\omega$ and $\eta$ decay to $\pipi\piz$. 
Its contribution, estimated with MC simulation, is shown by the
histogram in Fig.~\ref{neveta2pi0}(a) and by the triangle symbols in
Fig.~\ref{neveta2pi0}(b).

We also have peaking background from the general $uds$ reactions,
shown by squares in Fig.~\ref{neveta2pi0}(b).
And finally, we remove events from the $\omega\piz\eta$ final state (up-down triangles).

The  $\epem\to\eta\pipi\ppz$ cross section, corrected
for the $\eta\to\pipi\piz$ branching fraction,
is shown in Fig.~\ref{xs_2pi0eta} and tabulated in 0.1 GeV bins in
Table~\ref{eta2pi0_table}.
The uncertainties are statistical only.
The systematic uncertainties are about 10\%.
We are in good
agreement with a recent measurement of this cross
section~\cite{isr2pi3pi0} in the $\eta\to\gamma\gamma$ decay mode. 
\begin{figure}[t]
\begin{center}
\includegraphics[width=0.5\linewidth]{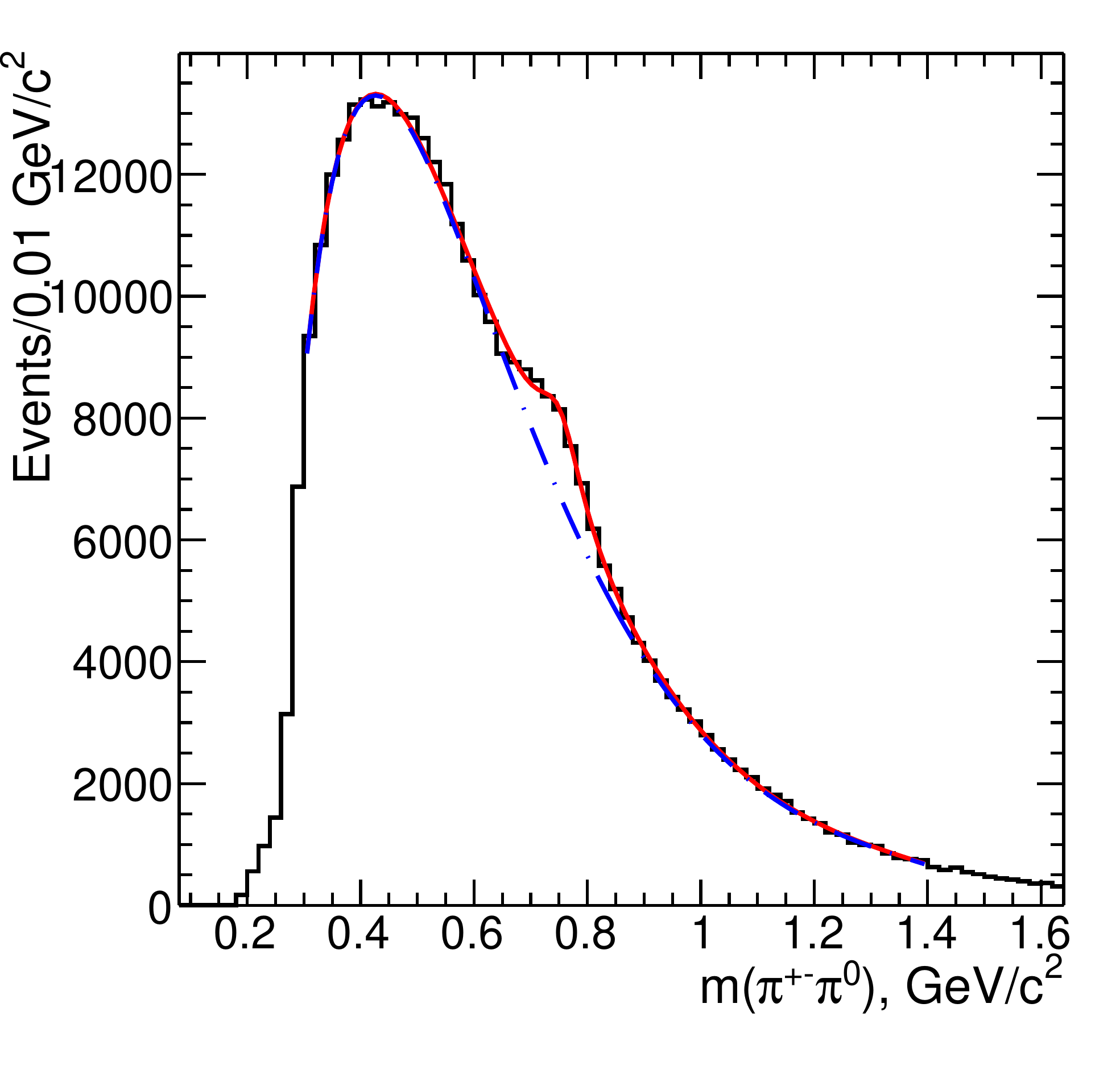}
\put(-50,100){\makebox(0,0)[lb]{\bf(a)}}
\includegraphics[width=0.5\linewidth]{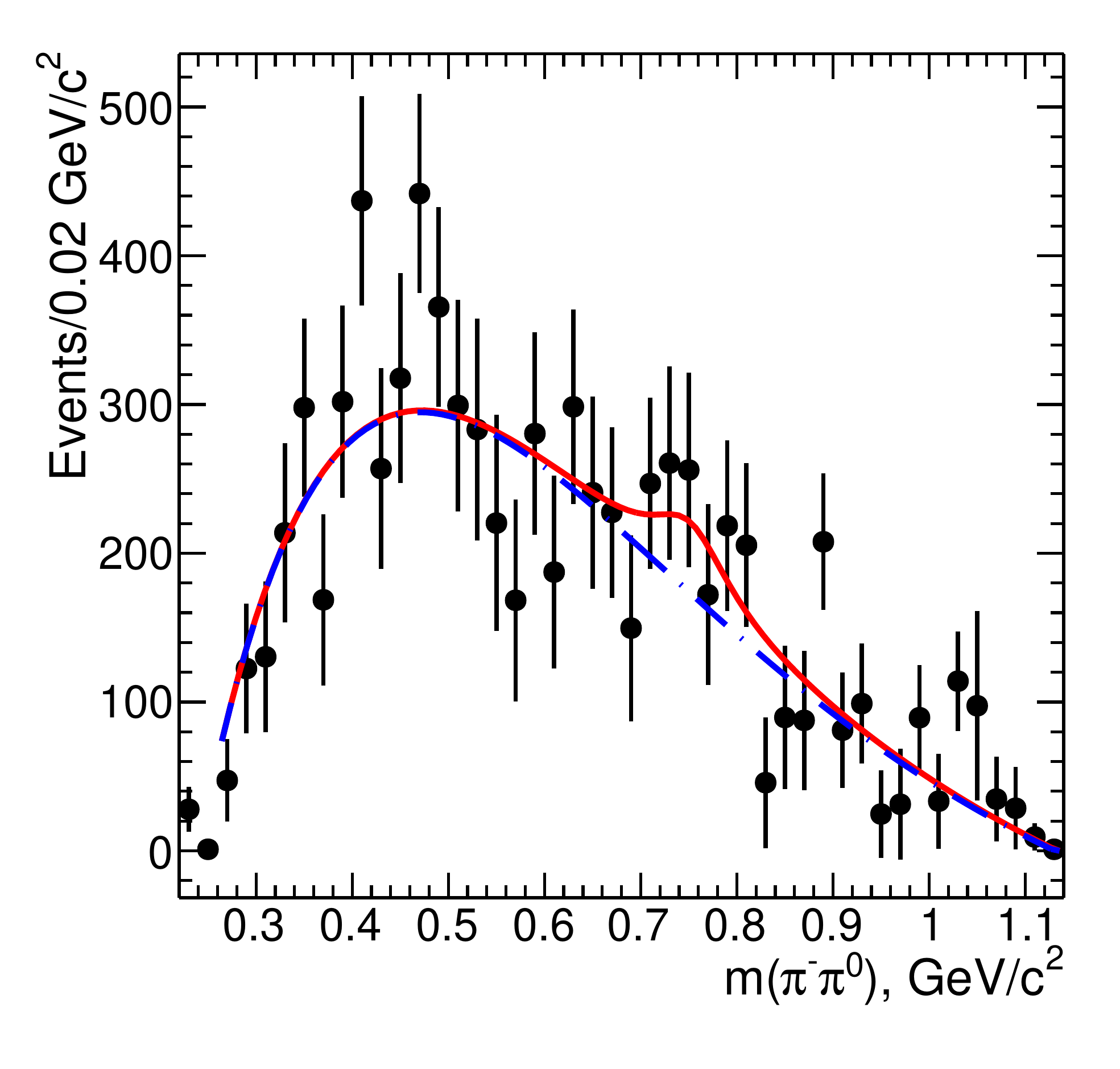}
\put(-50,100){\makebox(0,0)[lb]{\bf(b)}}\\
\vspace{-0.3cm}
\includegraphics[width=0.9\linewidth]{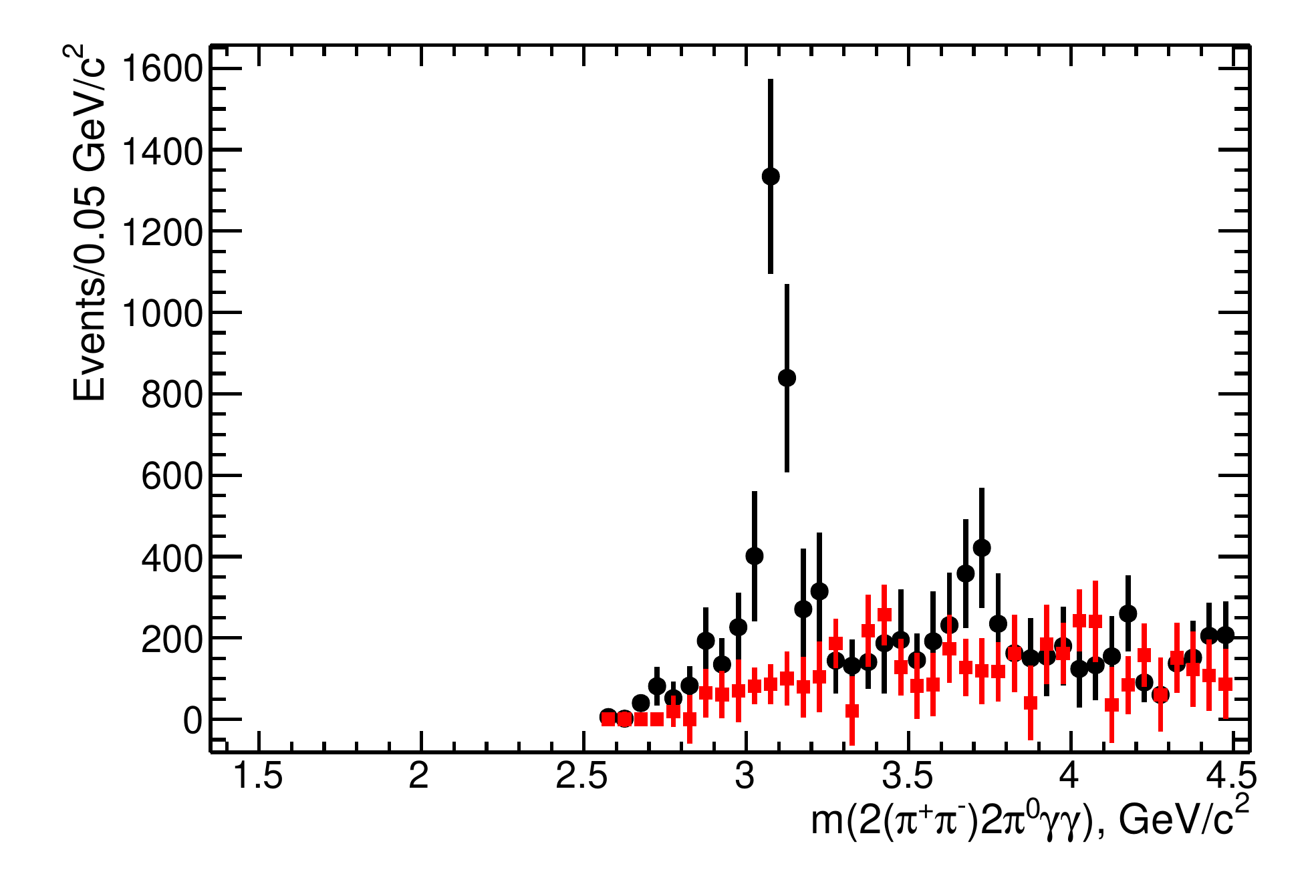}
\put(-50,120){\makebox(0,0)[lb]{\bf(c)}}
\vspace{-0.5cm}
\caption{Mass distributions for the $\rho^{\pm}\pi^{\mp}\pipi\ppz$ intermediate state:
  (a) The $\pi^{\pm}\piz$ invariant mass for data. 
The dashed curve shows the fit to the combinatorial background. The solid curve 
is the sum of the background curve and the BW function for the
$\rho^{\pm}$.
(b) The result of the $\rho^+$ fit in bins of 0.02~\gevcc
in the $\rho^-$ mass.
(c) Number of events in bins of \Ecm from the 
$\rho^{\pm}\to\pi^{\pm}\piz$ (circles) intermediate states. The squares show the 
event numbers obtained  from $uds$ production. 
}
\label{pipi0slices}
\end{center}
\end{figure}
\begin{figure}[h]
\begin{center}
\includegraphics[width=0.95\linewidth]{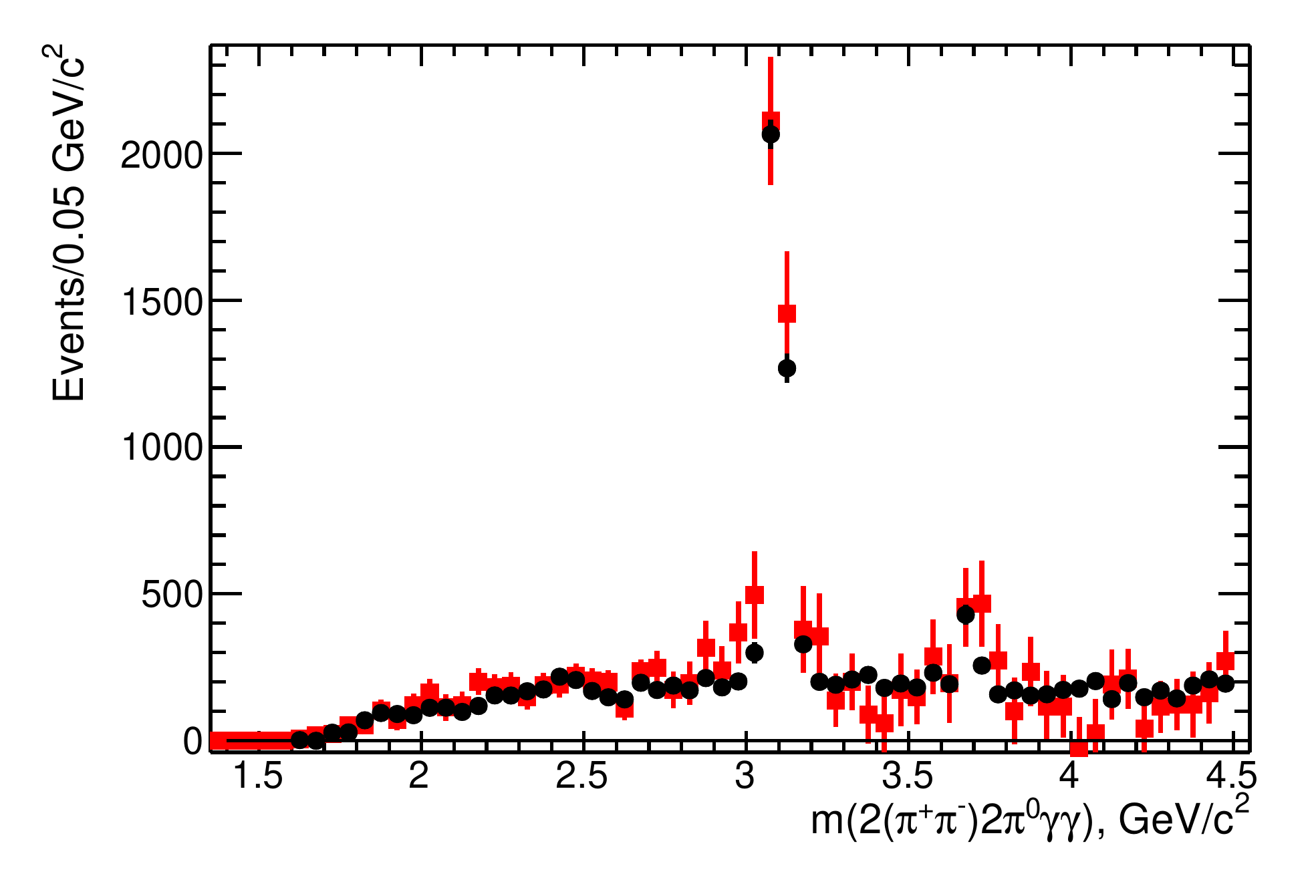}
\vspace{-0.5cm}
\caption{The $2(\pipi)2\piz\gamma\gamma$ mass distribution summed over the intermediate states.
  The circles show the number of events, determined from
  the $\piz$ fit.
The squares show the sum of events with $\eta$, $\omega$, and $\rho$ production,
 the latter corrected for
the $\rho^+\rho^-$ production.
}
\label{sumallev}
\end{center}
\end{figure}
\subsubsection{\bf\boldmath The $\rho(770)^{\pm}\pi^{\mp}\pipi\ppz$ intermediate state}
\label{sec:rhoselectpi0}
A similar approach is followed to study events
with a $\rho^{\pm}$ meson in the intermediate state.
Because the $\rho$ meson is broad, a BW
function is used to describe the signal shape.
There are twelve candidate $\rho^\pm$ entries per event, leading to
a large combinatorial background.
To extract the contribution of the $\rho^{\pm}\pi^{\mp}\pipi\ppz$ intermediate state we
fit the events in Fig.~\ref{pipi0vs7pi}(a) with a BW function
to describe the signal and a polynomial to describe the background.
The parameters of the $\rho$ resonance are taken from Ref.~\cite{PDG}.
The result of the fit  is shown in Fig.~\ref{pipi0slices}(a). 
We obtain $9138\pm371$ $\rho^{\pm}\pi^{\mp}\pipi\ppz$ events. The
distribution of these events vs the seven-pion invariant mass is shown
by the circle symbols in Fig.~\ref{pipi0slices}(c), while a
similar fit for the $uds$ simulation is shown by squares.
The $uds$ background is dominant in all energy regions except for $J/\psi$ and $\psi(2S)$.

In these events more than one $\rho^{\pm}$ per event can be expected, indicating
a significant production
of $J/\psi\to\rho^+\rho^-\pipi\piz$.
To determine the rate of $\rho^+\rho^-\pipi\piz$ events in the
$J/\psi$ decays,
we perform a fit to determine the number
of $\rho^+$ in intervals of 0.02~\gevcc
in the $\pi^-\piz$ distribution of Fig.~\ref{pipi0vs7pi}(b) for events
within $\pm$0.1\gevcc of the $J/\psi$ mass.
The result is shown in Fig.~\ref{pipi0slices}(b).
Indeed, a small $\rho^+$ peak with $415\pm340$ events is observed, compared to
2844 events in the $J/\psi$ peak region, corresponding to about 20\% of all
decays with one or two $\rho^{\pm}$. However, the uncertainty in this
estimate is almost at the same level.

The charmonium region for all intermediate states is discussed below.
\input{xs6pieta_table}

\subsection{\bf\boldmath The sum of intermediate states}
\label{sec:sum}

We consider whether the $2(\pipi)3\piz$ channel contains other intermediate state
contributions.
The circle symbols in Fig.~\ref{sumallev}  show the
total number of $2(\pipi)3\piz$ events,
repeated from Fig.~\ref{nev_4pi3pi0_data}.
We perform a sum of 
the number of $\eta2(\pipi)$, $\omega\piz\eta$, $\eta\pipi\ppz$,
$\omega\pipi\ppz$, and $\rho^{\pm}\pi^{\mp}\pipi\ppz$ intermediate state
events, found as described in the previous sections,
and show this sum by the square symbols in Fig.~\ref{sumallev}.
Based on the results of our study of correlated $\rho^+\rho^-$
production, we scale the number of events found from the
fit to the $\rho$ peak so that it corresponds to the number
of events with either a single $\rho^{\pm}$ or with a $\rho^+\rho^-$
pair.
This summed curve
is seen to be in agreement with the total number of
 $2(\pipi)3\piz$ events; we conclude there is no significant contribution from
 other (unobserved) intermediate states.

\begin{figure}[t]
  \begin{center}
    \vspace{-0.4cm}
\includegraphics[width=0.95\linewidth]{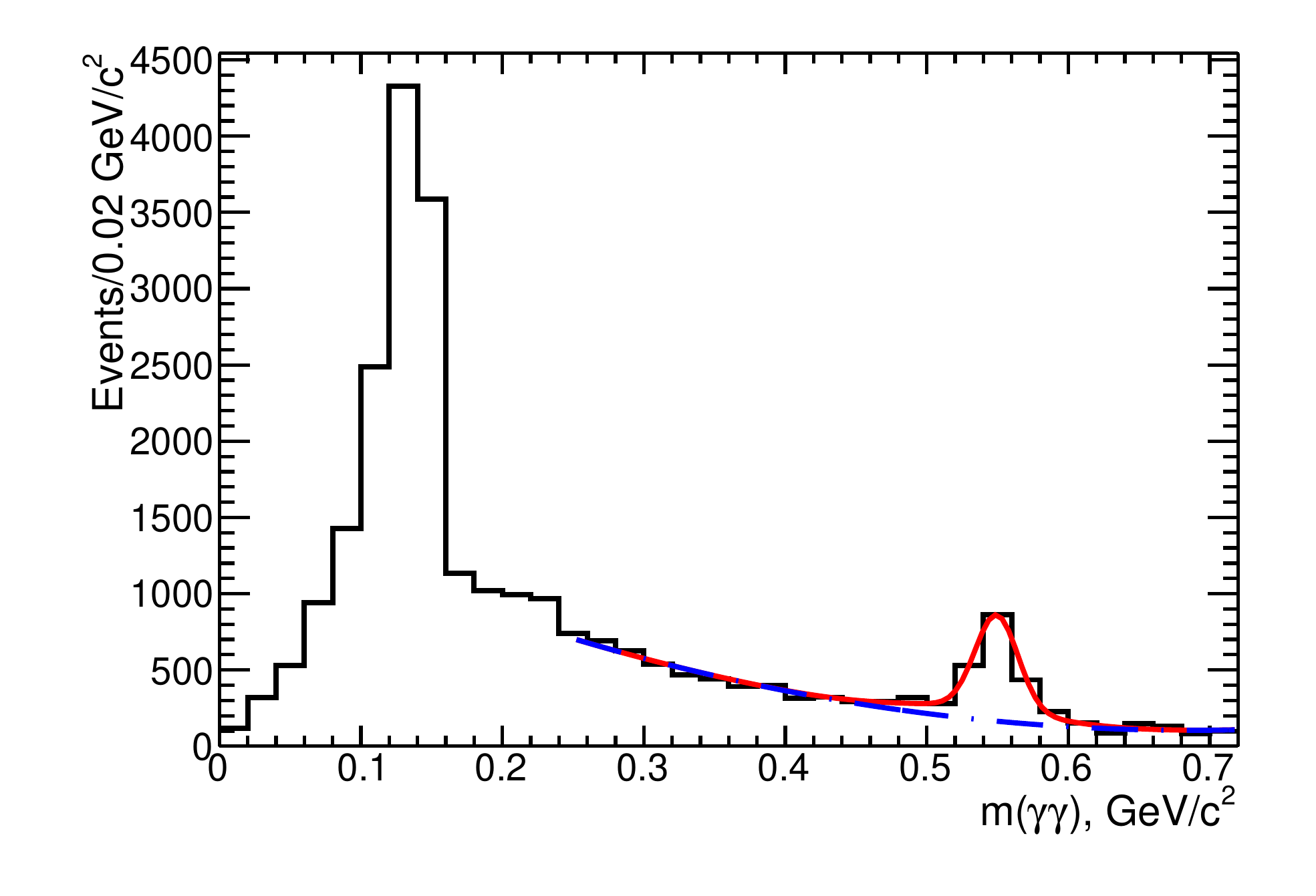}
\put(-50,120){\makebox(0,0)[lb]{\bf(a)}}\\
\vspace{-0.4cm}
\includegraphics[width=0.95\linewidth]{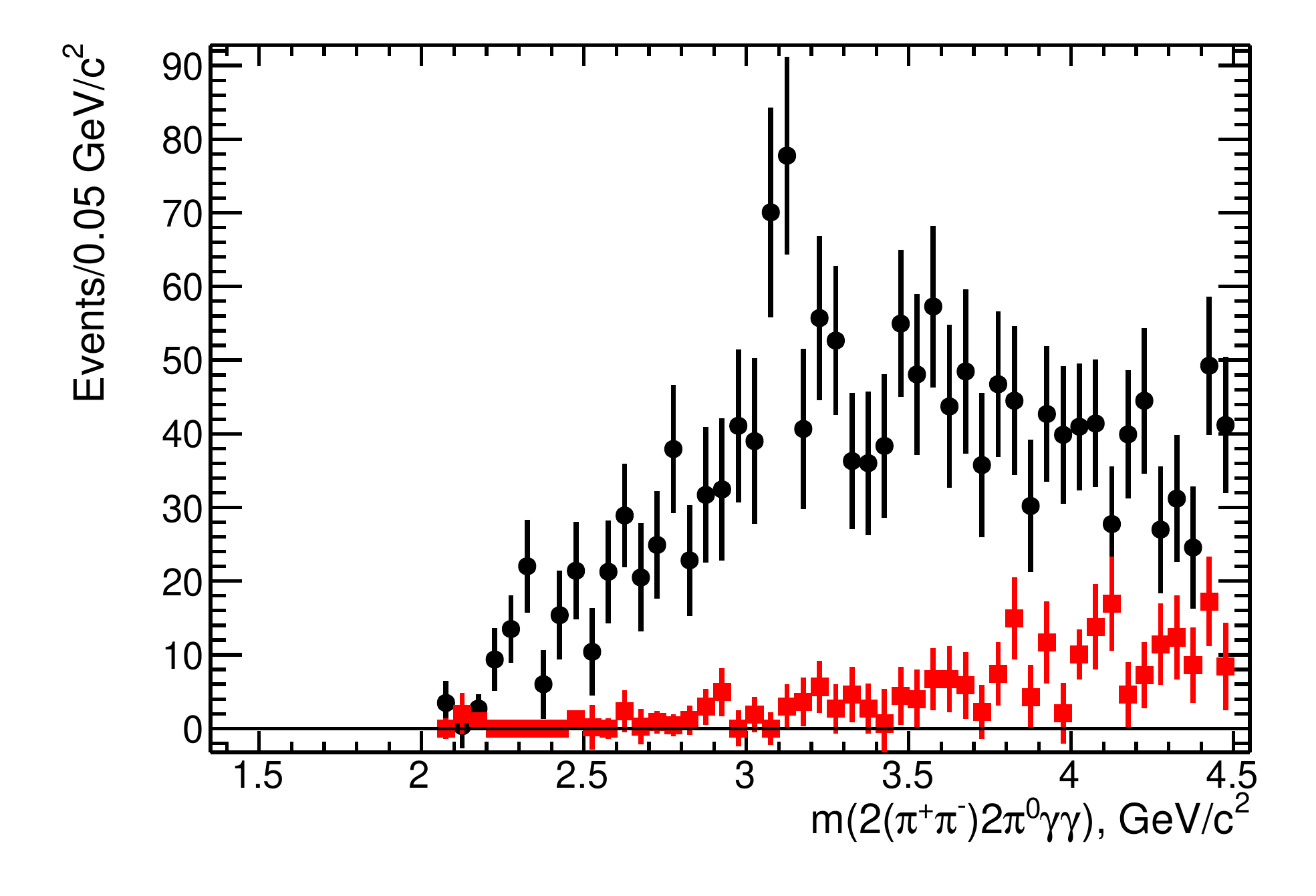}
\put(-50,120){\makebox(0,0)[lb]{\bf(b)}}
\vspace{-0.6cm}
\caption{ Mass distributions for the $2(\pipi)2\piz\eta$ final state.
  (a) The third-photon-pair invariant mass for data.
  The dashed curve shows the fitted background. 
  The solid curve
  shows the sum of background and the two-Gaussian fit function used to obtain the number of events
  with an $\eta$.
(b) The invariant-mass distribution for the
$2(\pipi)2\piz\eta$ events obtained from the $\eta$ signal fit. The
contribution of the $uds$ background events is shown by the squares.
}
\label{meta_data_fit}
\end{center}
\end{figure}
\begin{figure}[t]
  \begin{center}
    \vspace{-0.3cm}
\includegraphics[width=0.95\linewidth]{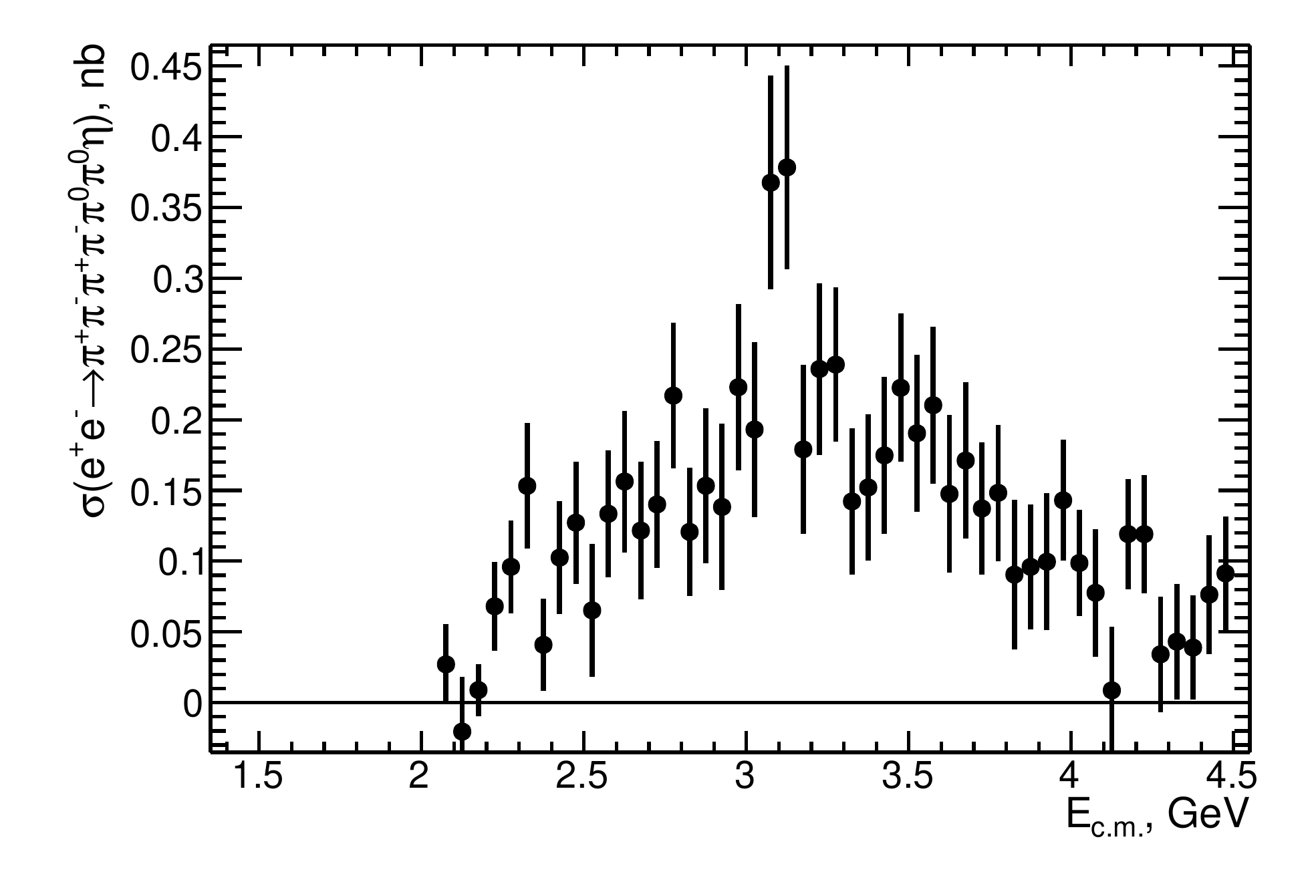}
\vspace{-0.7cm}
\caption{
Energy dependent  cross section
for   $\epem\to2(\pipi)\ppz\eta$.
The uncertainties are statistical only.
}
\label{2pi2pi0eta_ee_babar}
\end{center}
\end{figure}
\section{\bf\boldmath The $2(\pipi)2\piz\eta$ final state}
\subsection{Determination of  the number of events }
The analogous approach to that described above for
$\epem\to2(\pipi)\ppz\piz$ events is used to study
$\epem\to2(\pipi)\ppz\eta$ events.
We fit the $\eta$ signal in the third-photon-pair
invariant-mass distribution (cf., Fig.~\ref{4pi3pi0_chi2_all}) with
the sum of
two Gaussians with a common mean, while the relatively smooth 
background is described by a second-order polynomial function, as shown
in Fig.~\ref{meta_data_fit}(a).  We obtain $1651\pm50$ events.
Figure~\ref{meta_data_fit}(b) shows the mass distribution of these events.
\subsection{Peaking background}\label{sec:udsbkg2}
The major background producing an $\eta$ peak is the
non-ISR background, in particular $\epem\to2(\pipi)\ppz\piz\eta$
when one of the neutral pions decays asymmetrically, producing a 
photon interpreted as ISR.  The $\eta$ peak from the $uds$ simulation
is visible in Fig.~\ref{udsbkg}.
  We fit the $\eta$
peak in the $uds$ simulation in intervals of 0.05~\gevcc in
$m(2(\pipi)\ppz\gamma\gamma)$. The results are shown by the squares in Fig.~\ref{meta_data_fit} (b).

To normalize the $uds$ simulation,
we form the diphoton invariant-mass distribution of
 the ISR candidate with all the remaining photons in the event.
Comparing the number of events in the $\piz$ peaks
in data and $uds$ simulation, we assign
a scale factor of $1.5\pm0.2$ to the simulation and subtract these
events from the data distribution.
\subsection{\boldmath Cross section for $\epem\to2(\pipi)\ppz\eta$}
\label{2pi2pi0eta}
The cross section for $\epem\to2(\pipi)\ppz\eta$ is determined
using Eq.~(\ref{xseq}).  The results are shown in
Fig.~\ref{2pi2pi0eta_ee_babar} and
listed in Table~\ref{6pieta_table}. These are the first results for
this process. 
The systematic uncertainties and corrections
are the same as those presented in Table~\ref{error_tab}
except that the uncertainty in the detection efficiency increases to 13\%.

The cross section is approximately zero until well above
2 \gev and so is not useful in
the vacuum polarization calculations;
we have not yet performed a study of intermediate states for this process.

\section{\bf\boldmath The $J/\psi$ region}
\subsection{\bf\boldmath The $2(\pipi)3\piz$  final state}
\begin{figure}
\includegraphics[width=0.50\linewidth]{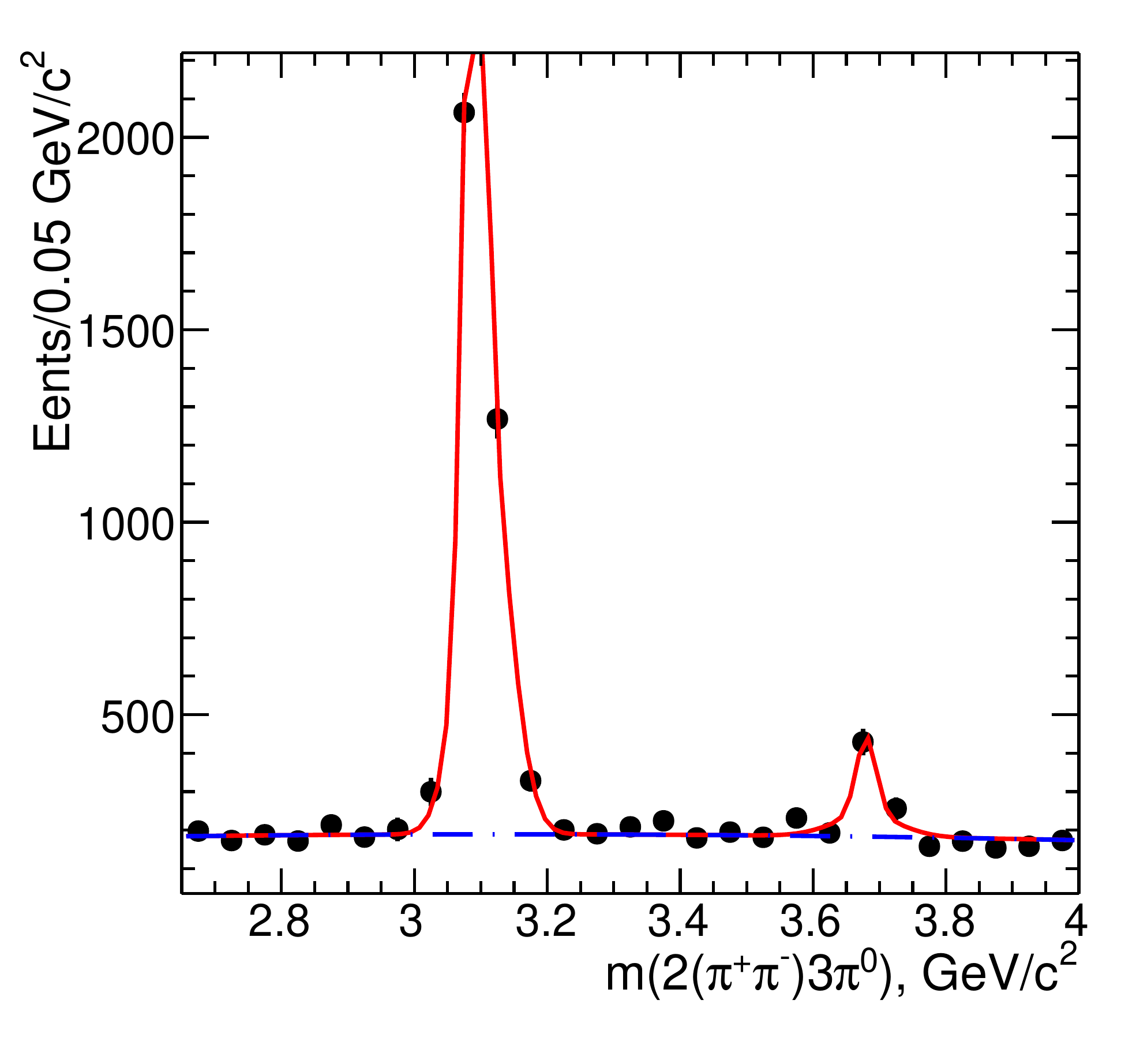}
\put(-40,80){\makebox(0,0)[lb]{\bf(a)}}
\includegraphics[width=0.49\linewidth]{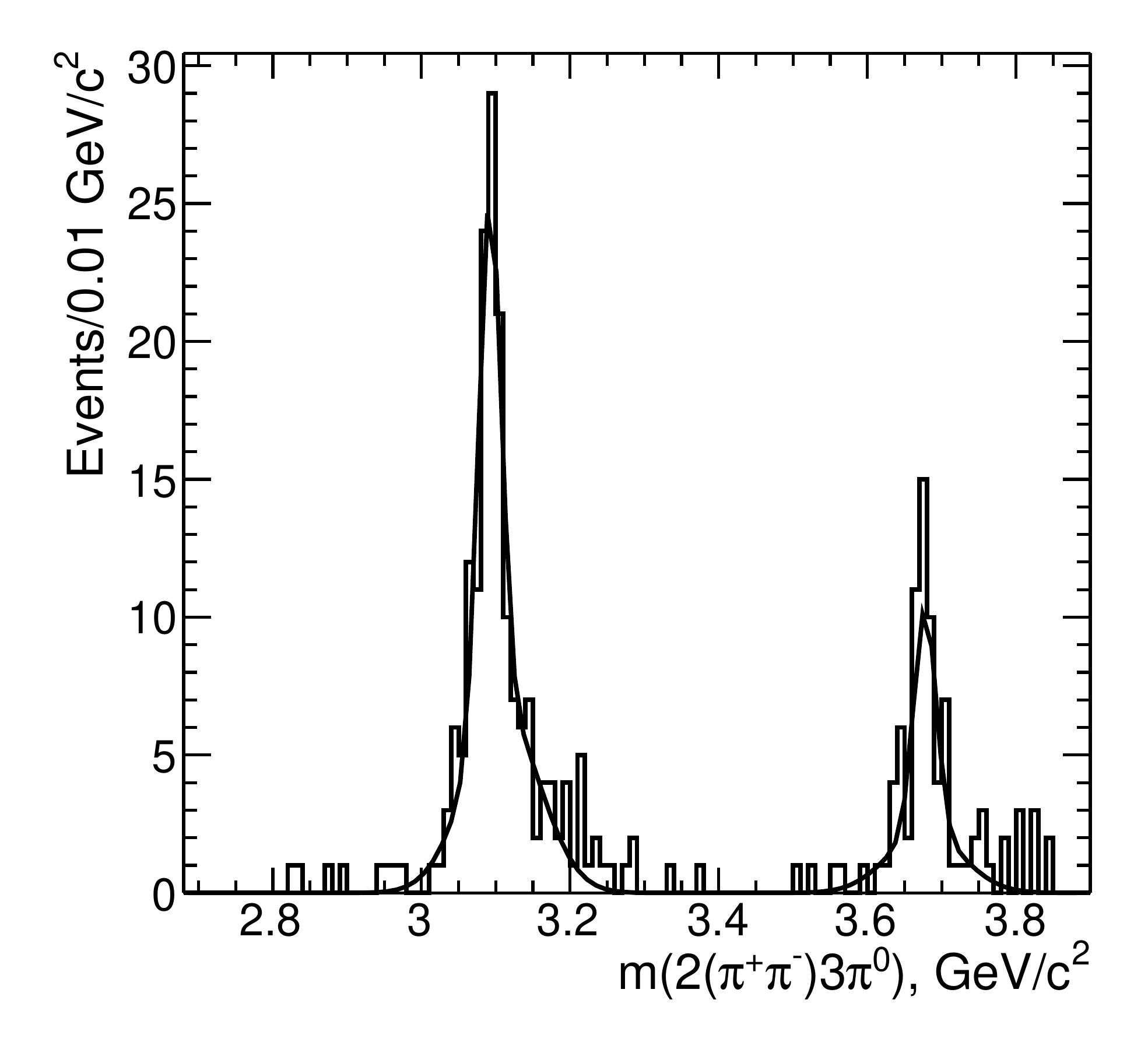}
\put(-40,80){\makebox(0,0)[lb]{\bf(b)}}
\vspace{-0.5cm}
\caption{(a)
The $2(\pipi)3\piz$ mass distribution for ISR-produced
$\epem\to2(\pipi)\ppz\piz$ events in the $J/\psi$--$\psi(2S)$
        region.  
(b) The MC-simulated signals. The curves show the fit functions
described in the text.
}
\label{jpsi}
\end{figure}

Figure~\ref{jpsi}(a) shows an expanded view of the charmonium region
from Fig.~\ref{nev_4pi3pi0_data}, which has large contributions from
the $J/\psi$ and $\psi(2S)$ decays to seven pions.
The non-resonant background distribution is  flat in this region.

The observed peak shapes are not purely Gaussian because of radiation
effects and resolution, 
as is also seen in the simulated signal distributions
shown in Fig.~\ref{jpsi}(b).  The sum of two Gaussians with a common
mean is used to describe each peak.
We obtain $3391\pm101$  $J/\psi$ events and $290\pm40$
$\psi(2S)$ events.
Using these results for the number of events, the detection
efficiency, and the ISR luminosity,
we determine the product:
\begin{eqnarray}
  B_{J/\psi\to 7\pi}\cdot\Gamma^{J/\psi}_{ee}
  = \frac{N(J/\psi\to 2(\pipi) 3\piz)\cdot m_{J/\psi}^2}%
           {6\pi^2\cdot d{\cal L}/dE\cdot\epsilon^{\rm
  MC}\cdot\epsilon^{\rm corr}\cdot C} \label{jpsieq}\\
  = (345\pm 10\pm  50)~\ev\ ,\nonumber
\end{eqnarray}
where $\Gamma^{J/\psi}_{ee}$ is
the electronic width, $d{\cal L}/dE =
  180~\invnb/\mev$ is the ISR luminosity 
at the $J/\psi$ mass $m_{J/\psi}$, $\epsilon^{\rm MC} = 0.027\pm0.002$ is the detection
efficiency from simulation, $\epsilon^{\rm corr}=0.85$ is the correction, discussed in 
Sec.~\ref{sec:Systematics},
and  $C = 3.894\times 10^{11}~\nb\mev^2$ is a
conversion constant ~\cite{PDG}. We estimate the systematic uncertainty for this
region to be 15\%.  
The subscript ``$7\pi$'' for the branching fraction refers
to the $2(\pipi)3\piz$ final state exclusively.

Using $\Gamma^{J/\psi}_{ee} =5.53\pm0.10~\kev$ ~\cite{PDG}, we obtain
$B_{J/\psi\to 7\pi} = (6.2\pm 0.2\pm 0.9)\times 10^{-2}$: no
other measurements for this channel exist. It is the largest decay mode
of the $J/\psi$ measured so far.

Using Eq.(\ref{jpsieq}) and the result $d{\cal L}/dE =
  228~\invnb/\mev$  at the $\psi(2S)$ mass, we obtain:
\begin{eqnarray*}
  B_{\psi(2S)\to 7\pi}\cdot\Gamma^{\psi(2S)}_{ee}
  &=& (33\pm5\pm5)~\ev\ .
\end{eqnarray*}
With $\Gamma^{\psi(2S)}_{ee} =2.33\pm0.04~\kev$ ~\cite{PDG}  we
find $B_{\psi(2S)\to 7\pi} = (1.4\pm 0.2\pm 0.2)\times
10^{-2}$. For this channel also, no previous result exists.

\begin{figure}[t]
\includegraphics[width=0.50\linewidth]{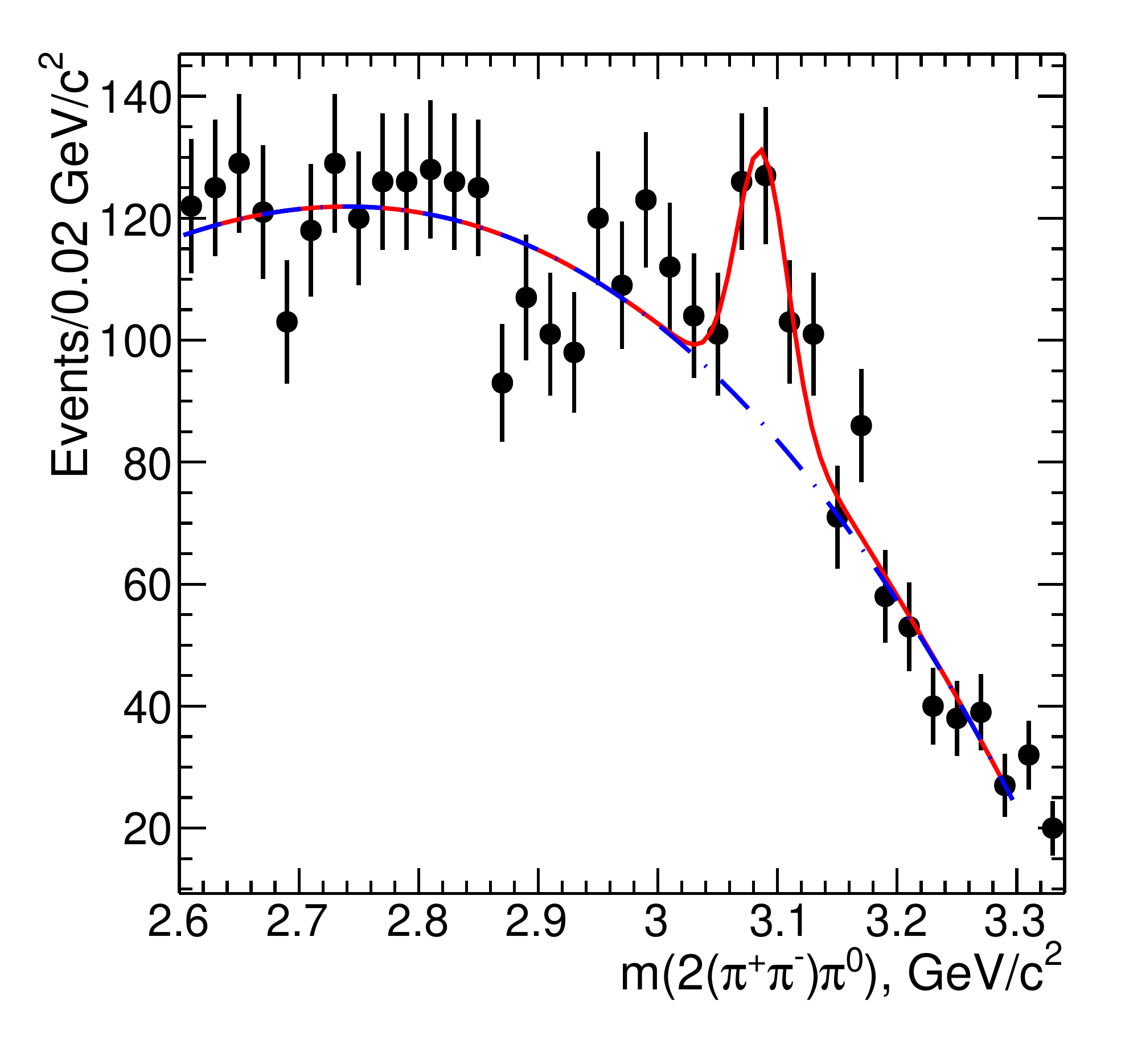}
\put(-30,80){\makebox(0,0)[lb]{\bf(a)}}
\includegraphics[width=0.48\linewidth]{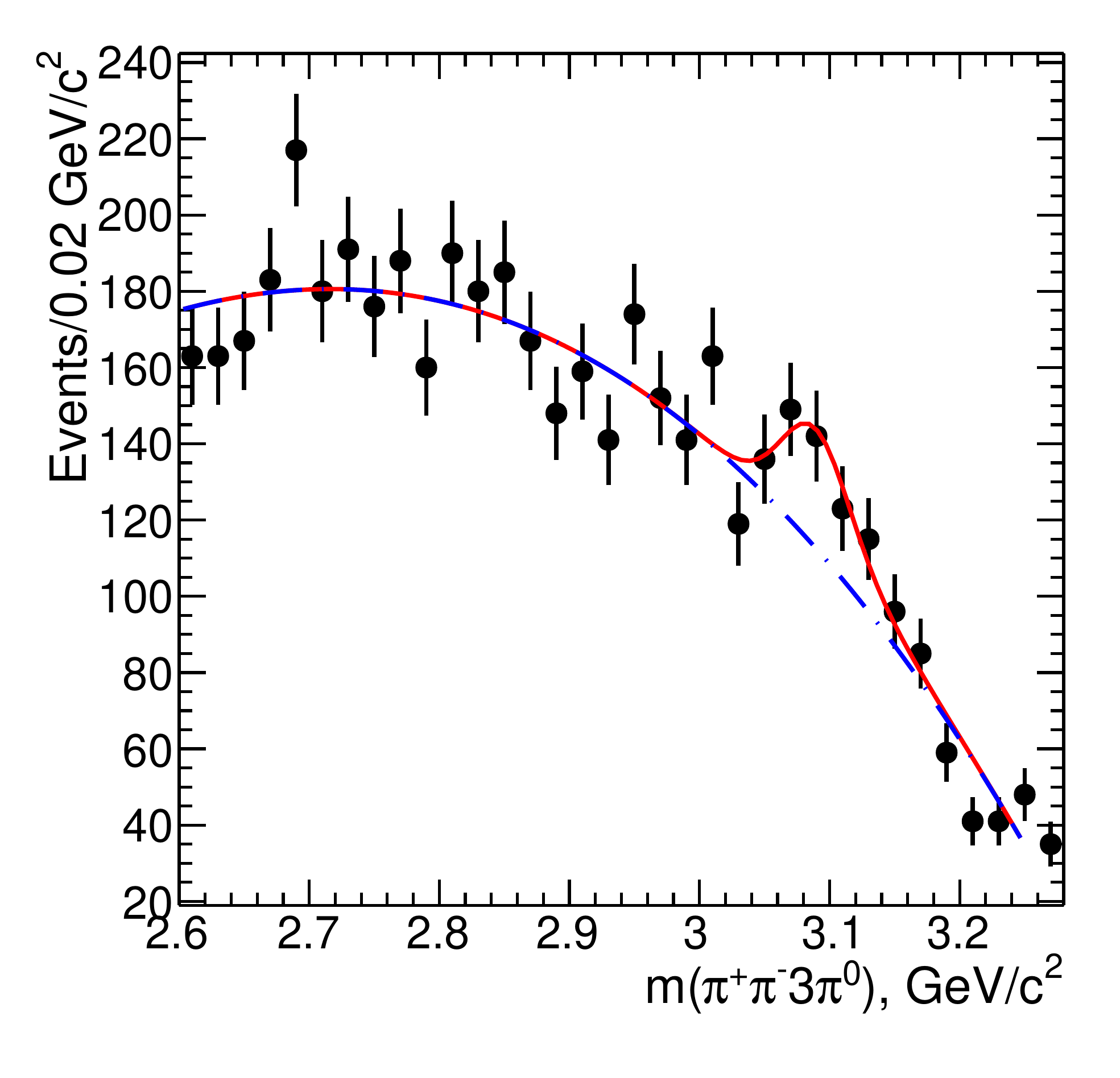}
\put(-30,80){\makebox(0,0)[lb]{\bf(b)}}
\vspace{-0.5cm}
\caption{
The $2(\pipi)\piz$ invariant mass (a) and  $\pipi3\piz$ invariant mass
(b) for events with a seven-pion invariant mass within $\pm100$~\mevcc  of the $\psi(2S)$.
The curves show the fit functions for all events
(solid) and the contribution of the background (dashed).
}
\label{psi2s_chain}
\end{figure}    
\begin{figure}[b]
\includegraphics[width=0.49\linewidth]{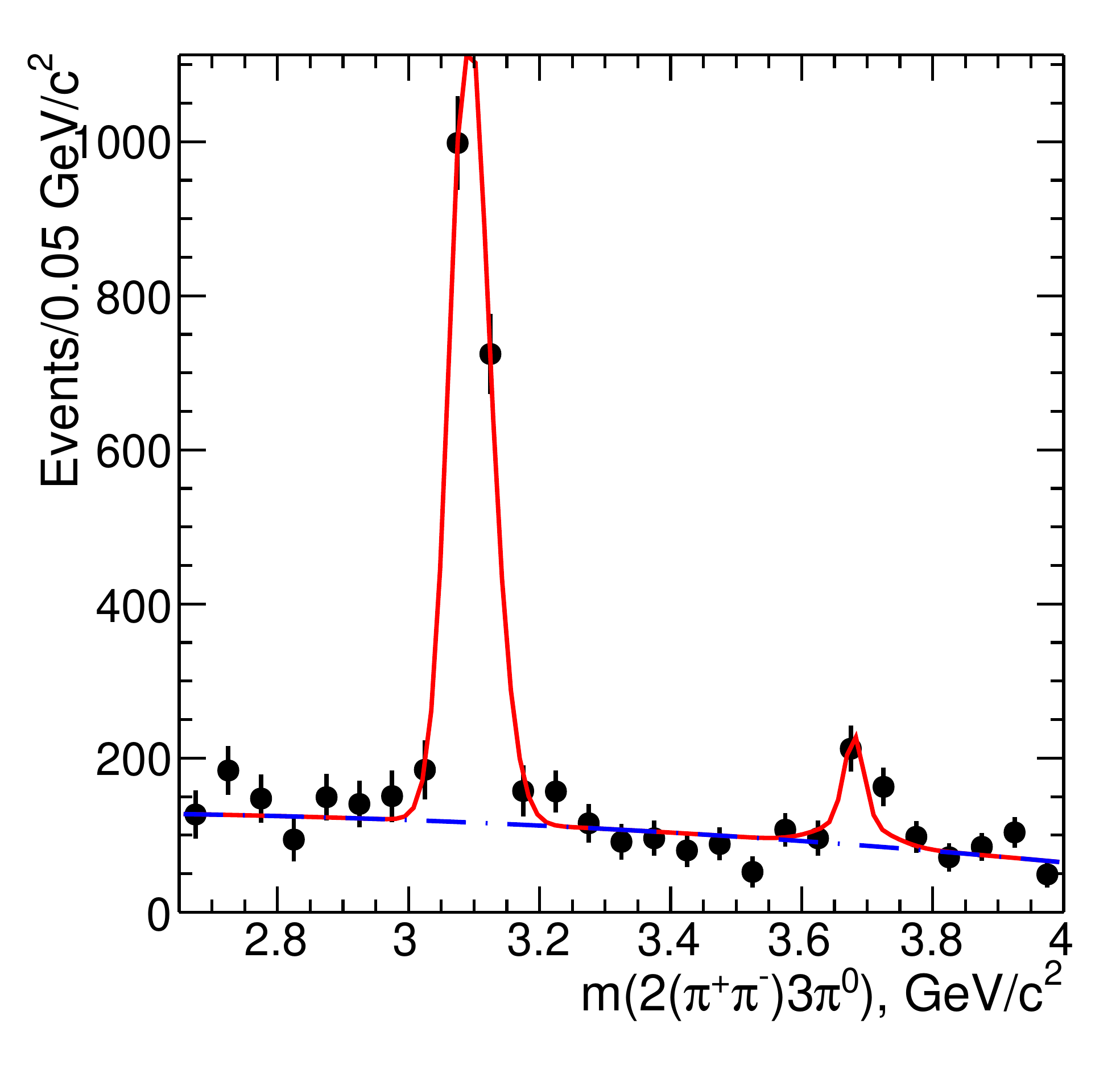}
\put(-40,80){\makebox(0,0)[lb]{\bf(a)}}
\includegraphics[width=0.49\linewidth]{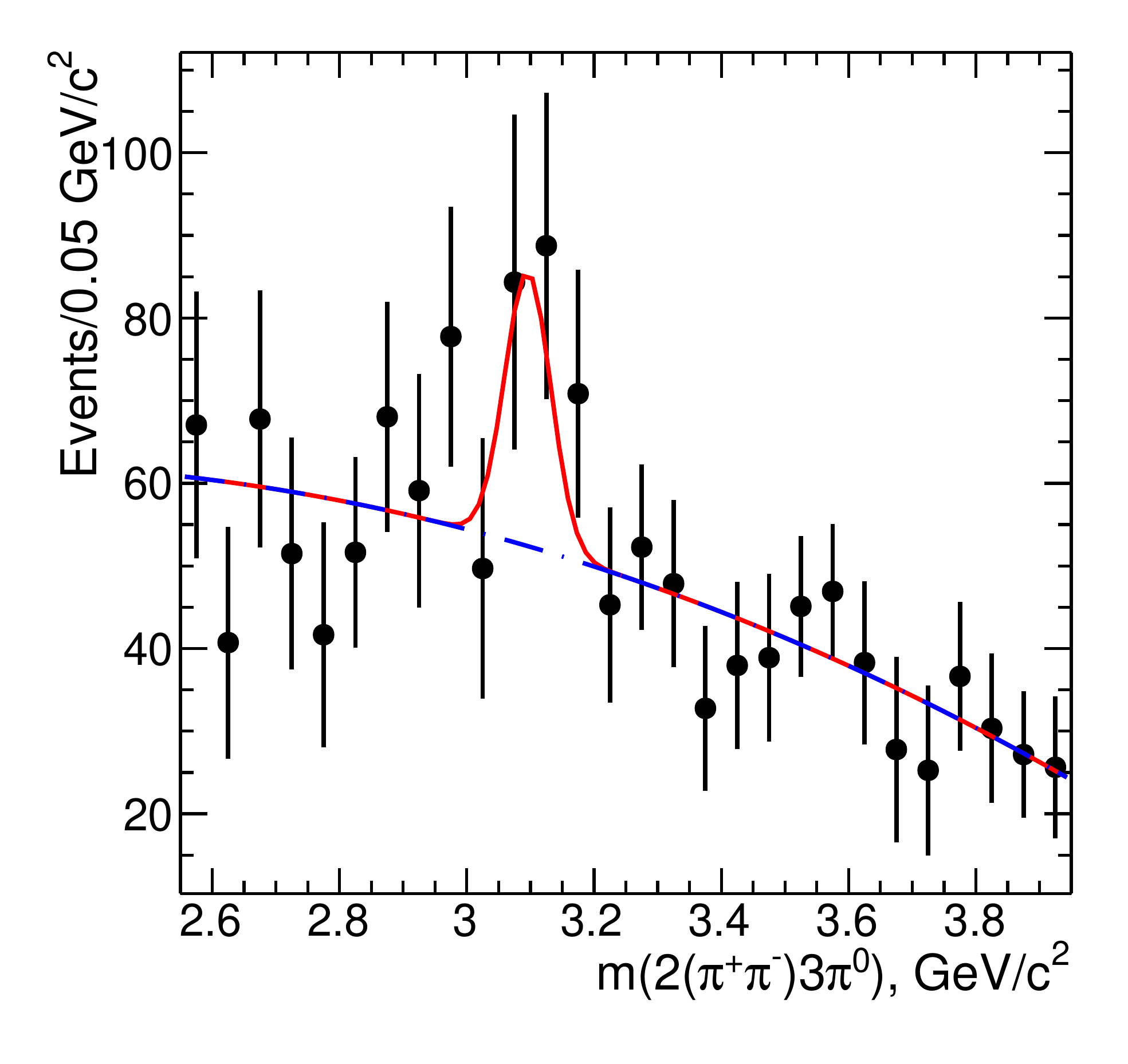}
\put(-40,80){\makebox(0,0)[lb]{\bf(b)}}
\vspace{-0.5cm}
\caption{
The seven-pion invariant mass for events with a three-pion invariant mass
in the $\omega(792)$ (a) or $\eta$ (b) mass regions.
The curves show the fit functions
described in the text.
}
\label{etaomega2pi}
\end{figure}

\begin{figure*}[t]
\includegraphics[width=0.33\linewidth]{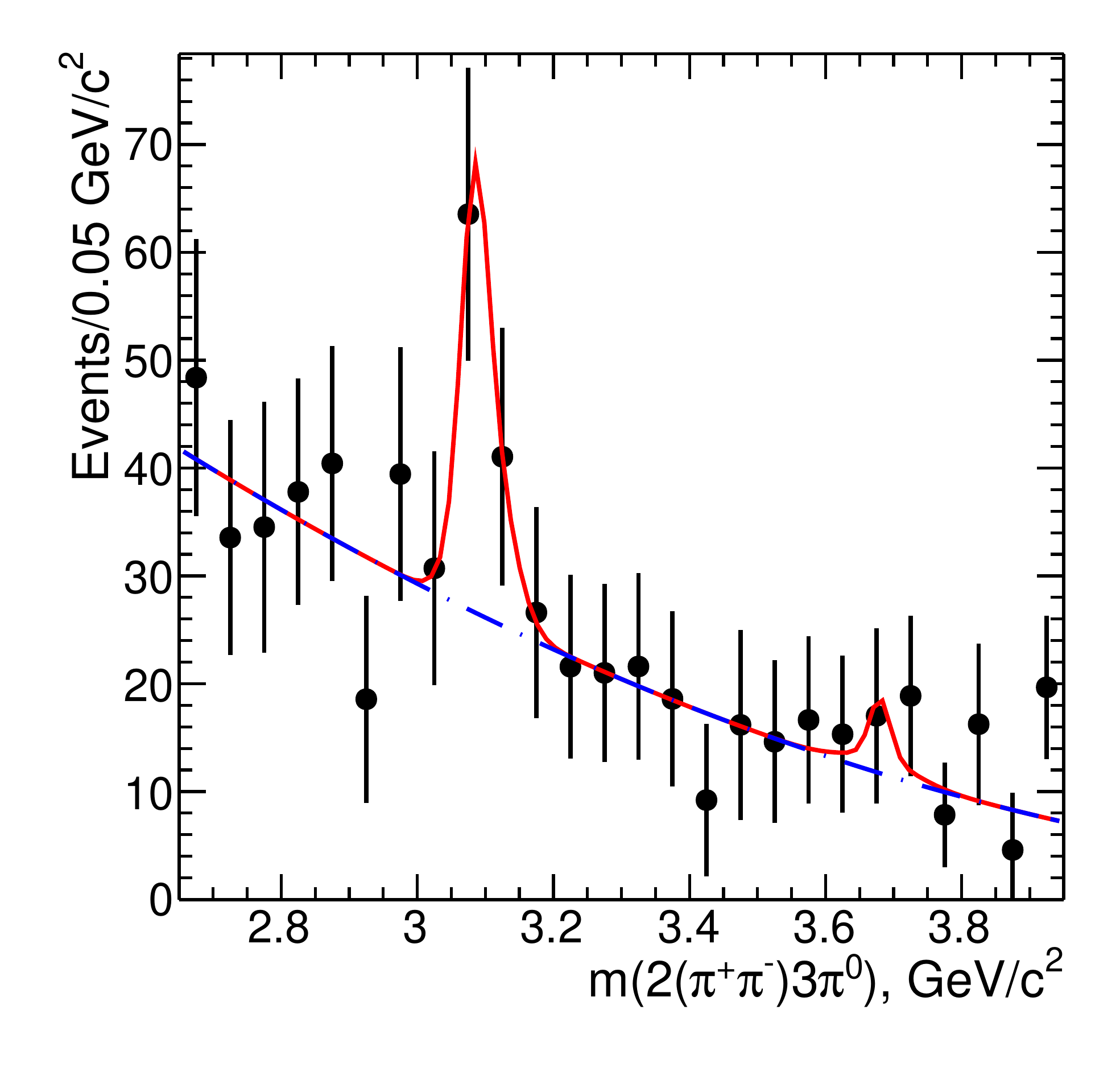}
\put(-50,120){\makebox(0,0)[lb]{\bf(a)}}
\includegraphics[width=0.33\linewidth]{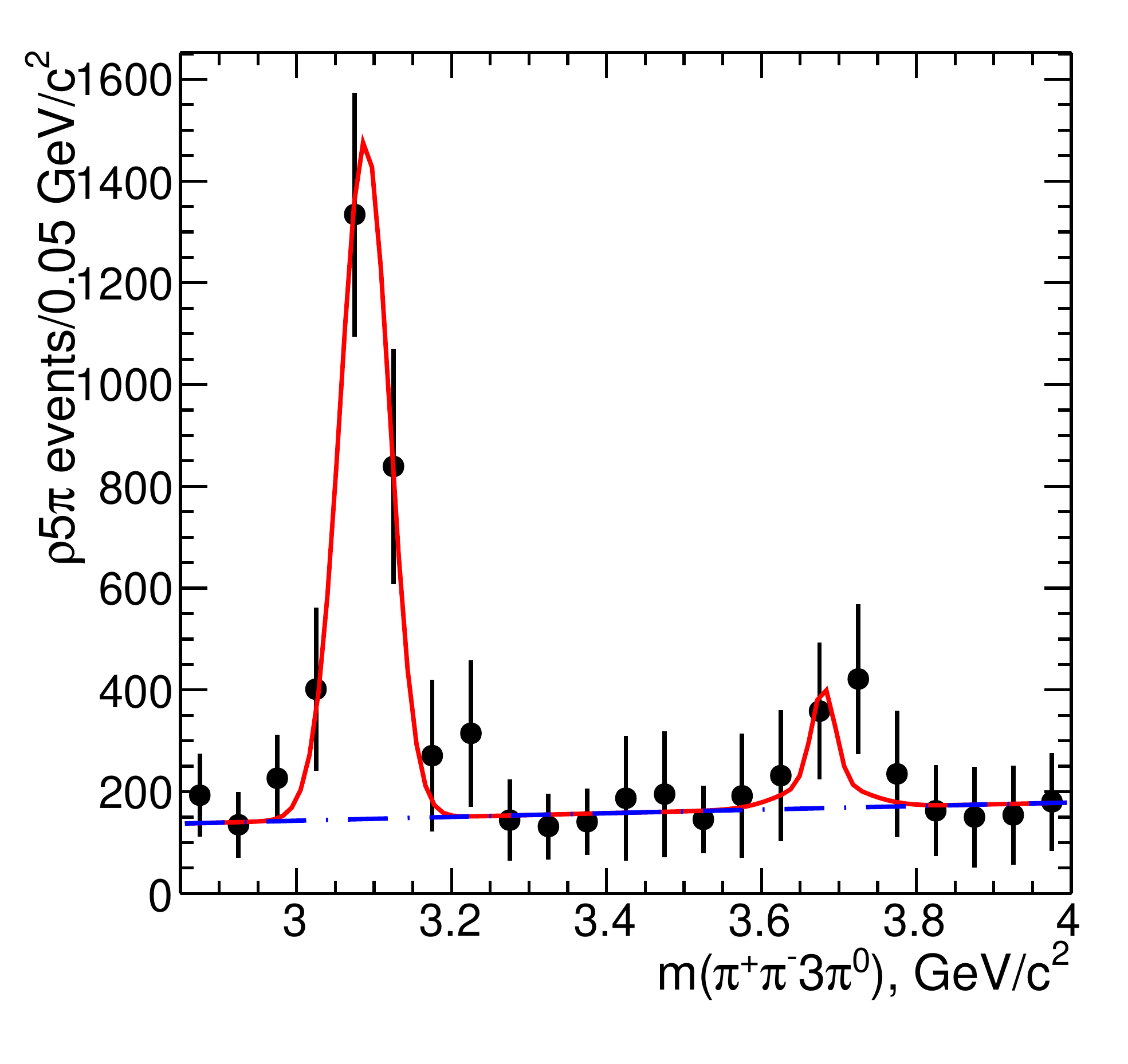}
\put(-50,120){\makebox(0,0)[lb]{\bf(b)}}
\includegraphics[width=0.33\linewidth]{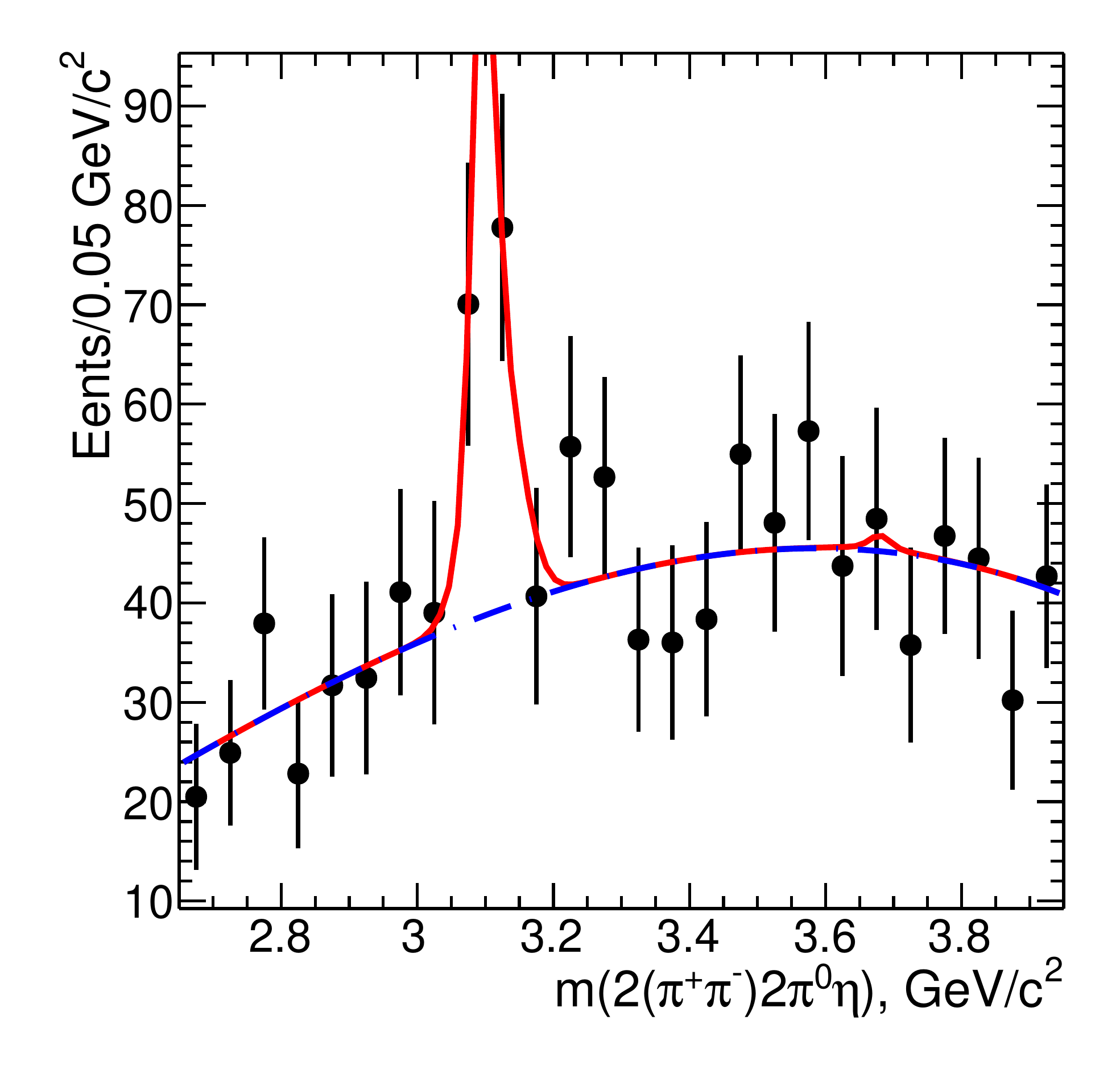}
\put(-50,120){\makebox(0,0)[lb]{\bf(c)}}
\vspace{-0.5cm}
\caption{
The $J/\psi$ region for  $2(\pipi)3\piz$ events for selection
of (a) the $\eta\pipi\pipi$ and 
(b)  the $\rho^{\pm}\pi^{\mp}\pipi\piz$ intermediate states. (c) The
$J/\psi$ region for   $2(\pipi)2\piz\eta$ events.
The curves show the fit
functions described in the text.
}
\label{4pietaatjpsi}
\end{figure*}    

The $\psi(2S)$ peak partly corresponds to the decay chain
$\psi(2S)\to J/\psi\ppz$ or $\psi(2S)\to J/\psi\pipi$, with $J/\psi$ decaying to
five pions. We select the $2(\pipi)3\piz$ events in the $\pm100$\mevcc
window around the $\psi(2S)$ mass and calculate $2(\pipi)\piz$ and
$\pipi3\piz$ invariant masses, shown in Fig.~\ref{psi2s_chain}(a) and
Fig.~\ref{psi2s_chain}(b), respectively.
 Clear signals from the above decay chains are seen. 
Performing a fit to these distributions yields
$130\pm21$ $\psi(2S)\to J/\psi\ppz\to2(\pipi)3\piz$
events and $114\pm27$ $\psi(2S)\to J/\psi\pipi\to2(\pipi)3\piz$
events.
In conjunction with the detection efficiency and
ISR luminosity, this yields:
\begin{eqnarray*}
  B_{\psi(2S)\to J/\psi\ppz}\cdot B_{J/\psi\to2(\pipi)\pi^0}\cdot\Gamma^{\psi(2S)}_{ee}
  &=&\\ (14.8\pm2.6\pm2.2)~\ev\ ,\\
    B_{\psi(2S)\to J/\psi\pipi}\cdot B_{J/\psi\to\pipi3\pi^0}\cdot\Gamma^{\psi(2S)}_{ee}
  &=&\\ (19.2\pm4.5\pm3.2)~\ev\ .
\end{eqnarray*}
With $\Gamma^{\psi(2S)}_{ee}$as stated above  and $B_{\psi(2S)\to J/\psi\ppz} = 0.1824\pm
0.0031$, $B_{\psi(2S)\to J/\psi\pipi} = 0.3468\pm0.0030$~\cite{PDG}, we obtain $B_{J/\psi\to2(\pipi)\pi^0} =
(3.47\pm 0.61\pm 0.52)\%$ and  $B_{J/\psi\to\pipi3\pi^0} =
(2.38\pm 0.56\pm 0.36)\%$.
These results are 
in  agreement with the  PDG values  
$B_{J/\psi\to2(\pipi)\pi^0} = (3.73\pm 0.32)\% ~S=1.4$~\cite{PDG}
and $B_{J/\psi\to\pipi3\pi^0} = (2.71\pm 0.29)\%$~\cite{isr2pi3pi0}.
Only \babar~ measurements are listed in PDG~\cite{PDG} for the last channel.

\subsubsection{\bf\boldmath The $\omega\pipi\ppz$,  $\eta\pipi\ppz$ intermediate states}

Figure~\ref{etaomega2pi}(a) shows 
an expanded view of Fig.~\ref{nevomega2pi0} with the $2(\pipi)3\piz$ mass
distribution for events obtained by a fit to the $\pipi\piz$
mass distribution to select events with an $\omega$.
The two-Gaussian fit, implemented as described above, yields $1619\pm92$ and $159\pm35$ events for
the $J/\psi$ and $\psi(2S)$, respectively. 
Using Eq.(\ref{jpsieq}) we obtain:
\begin{eqnarray*}
  B_{J/\psi\to\omega\pipi\ppz}\cdot B_{\omega\to\pipi\pi^0}\cdot\Gamma^{J/\psi}_{ee}
  &=&\\ (165\pm9\pm25)\ev\ ,\\
  B_{\psi(2S)\to\omega\pipi\ppz}\cdot B_{\omega\to\pipi\pi^0}\cdot\Gamma^{\psi(2S)}_{ee}
  &=& \\(18\pm4\pm3)\ev\ .
\end{eqnarray*} 
Using $B_{\omega\to\pipi\pi^0} = 0.891$ and the value of $\Gamma_{ee}$ from
Ref.~\cite{PDG}, we
obtain $B_{J/\psi\to\omega\pipi\ppz} = (3.3\pm 0.2\pm 0.5)\times 10^{-2}$
and $B_{\psi(2S)\to\omega\pipi\ppz} = (0.87\pm 0.19\pm 0.15)\times
10^{-2}$.
There are no other measurements of these decays.

Similarly, an expanded view of Fig.~\ref{neveta2pi0} is shown in
Fig.~\ref{etaomega2pi}(b) for   $2(\pipi)3\piz$ events with an $\eta$ signal
in the $\pipi\piz$ invariant mass. The fit yields $60\pm41$ events
corresponding to
\begin{eqnarray*}
 B_{J/\psi\to\eta\pipi\ppz}\cdot B_{\eta\to\pipi\pi^0}\cdot\Gamma^{J/\psi}_{ee}
  &=&\\ (6\pm4\pm1)\ev\ ,
\end{eqnarray*} 
which corresponds to $B_{J/\psi\to\eta\pipi\ppz} = (4.8\pm 3.2\pm
0.8)\times 10^{-3}$. We obtain  reasonable agreement with the only
available  result  $B_{J/\psi\to\eta\pipi\ppz} =
(2.3\pm0.5)\times 10^{-3}$~\cite{isr2pi3pi0}, obtained by \babar~ in
the $\eta\to\gamma\gamma$ decay mode. 

Note  that the $J/\psi$ decay to the $\omega\piz\eta$ mode is almost
ten times smaller,  $(0.34\pm0.17)\times 10^{-3}$~\cite{isr2pi3pi0},
and cannot be extracted from our data.

\subsubsection{\bf\boldmath The $\eta\pipi\pipi$ intermediate state}

An expanded view of Fig.~\ref{neveta2pi} is shown in
Fig.~\ref{4pietaatjpsi}(a).   The fit yields $55\pm25$ events
corresponding to
\begin{eqnarray*}
  B_{J/\psi\to\eta2(\pipi)}\cdot B_{\eta\to3\piz}\cdot\Gamma^{J/\psi}_{ee}
  &=&\\ (5.6\pm2.6\pm0.8)\ev\ ,\\
B_{J/\psi\to\eta2(\pipi)} = (2.6\pm 1.2\pm0.5)\times 10^{-3}.
\end{eqnarray*} 
The result is in agreement with world average value $(2.26\pm0.28)\times
10^{-3}$~\cite{PDG}.
We only can set an upper limit for the $\psi(2S)\to\eta2(\pipi)$
decay: we observe $<20$ events corresponding to
$B_{\psi(2S)\to\eta2(\pipi)}< 2.4\times 10^{-3}$ at 90\% C.L.,
which is consistent with
the world average value $1.2\pm0.6\times 10^{-3}$~\cite{PDG}.

\begin{table*}[tbh]
\caption{
  Summary of the $J/\psi$ and $\psi(2S)$ branching fractions.
  }
\label{jpsitab}
\begin{tabular}{r@{$\cdot$}l  r@{.}l@{$\pm$}l@{$\pm$}l 
                              r@{.}l@{$\pm$}l@{$\pm$}l
                              r@{.}l@{$\pm$}l } 
\multicolumn{2}{c}{Measured} & \multicolumn{4}{c}{Measured}    &  
\multicolumn{7}{c}{$J/\psi$ or $\psi(2S)$ Branching Fraction  (10$^{-3}$)}\\
\multicolumn{2}{c}{Quantity} & \multicolumn{4}{c}{Value (\ev)} &
\multicolumn{4}{c}{Calculated, this work}    & 
\multicolumn{3}{c}{PDG~\cite{PDG}} \\
\hline
$\Gamma^{J/\psi}_{ee}$  &  $\BR_{J/\psi \to \pipi\pipi\ppz\piz}$  &
  345& 0 & 10.0 & 50.0  &   ~~~62&0 & 2.0 & 9.0  &   \multicolumn{3}{c}{no entry}  \\

$\Gamma^{J/\psi}_{ee}$  &  $\BR_{J/\psi  \to \omega\pipi\ppz}
                           \cdot \BR_{\omega    \to \pipi\piz}$  &
  165&0 & 9.0 & 25.0  &   33&0 & 2.0 & 5.0  &    \multicolumn{3}{c}{no entry}   \\

$\Gamma^{J/\psi}_{ee}$  &  $\BR_{J/\psi  \to \eta\pipi\ppz}
                           \cdot \BR_{\eta   \to \pipi\piz}$  &
  6&0 & 4.0 & 1.0  &   4&8 & 3.2 & 0.8  & ~~~~2&3 & 0.5  \\
  

  $\Gamma^{J/\psi}_{ee}$  &  $\BR_{J/\psi  \to \pipi\pipi\eta}
                            \cdot\BR_{\eta   \to \ppz\piz}$ &
   5&6& 2.6& 0.8 &   2&6 & 1.2 & 0.5  & ~~~~2&26 & 0.28 \\
  
$\Gamma^{J/\psi}_{ee}$  &  $\BR_{J/\psi  \to  \rho^{\pm}\pi^{\mp}\pipi\ppz}$&
  155&0 & 26.0 & 36.0  &   28&0 & 4.7 & 6.6  &  \multicolumn{3}{c}{no entry} \\

$\Gamma^{J/\psi}_{ee}$  &  $\BR_{J/\psi  \to \rho^+\rho^-\pipi\piz}  $  &
   32&0& 13.0& 15.0 &   5&7 & 2.4 & 2.7  &    \multicolumn{3}{c}{no entry}\\

  $\Gamma^{J/\psi}_{ee}$  &  $\BR_{J/\psi  \to \pipi\pipi\ppz\eta}
                            \cdot\BR_{\eta   \to \gamma\gamma}$ &
   9&1& 2.6& 1.4 &   4&2 & 1.2 & 0.6  &  \multicolumn{3}{c}{no entry} \\

$\Gamma^{\psi(2S)}_{ee}$  &  $\BR_{\psi(2S) \to\pipi\pipi\ppz\piz} $  &
   33&0& 5.0& 5.0 &   14&0  & 2.0  & 2.0   &  \multicolumn{3}{c}{no entry} \\

$\Gamma^{\psi(2S)}_{ee}$  &  $\BR_{\psi(2S) \to J/\psi\ppz} 
                             \cdot \BR_{J/\psi \to \pipi\pipi\piz}  $  &
   14&8& 2.6& 2.2 &   34&7  & 6.1  & 5.2   &   ~~~~~~33&7 & 2.6 \\

$\Gamma^{\psi(2S)}_{ee}$  &  $\BR_{\psi(2S) \to J/\psi\pipi} 
                             \cdot \BR_{J/\psi \to \pipi\ppz\piz}  $  &
   19&2& 4.5& 3.2 &   23&8  & 5.6  & 3.6   &   ~~~~~~27&1 & 2.9 \\
  
$\Gamma^{\psi(2S)}_{ee}$  &  $\BR_{\psi(2S) \to \omega\pipi\ppz}
                        \cdot \BR_{\omega     \to \pipi\piz} $  &
   18&0& 4.0& 3.0 &   8&7 & 1.9 & 1.5  &   \multicolumn{3}{c}{no entry}\\

 $\Gamma^{\psi(2S)}_{ee}$  &  $\BR_{\psi(2S) \to \pipi\pipi\ppz\eta} \cdot \BR_{\eta   \to \gamma\gamma}$  &
   ~~~~$<$1 & \multicolumn{3}{l}{9 at 90\% C.L.} & ~~~~ $<$2 &
                                                     \multicolumn{3}{l}{0
                                                     at 90\% C.L.} &      \multicolumn{3}{c}{no entry}\\ 

$\Gamma^{\psi(2S)}_{ee}$  &  $\BR_{\psi(2S) \to \pipi\pipi\eta} \cdot \BR_{\eta   \to \ppz\piz}$  &
   ~~~~$<$2 & \multicolumn{3}{l}{3 at 90\% C.L.} & ~~~~ $<$2 &
                                                     \multicolumn{3}{l}{4
                                                     at 90\% C.L.} &       ~~~~~~1&2 & 0.6 \\

\hline
\end{tabular}
\end{table*}

\subsubsection{\bf\boldmath The $\rho^{\pm}\pi^{\mp}\pipi\ppz$ intermediate state}

Figure~\ref{4pietaatjpsi}(b) shows 
an expanded view of Fig.~\ref{sumallev}(a) (circles) for the $2(\pipi)3\piz$ mass
for events obtained from the fit to the $\rho$ signal in the $\pi^{\pm}\piz$ mass.
The two-Gaussian fit yields $2149\pm363$ and $266\pm157$ events  for
the $J/\psi$ and $\psi(2S)$, respectively. 

As shown in Sec.~\ref{sec:rhoselectpi0} about 20\% of these events
arise from 
the $J/\psi\to\rho^{\pm}\rho^{\mp}\pipi\piz$ decays.
We estimate the number of  $J/\psi$ decays to single-
and double-$\rho$ to be  $1526\pm258\pm279$ and $312\pm129\pm140$, respectively.
The second uncertainty is
due to the fraction of
$\rho^+\rho^-$ events,  given above.
We obtain:
\begin{eqnarray*}
  B_{J/\psi\to\rho^{\pm}\pi^{\mp}\pipi\ppz}\cdot\Gamma^{J/\psi}_{ee}
  &=& (155\pm26\pm28\pm22)\ev\ ,\\
  B_{J/\psi\to\rho^+\rho^-\pipi\piz}\cdot\Gamma^{J/\psi}_{ee}
  &=& (32\pm13\pm14\pm5)\ev\ .
\end{eqnarray*} 
Dividing by the value of $\Gamma_{ee}$ from Ref.~\cite{PDG}
then yields:
\begin{eqnarray*}
  B_{J/\psi\to\rho^{\pm}\pi^{\mp}\pipi\ppz}
  &=& (2.8\pm0.47\pm0.51\pm0.42)\times10^{-2} ,\\
  B_{J/\psi\to\rho^+\rho^-\pipi\piz}
  &=& (0.57\pm0.24\pm0.25\pm0.09)\times10^{-2} ,
\end{eqnarray*} 
where the third uncertainty is associated with the
procedure used to determine
the correlated $\rho^+\rho^-$ rate.
No other measurements for these processes exist.

For the $\psi(2S)\to\rho^{\pm}\pi^{\mp}\pipi\ppz$ decay we find
$266\pm157$ events. We cannot extract and estimate
the size of
the contribution of double-$\rho$ events, so we do not calculate
branching fractions.

\subsection{\bf\boldmath The $2(\pipi)2\piz\eta$  final state}

The expanded view of Fig.~\ref{meta_data_fit}(b) is shown in
Fig.~\ref{4pietaatjpsi}(c).   The fit yields $90\pm26$ for the $J/\psi\to2(\pipi)2\piz\eta$ events
corresponding to
\begin{eqnarray*}
  B_{J/\psi\to\eta2(\pipi)2\piz}\cdot B_{\eta\to\gamma\gamma}\cdot\Gamma^{J/\psi}_{ee}
  &=&\\ (9.1\pm2.6\pm1.4)\ev\ ,\\
B_{J/\psi\to\eta2(\pipi)2\piz} = (4.2\pm 1.2\pm0.6)\times 10^{-3}.
\end{eqnarray*}
We set an upper limit $\psi(2S)\to\eta2(\pipi)2\piz$
decay: we observe $<18$ events at 90\% C.L. corresponding to
$B_{\psi(2S)\to\eta2(\pipi)2\piz}< 2.0\times 10^{-3}$.

There are no previous results for these final states.

\subsection{Summary of the charmonium region study}
The rates of  $J/\psi$ and $\psi(2S)$ decays to $2(\pipi)3\piz$,
$2(\pipi)2\piz\eta$ and several intermediate  final
states have been measured. 
The measured products and calculated 
branching fractions are summarized in Table~\ref{jpsitab}
together with the available PDG values for comparison.

\section{Summary}
\label{sec:Summary}
\noindent
The excellent performance of the \babar~ detector for photon energy and
charged-particle resolution, together with its strong particle identification
capabilities, allow
the reconstruction of the
$2(\pipi)3\piz$ and $2(\pipi)2\piz\eta$ final 
states from threshold up to 4.5~\gev via the ISR process.

The analysis shows that the effective luminosity and efficiency have been understood
with 10--13\% accuracy.
The cross section measurements for the $\epem\to2(\pipi)3\piz$
and the $\epem\to2(\pipi)2\piz\eta$ reactions
has been measured for the first time.

The selected multi-hadronic final states in the broad range of accessible
energies provide new information on hadron spectroscopy. The  
observed $\epem\to \omega\pipi\ppz$,  $\epem\to\eta\pipi\ppz$, and $\epem\to \eta2(\pipi)$  cross
sections provide  additional information for the hadronic contribution
calculation of the muon $g-2$.  

The initial-state radiation events also allow a study of $J/\psi$ and
$\psi(2S)$ production and a measurement of the corresponding products of
the decay branching fractions and $\epem$ width for most of
the studied channels, the majority of them for the first time.


\section{Acknowledgments}
\label{sec:Acknowledgments}

\input acknowledgements

\newpage

\input {biblio_PRD}
\end{document}

%% file: authors_jan2021_frozen.tex
\author{J.~P.~Lees}
\author{V.~Poireau}
\author{V.~Tisserand}
\affiliation{Laboratoire d'Annecy-le-Vieux de Physique des Particules (LAPP), Universit\'e de Savoie, CNRS/IN2P3,  F-74941 Annecy-Le-Vieux, France}
\author{E.~Grauges}
\affiliation{Universitat de Barcelona, Facultat de Fisica, Departament ECM, E-08028 Barcelona, Spain }
\author{A.~Palano}
\affiliation{INFN Sezione di Bari, I-70126 Bari, Italy}
\author{G.~Eigen}
\affiliation{University of Bergen, Institute of Physics, N-5007 Bergen, Norway }
\author{D.~N.~Brown}
\author{Yu.~G.~Kolomensky}
\affiliation{Lawrence Berkeley National Laboratory and University of California, Berkeley, California 94720, USA }
\author{M.~Fritsch}
\author{H.~Koch}
\author{T.~Schroeder}
\affiliation{Ruhr Universit\"at Bochum, Institut f\"ur Experimentalphysik 1, D-44780 Bochum, Germany }
% ...
\collaboration{The \babar\ Collaboration}
\noaffiliation

%% file: xs4pi3pi0_table.tex
\begin{table*}
\caption{Summary of the $\epem\to 2(\pipi)3\piz$ 
cross section measurement. The uncertainties are statistical only.}
\label{4pi3pi0_tab}
\begin{tabular}{c c c c c c c c c c}
$E_{\rm c.m.}$, GeV & $\sigma$, nb  
& $E_{\rm c.m.}$, GeV & $\sigma$, nb 
& $E_{\rm c.m.}$, GeV & $\sigma$, nb 
& $E_{\rm c.m.}$, GeV & $\sigma$, nb  
& $E_{\rm c.m.}$, GeV & $\sigma$, nb  
\\
\hline
1.575 & 0.00 $\pm$ 0.01 &2.175 & 0.84 $\pm$ 0.15 &2.775 & 0.96 $\pm$ 0.14 &3.375 & 0.78 $\pm$ 0.12 &3.975 & 0.46 $\pm$ 0.09 \\ 
1.625 & 0.02 $\pm$ 0.01 &2.225 & 1.06 $\pm$ 0.11 &2.825 & 0.88 $\pm$ 0.13 &3.425 & 0.59 $\pm$ 0.09 &4.025 & 0.44 $\pm$ 0.09 \\ 
1.675 & 0.00 $\pm$ 0.02 &2.275 & 1.07 $\pm$ 0.14 &2.875 & 1.12 $\pm$ 0.13 &3.475 & 0.70 $\pm$ 0.11 &4.075 & 0.42 $\pm$ 0.09 \\ 
1.725 & 0.26 $\pm$ 0.06 &2.325 & 1.11 $\pm$ 0.12 &2.925 & 0.88 $\pm$ 0.13 &3.525 & 0.67 $\pm$ 0.10 &4.125 & 0.32 $\pm$ 0.07 \\ 
1.775 & 0.25 $\pm$ 0.07 &2.375 & 1.14 $\pm$ 0.14 &2.975 & 1.02 $\pm$ 0.17 &3.575 & 0.73 $\pm$ 0.12 &4.175 & 0.56 $\pm$ 0.08 \\ 
1.825 & 0.62 $\pm$ 0.09 &2.425 & 1.39 $\pm$ 0.16 &3.025 & 1.49 $\pm$ 0.20 &3.625 & 0.63 $\pm$ 0.11 &4.225 & 0.31 $\pm$ 0.08 \\ 
1.875 & 0.82 $\pm$ 0.14 &2.475 & 1.21 $\pm$ 0.16 &3.075 & 10.76 $\pm$ 0.26 &3.675 & 1.53 $\pm$ 0.15 &4.275 & 0.35 $\pm$ 0.06 \\ 
1.925 & 0.73 $\pm$ 0.09 &2.525 & 1.01 $\pm$ 0.16 &3.125 & 6.30 $\pm$ 0.26 &3.725 & 0.81 $\pm$ 0.13 &4.325 & 0.23 $\pm$ 0.07 \\ 
1.975 & 0.69 $\pm$ 0.10 &2.575 & 0.84 $\pm$ 0.14 &3.175 & 1.44 $\pm$ 0.15 &3.775 & 0.31 $\pm$ 0.11 &4.375 & 0.42 $\pm$ 0.06 \\ 
2.025 & 0.90 $\pm$ 0.15 &2.625 & 0.82 $\pm$ 0.11 &3.225 & 0.90 $\pm$ 0.11 &3.825 & 0.53 $\pm$ 0.10 &4.425 & 0.45 $\pm$ 0.07 \\ 
2.075 & 0.88 $\pm$ 0.14 &2.675 & 1.02 $\pm$ 0.15 &3.275 & 0.67 $\pm$ 0.12 &3.875 & 0.29 $\pm$ 0.09 &4.475 & 0.30 $\pm$ 0.07 \\ 
2.125 & 0.70 $\pm$ 0.16 &2.725 & 0.95 $\pm$ 0.15 &3.325 & 0.82 $\pm$0.12 &3.925 & 0.46 $\pm$ 0.09 &  &   \\ 
\hline
\end{tabular}
\end{table*}

%% file: xs4pieta_table.tex
\begin{table*}
\caption{Summary of the $\epem\to\eta2(\pipi)$ 
cross section measurement. The uncertainties are statistical only.}
\label{2pieta_table}
\begin{tabular}{c c c c c c c c c c}
$E_{\rm c.m.}$, GeV & $\sigma$, nb  
& $E_{\rm c.m.}$, GeV & $\sigma$, nb 
& $E_{\rm c.m.}$, GeV & $\sigma$, nb 
& $E_{\rm c.m.}$, GeV & $\sigma$, nb  
& $E_{\rm c.m.}$, GeV & $\sigma$, nb  
\\
\hline
1.575 & 0.00 $\pm$ 0.00 &2.175 & 1.48 $\pm$ 0.33 &2.775 & 0.62 $\pm$ 0.21 &3.375 & 0.27 $\pm$ 0.12 &3.975 & 0.08 $\pm$ 0.07 \\ 
1.625 & 0.00 $\pm$ 0.25 &2.225 & 1.11 $\pm$ 0.30 &2.825 & 0.66 $\pm$ 0.18 &3.425 & 0.13 $\pm$ 0.10 &4.025 & 0.00 $\pm$ 0.07 \\ 
1.675 & 0.16 $\pm$ 0.12 &2.275 & 1.98 $\pm$ 0.33 &2.875 & 0.70 $\pm$ 0.19 &3.475 & 0.23 $\pm$ 0.12 &4.075 & 0.03 $\pm$ 0.09 \\ 
1.725 & 0.00 $\pm$ 0.28 &2.325 & 0.92 $\pm$ 0.26 &2.925 & 0.31 $\pm$ 0.16 &3.525 & 0.20 $\pm$ 0.10 &4.125 & 0.04 $\pm$ 0.05 \\ 
1.775 & 0.19 $\pm$ 0.20 &2.375 & 1.12 $\pm$ 0.28 &2.975 & 0.66 $\pm$ 0.20 &3.575 & 0.23 $\pm$ 0.11 &4.175 & 0.07 $\pm$ 0.06 \\ 
1.825 & 0.27 $\pm$ 0.22 &2.425 & 1.51 $\pm$ 0.30 &3.025 & 0.50 $\pm$ 0.18 &3.625 & 0.21 $\pm$ 0.10 &4.225 & 0.09 $\pm$ 0.06 \\ 
1.875 & 0.19 $\pm$ 0.25 &2.475 & 1.28 $\pm$ 0.27 &3.075 & 1.02 $\pm$ 0.22 &3.675 & 0.23 $\pm$ 0.11 &4.275 & 0.07 $\pm$ 0.04 \\ 
1.925 & 0.53 $\pm$ 0.25 &2.525 & 0.54 $\pm$ 0.22 &3.125 & 0.65 $\pm$ 0.19 &3.725 & 0.25 $\pm$ 0.10 &4.325 & 0.00 $\pm$ 0.05 \\ 
1.975 & 0.20 $\pm$ 0.29 &2.575 & 0.98 $\pm$ 0.24 &3.175 & 0.41 $\pm$ 0.15 &3.775 & 0.10 $\pm$ 0.06 &4.375 & 0.05 $\pm$ 0.05 \\ 
2.025 & 0.88 $\pm$ 0.31 &2.625 & 0.84 $\pm$ 0.19 &3.225 & 0.33 $\pm$ 0.13 &3.825 & 0.20 $\pm$ 0.09 &4.425 & 0.01 $\pm$ 0.01 \\ 
2.075 & 0.72 $\pm$ 0.30 &2.675 & 0.90 $\pm$ 0.24 &3.275 & 0.32 $\pm$ 0.12 &3.875 & 0.06 $\pm$ 0.07 &4.475 & 0.02 $\pm$ 0.02 \\ 
2.125 & 1.10 $\pm$ 0.30 &2.725 & 0.61 $\pm$ 0.20 &3.325 & 0.32 $\pm$ 0.13 &3.925 & 0.24 $\pm$ 0.08 &&  \\

\hline
\end{tabular}
\end{table*}

%% file: xsomegapi0eta_table.tex
\begin{table*}
\caption{Summary of the $\epem\to\omega\piz\eta$ 
cross section measurement. The uncertainties are statistical only.}
\label{ompi0eta_table}
\begin{tabular}{c c c c c c c c c c}
$E_{\rm c.m.}$, GeV & $\sigma$, nb  
& $E_{\rm c.m.}$, GeV & $\sigma$, nb 
& $E_{\rm c.m.}$, GeV & $\sigma$, nb 
& $E_{\rm c.m.}$, GeV & $\sigma$, nb  
& $E_{\rm c.m.}$, GeV & $\sigma$, nb  
\\
  \hline
1.575 & 0.00 $\pm$ 0.00 &2.175 & 0.70 $\pm$ 0.39 &2.775 & 0.55 $\pm$ 0.17 &3.375 & 0.03 $\pm$ 0.08 &3.975 & 0.04 $\pm$ 0.04 \\ 
1.625 & -0.01 $\pm$ 0.09 &2.225 & 0.76 $\pm$ 0.42 &2.825 & 0.15 $\pm$ 0.18 &3.425 & 0.18 $\pm$ 0.10 &4.025 & 0.08 $\pm$ 0.05 \\ 
1.675 & 0.17 $\pm$ 0.08 &2.275 & 0.37 $\pm$ 0.39 &2.875 & 0.19 $\pm$ 0.13 &3.475 & 0.14 $\pm$ 0.07 &4.075 & 0.06 $\pm$ 0.07 \\ 
1.725 & 0.53 $\pm$ 0.39 &2.325 & 0.68 $\pm$ 0.34 &2.925 & 0.23 $\pm$ 0.15 &3.525 & -0.01 $\pm$ 0.02 &4.125 & 0.01 $\pm$ 0.04 \\ 
1.775 & 1.21 $\pm$ 0.44 &2.375 & 0.43 $\pm$ 0.24 &2.975 & 0.07 $\pm$ 0.13 &3.575 & 0.02 $\pm$ 0.07 &4.175 & 0.01 $\pm$ 0.02 \\ 
1.825 & 1.69 $\pm$ 0.51 &2.425 & 0.25 $\pm$ 0.25 &3.025 & -0.00 $\pm$ 0.08 &3.625 & 0.03 $\pm$ 0.07 &4.225 & 0.06 $\pm$ 0.05 \\ 
1.875 & 1.63 $\pm$ 0.55 &2.475 & 0.94 $\pm$ 0.31 &3.075 & 0.61 $\pm$ 0.16 &3.675 & 0.04 $\pm$ 0.10 &4.275 & 0.03 $\pm$ 0.03 \\ 
1.925 & 1.78 $\pm$ 0.54 &2.525 & 0.09 $\pm$ 0.19 &3.125 & 0.23 $\pm$ 0.17 &3.725 & 0.09 $\pm$ 0.07 &4.325 & 0.02 $\pm$ 0.03 \\ 
1.975 & 1.09 $\pm$ 0.51 &2.575 & 0.23 $\pm$ 0.19 &3.175 & 0.13 $\pm$ 0.09 &3.775 & 0.04 $\pm$ 0.04 &4.375 & 0.03 $\pm$ 0.03 \\ 
2.025 & 1.35 $\pm$ 0.53 &2.625 & 0.41 $\pm$ 0.19 &3.225 & 0.09 $\pm$ 0.10 &3.825 & -0.00 $\pm$ 0.05 &4.425 & 0.04 $\pm$ 0.03 \\ 
2.075 & 1.88 $\pm$ 0.54 &2.675 & 0.25 $\pm$ 0.17 &3.275 & 0.05 $\pm$ 0.10 &3.875 & 0.09 $\pm$ 0.07 &4.475 & 0.01 $\pm$ 0.03 \\ 
2.125 & 1.35 $\pm$ 0.47 &2.725 & 0.60 $\pm$ 0.19 &3.325 & 0.13 $\pm$ 0.09 &3.925 & 0.03 $\pm$ 0.03 &&  \\

\hline
\end{tabular}
\end{table*}

%% file: xs2pi2pi0omega_table.tex
\begin{table*}
\caption{Summary of the $\epem\to\pipi\ppz\omega$ 
cross section measurement. The uncertainties are statistical only.}
\label{omega2pi0_table}
\begin{tabular}{c c c c c c c c c c}
$E_{\rm c.m.}$, GeV & $\sigma$, nb  
& $E_{\rm c.m.}$, GeV & $\sigma$, nb 
& $E_{\rm c.m.}$, GeV & $\sigma$, nb 
& $E_{\rm c.m.}$, GeV & $\sigma$, nb  
& $E_{\rm c.m.}$, GeV & $\sigma$, nb  
\\
\hline
1.575 & 0.00 $\pm$ 0.00 &2.175 & 0.53 $\pm$ 0.30 &2.775 & 0.48 $\pm$ 0.21 &3.375 & 0.31 $\pm$ 0.13 &3.975 & 0.10 $\pm$ 0.08 \\ 
1.625 & 0.01 $\pm$ 0.06 &2.225 & 0.86 $\pm$ 0.30 &2.825 & 0.18 $\pm$ 0.20 &3.425 & 0.15 $\pm$ 0.13 &4.025 & 0.08 $\pm$ 0.08 \\ 
1.675 & 0.02 $\pm$ 0.09 &2.275 & 0.20 $\pm$ 0.29 &2.875 & 0.59 $\pm$ 0.20 &3.475 & 0.16 $\pm$ 0.12 &4.075 & 0.07 $\pm$ 0.09 \\ 
1.725 & 0.19 $\pm$ 0.17 &2.325 & 0.28 $\pm$ 0.26 &2.925 & 0.54 $\pm$ 0.20 &3.525 & 0.08 $\pm$ 0.12 &4.125 & 0.08 $\pm$ 0.09 \\ 
1.775 & 0.36 $\pm$ 0.21 &2.375 & 0.73 $\pm$ 0.26 &2.975 & 0.68 $\pm$ 0.21 &3.575 & 0.38 $\pm$ 0.12 &4.175 & 0.04 $\pm$ 0.07 \\ 
1.825 & 0.18 $\pm$ 0.23 &2.425 & 0.41 $\pm$ 0.27 &3.025 & 0.83 $\pm$ 0.24 &3.625 & 0.27 $\pm$ 0.13 &4.225 & 0.15 $\pm$ 0.07 \\ 
1.875 & 0.71 $\pm$ 0.26 &2.475 & 0.82 $\pm$ 0.25 &3.075 & 5.46 $\pm$ 0.36 &3.675 & 0.77 $\pm$ 0.15 &4.275 & 0.13 $\pm$ 0.06 \\ 
1.925 & 0.35 $\pm$ 0.27 &2.525 & 0.77 $\pm$ 0.25 &3.125 & 3.88 $\pm$ 0.30 &3.725 & 0.58 $\pm$ 0.13 &4.325 & 0.19 $\pm$ 0.07 \\ 
1.975 & 0.65 $\pm$ 0.30 &2.575 & 0.56 $\pm$ 0.23 &3.175 & 0.61 $\pm$ 0.19 &3.775 & 0.30 $\pm$ 0.11 &4.375 & 0.11 $\pm$ 0.06 \\ 
2.025 & 0.53 $\pm$ 0.32 &2.625 & 0.22 $\pm$ 0.21 &3.225 & 0.61 $\pm$ 0.16 &3.825 & 0.13 $\pm$ 0.10 &4.425 & 0.04 $\pm$ 0.06 \\ 
2.075 & 0.46 $\pm$ 0.32 &2.675 & 0.51 $\pm$ 0.22 &3.275 & 0.33 $\pm$ 0.15 &3.875 & 0.19 $\pm$ 0.09 &4.475 & 0.20 $\pm$ 0.07 \\ 
2.125 & 0.38 $\pm$ 0.32 &2.725 & 0.69 $\pm$ 0.22 &3.325 & 0.14 $\pm$0.14 &3.925  & 0.33 $\pm$ 0.09 & & \\ 
\hline
\end{tabular}
\end{table*}

%% file: xs2pi2pi0eta_table.tex
\begin{table*}
\caption{Summary of the $\epem\to\pipi\ppz\eta$ 
cross section measurement. The uncertainties are statistical only.}
\label{eta2pi0_table}
\begin{tabular}{c c c c c c c c c c}
$E_{\rm c.m.}$, GeV & $\sigma$, nb  
& $E_{\rm c.m.}$, GeV & $\sigma$, nb 
& $E_{\rm c.m.}$, GeV & $\sigma$, nb 
& $E_{\rm c.m.}$, GeV & $\sigma$, nb  
& $E_{\rm c.m.}$, GeV & $\sigma$, nb  
\\
  \hline
1.650 & 0.34 $\pm$ 0.23 &2.250 & 1.35 $\pm$ 0.48 &2.850 & 1.26 $\pm$ 0.25 &3.450 & 0.54 $\pm$ 0.17 &4.050 & 0.03 $\pm$ 0.11 \\ 
1.750 & 0.32 $\pm$ 0.45 &2.350 & 1.41 $\pm$ 0.40 &2.950 & 1.42 $\pm$ 0.28 &3.550 & 0.66 $\pm$ 0.18 &4.150 & 0.11 $\pm$ 0.09 \\ 
1.850 & 0.91 $\pm$ 0.56 &2.450 & 1.12 $\pm$ 0.39 &3.050 & 0.95 $\pm$ 0.32 &3.650 & 0.34 $\pm$ 0.18 &4.250 & -0.01 $\pm$ 0.08 \\ 
1.950 & 0.89 $\pm$ 0.57 &2.550 & 1.52 $\pm$ 0.36 &3.150 & 1.19 $\pm$ 0.30 &3.750 & 0.20 $\pm$ 0.16 &4.350 & 0.10 $\pm$ 0.09 \\ 
2.050 & 1.75 $\pm$ 0.59 &2.650 & 1.00 $\pm$ 0.31 &3.250 & 0.70 $\pm$ 0.20 &3.850 & 0.24 $\pm$ 0.13 &4.450 & 0.06 $\pm$ 0.12 \\ 
2.150 & 1.49 $\pm$ 0.54 &2.750 & 0.55 $\pm$ 0.29 &3.350 & 0.39 $\pm$ 0.18 &3.950 & 0.22 $\pm$ 0.13 & & \\ 
\hline
\end{tabular}
\end{table*}

%% file: xs6pieta_table.tex
\begin{table*}
\caption{Summary of the $\epem\to 2(\pipi)\ppz\eta$ 
cross section measurement. The uncertainties are statistical only.}
\label{6pieta_table}
\begin{tabular}{c c c c c c c c c c}
$E_{\rm c.m.}$, GeV & $\sigma$, nb  
& $E_{\rm c.m.}$, GeV & $\sigma$, nb 
& $E_{\rm c.m.}$, GeV & $\sigma$, nb 
& $E_{\rm c.m.}$, GeV & $\sigma$, nb  
& $E_{\rm c.m.}$, GeV & $\sigma$, nb  
\\
\hline
2.075 & 0.03 $\pm$ 0.03 &2.575 & 0.13 $\pm$ 0.05 &3.075 & 0.37 $\pm$ 0.08 &3.575 & 0.21 $\pm$ 0.06 &4.075 & 0.08 $\pm$ 0.05 \\ 
2.125 & -0.02 $\pm$ 0.04 &2.625 & 0.16 $\pm$ 0.05 &3.125 & 0.38 $\pm$ 0.07 &3.625 & 0.15 $\pm$ 0.06 &4.125 & 0.01 $\pm$ 0.04 \\ 
2.175 & 0.01 $\pm$ 0.02 &2.675 & 0.12 $\pm$ 0.05 &3.175 & 0.18 $\pm$ 0.06 &3.675 & 0.17 $\pm$ 0.06 &4.175 & 0.12 $\pm$ 0.04 \\ 
2.225 & 0.07 $\pm$ 0.03 &2.725 & 0.14 $\pm$ 0.04 &3.225 & 0.24 $\pm$ 0.06 &3.725 & 0.14 $\pm$ 0.05 &4.225 & 0.12 $\pm$ 0.04 \\ 
2.275 & 0.10 $\pm$ 0.03 &2.775 & 0.22 $\pm$ 0.05 &3.275 & 0.24 $\pm$ 0.05 &3.775 & 0.15 $\pm$ 0.05 &4.275 & 0.03 $\pm$ 0.04 \\ 
2.325 & 0.15 $\pm$ 0.04 &2.825 & 0.12 $\pm$ 0.05 &3.325 & 0.14 $\pm$ 0.05 &3.825 & 0.09 $\pm$ 0.05 &4.325 & 0.04 $\pm$ 0.04 \\ 
2.375 & 0.04 $\pm$ 0.03 &2.875 & 0.15 $\pm$ 0.05 &3.375 & 0.15 $\pm$ 0.05 &3.875 & 0.10 $\pm$ 0.04 &4.375 & 0.04 $\pm$ 0.04 \\ 
2.425 & 0.10 $\pm$ 0.04 &2.925 & 0.14 $\pm$ 0.06 &3.425 & 0.17 $\pm$ 0.06 &3.925 & 0.10 $\pm$ 0.05 &4.425 & 0.08 $\pm$ 0.04 \\ 
2.475 & 0.13 $\pm$ 0.04 &2.975 & 0.22 $\pm$ 0.06 &3.475 & 0.22 $\pm$ 0.05 &3.975 & 0.14 $\pm$ 0.04 &4.475 & 0.09 $\pm$ 0.04 \\ 
2.525 & 0.07 $\pm$ 0.05 &3.025 & 0.19 $\pm$ 0.06 &3.525 & 0.19 $\pm$ 0.06 &4.025 & 0.10 $\pm$ 0.04 &&  \\ 
 
\hline
\end{tabular}
\end{table*}

%% file: acknowledgements.tex
We are grateful for the 
extraordinary contributions of our \pep2\ colleagues in
achieving the excellent luminosity and machine conditions
that have made this work possible.
The success of this project also relies critically on the 
expertise and dedication of the computing organizations that 
support \babar.
The collaborating institutions wish to thank 
SLAC for its support and the kind hospitality extended to them. 
This work is supported by the
US Department of Energy
and National Science Foundation, the
Natural Sciences and Engineering Research Council (Canada),
the Commissariat \`a l'Energie Atomique and
Institut National de Physique Nucl\'eaire et de Physique des Particules
(France), the
Bundesministerium f\"ur Bildung und Forschung and
Deutsche Forschungsgemeinschaft
(Germany), the
Istituto Nazionale di Fisica Nucleare (Italy),
the Foundation for Fundamental Research on Matter (The Netherlands),
the Research Council of Norway, the
Ministry of Education and Science of the Russian Federation, 
Ministerio de Econom\'{\i}a y Competitividad (Spain), the
Science and Technology Facilities Council (United Kingdom),
and the Binational Science Foundation (U.S.-Israel).
Individuals have received support from 
the Marie-Curie IEF program (European Union) and the A. P. Sloan Foundation (USA). 
